\documentclass[longauth]{aa} 
\usepackage{graphicx}
\usepackage{lscape}
\usepackage{booktabs}
\usepackage{longtable}
\usepackage{multicol}
\usepackage{multirow}
\usepackage{txfonts}
\usepackage{natbib}
\usepackage{subfigure}
\usepackage{amsmath}
\usepackage{color}
\usepackage{url}
\usepackage{ifthen}
\usepackage[pdfpagemode=UseThumbs,colorlinks=true,bookmarks=true]{hyperref}
\hypersetup{citecolor={blue},linkcolor={blue},urlcolor={blue}}
\bibpunct{(}{)}{;}{a}{}{,}

\begin{document} 

    \title{Tracing kinematic (mis)alignments in CALIFA merging galaxies}
        \subtitle{Stellar and ionized gas kinematic orientations at every merger stage}

\titlerunning{Stellar and ionized gas kinematic at every merger stage}
   \author{J.K.~Barrera-Ballesteros \inst{\ref{iac},\ref{ull}}\fnmsep\thanks{\email{jkbb@iac.es}}
          \and
          B.~Garc\'\i a-Lorenzo\inst{\ref{iac},\ref{ull}}
          \and
                  J.~Falc\'{o}n-Barroso\inst{\ref{iac},\ref{ull}}
                  \and
                  G.~van~de~Ven\inst{\ref{mpia}}
                  \and
                  M.~Lyubenova\inst{\ref{mpia},\ref{gron}}
                  \and
                  V.~Wild\inst{\ref{StAnd}} 
                  \and
          J.~M\'{e}ndez-Abreu\inst{\ref{StAnd}} 
                  \and
                  S.\,F.~S\'anchez\inst{\ref{unam}} 
              \and
          I.~Marquez\inst{\ref{iaa}}
          \and
          J.~Masegosa\inst{\ref{iaa}}
          \and
                  A.~Monreal-Ibero \inst{\ref{aip}, \ref{CNRS}}
                  \and
          B.~Ziegler\inst{\ref{viena}}
          \and
          A.~del~Olmo\inst{\ref{iaa}}
          \and
          L.~Verdes-Montenegro\inst{\ref{iaa}}
          \and
          R.~Garc\'\i a-Benito \inst{\ref{iaa}}
          \and
          B.~Husemann\inst{\ref{eso},\ref{aip}}
          \and
          D.~Mast \inst{\ref{ICRA}}
          \and 
                  C.~Kehrig\inst{\ref{iaa}}
          \and
          J.~Iglesias-Paramo\inst{\ref{iaa},\ref{almeria}}
          \and 
          R.\,A.~Marino\inst{\ref{ucm}}
          \and
          J.~A.~L.~Aguerri\inst{\ref{iac},\ref{ull}}
          \and
          C.\,J.\,~Walcher\inst{\ref{aip}}
          \and
                  J.~M.~V\'ilchez\inst{\ref{iaa}}
          \and          
          D.~J.~Bomans\inst{\ref{boc1},\ref{boc2}}
                  \and 
                  C.~Cortijo-Ferrero\inst{\ref{iaa}}
                  \and
                  R.~M.~Gonz\'alez~Delgado\inst{\ref{iaa}} 
                  \and
          J.~Bland-Hawthorn\inst{\ref{SIA}}
          \and
          D.~H.~McIntosh\inst{\ref{kansas}}
          \and
                         Simona Bekerait\.{e}\inst{\ref{aip}} 
                         \and
the  CALIFA Collaboration
          }
          
\institute{
\label{iac}Instituto de Astrof\'\i sica de Canarias (IAC), E-38205 La Laguna, Tenerife, Spain 
       \and
\label{ull}Depto. Astrof\'\i sica, Universidad de La Laguna (ULL), E-38206 La Laguna, Tenerife, Spain
\and
\label{mpia}Max-Planck-Institut f\"ur Astronomie, Heidelberg, Germany. 
\and
\label{gron} Kapteyn Astronomical Institute, University of Groningnen, Landleven 12, 9747 AD Groningen, the Netherlands
\and
\label{StAnd}School of Physics and Astronomy, University of St Andrews, North Haugh, St Andrews, KY16 9SS, U.K. (SUPA)
\and
\label{unam}Instituto de Astronom\'\i a,Universidad Nacional Auton\'oma de Mexico, A.P. 70-264, 04510, M\'exico,D.F.
\and
\label{iaa}Instituto de Astrof\'{\i}sica de Andaluc\'{\i}a (CSIC), Glorieta de la Astronom\'\i a s/n, Aptdo. 3004, E18080-Granada, Spain.
\and
\label{aip} Leibniz-Institut f\"ur Astrophysik Potsdam (AIP), An der Sternwarte 16, D-14482 Potsdam, Germany
\and
\label{CNRS} GEPI Observatoire de Paris, CNRS, Université Paris Diderot, Place
Jules Janssen, 92190 Meudon, France 
\and
\label{viena} University of Vienna, T\"urkenschanzstrasse 17, 1180 Vienna, Austria.
\and
\label{eso} European Southern Observatory (ESO), Karl-Schwarzschild-Str. 2, D-85748 Garching b. Muenchen, Germany
\and
\label{ICRA} Instituto de Cosmologia, Relatividade e Astrof\'{i}sica – ICRA, Centro Brasileiro de Pesquisas F\'{i}sicas, Rua Dr.Xavier Sigaud 150, CEP 22290-180, Rio de Janeiro, RJ, Brazil
\and
\label{almeria}Estaci\'{o}n Experimental de Zonas Aridas (CSIC), Ctra. de Sacramento s/n, La Ca\~{n}ada, Almer\'{\i}a, Spain.
\and
\label{ucm}CEI Campus Moncloa, UCM-UPM, Departamento de Astrof\'{i}sica y CC$.$ de la Atm\'{o}sfera, Facultad de CC$.$ F\'{i}sicas, Universidad Complutense de Madrid, Avda.\,Complutense s/n, 28040 Madrid, Spain.
   \and
   Astronomical Institute of the Ruhr-University Bochum, Universit\"atsstr. 150, 44580 Bochum, Germany \label{boc1}
   \and
   RUB Research Department 'Plasmas with Complex Interactions', Universit\"atsstr. 150, 44580 Bochum, Germany  \label{boc2}
         \and
         Sydney Institute for Astronomy, School of Physics A28, University of Sydney, NSW 2006, Australia  \label{SIA}
         \and
         Department of Physics \& Astronomy, University of Missouri-Kansas City, 5110 Rockhill Road, Kansas City, MO, USA \label{kansas}
}          
          
  \abstract
   {We present spatially resolved stellar and/or ionized gas kinematic properties for a sample of 103 interacting galaxies, tracing all merger stages: close companions, pairs with morphological signatures of interaction, and coalesced merger remnants. In order to distinguish kinematic properties caused by a merger event from those driven by internal processes, we compare our galaxies with a control sample of 80 non-interacting galaxies. We measure for both the stellar and the ionized gas components the major (projected) kinematic position angles (PA$_{\mathrm{kin}}$, approaching and receding) directly from the velocity distributions with no assumptions on the internal motions. This method also allow us to derive the deviations of the kinematic PAs from a straight line ($\delta$PA$_{\mathrm{kin}}$).
We find that around half of the interacting objects show morpho-kinematic PA misalignments that cannot be found in the control sample. In particular, we observe those misalignments in galaxies with morphological signatures of interaction.
On the other hand, the level of alignment between the approaching and receding sides for both samples is similar, with most of the galaxies displaying small misalignments.
Radial deviations of the kinematic PA orientation from a straight line in the stellar component measured by $\delta$PA$_{\mathrm{kin}}$ are large for both samples.
However, for a large fraction of interacting galaxies the ionized gas $\delta$PA$_{\mathrm{kin}}$ is larger than the typical values derived from isolated galaxies (48\%), indicating that this parameter is a good indicator with which to trace the impact of interaction and mergers in the internal motions of galaxies.
By comparing the stellar and ionized gas kinematic PA, we find that 42\% (28/66) of the interacting galaxies have misalignments larger than 16$^{\circ}$, compared to 10\% from the control sample.
Our results show the impact of interactions in the motion of stellar and ionized gas as well as the wide the variety of their spatially resolved kinematic distributions. This study also provides a local Universe benchmark for kinematic studies in merging galaxies at high redshift.
}
\keywords{galaxies: evolution - galaxies: kinematics - galaxies: interactions - galaxies: statistics}

   \maketitle
\section{Introduction}
\label{sec:Introduction}

Galaxy mergers play a crucial role in the (co-)evolution of stars and ionized gas in galaxies,  in particular their kinematics. From a morphological perspective only, the single-merger event is associated with close pairs, companions with clear signatures of interaction, and/or galaxies with signatures of a previous merger. Since gravitational forces shape the morphology of the galaxies as the merger evolves, kinematic departures from a pure rotational disk are expected in these systems. 

Over the past 50 years, the understanding of the galactic merger process has evolved tremendously. Early studies were based on observations of galaxies with peculiar features \citep{1966ApJS...14....1A}, and the seminal work by \citet{Tom_72} attempted to understand those peculiar features with rudimentary computational power. Nowadays, computational capability and a better understanding of the interstellar medium have provided us with a clearer understanding of the evolution of different galactic components across the merger event \citep[e.g., ][]{2005MNRAS.364.1105S,  2006ApJS..163....1H, 2007A&A...468...61D, 2008ApJS..175..390H, 2009AJ....137.3071B, 2011MNRAS.413.2860B}. Even though it is possible to study the kinematic properties of the entire merging sequence, numerical simulations in general focus on a given stage of the merger event  \citep[e.g., ][]{ 2003ApJ...597..893N, 2006ApJ...650..791C, 2009MNRAS.397.1202J, 2010ApJ...715L..88K, 2013AAS...22140501P}. This was the case for comparison with the available heterogeneous set of observational data. Studies of simulated line-of-sight velocity distributions or 2D velocity fields are becoming standard in the study of the velocity field distortions induced by merging \citep[e.g., ][]{2007A&A...473..761K, 2007MNRAS.376..997J, 2012arXiv1201.0885B, 2012arXiv1201.5918P}.

On the observational side, spatially resolved kinematics of merging galaxies have been analyzed in high- and low-redshift regimes. At high redshifts (z $\gtrsim$ 0.5), several surveys have studied the ionized gas velocity fields of star-forming galaxies, most of them showing strong interaction signatures \citep[e.g.,][]{2006A&A...455..107F,2006MNRAS.368.1631S,2008ApJ...682..231S,2009ApJ...706.1364F,2009ApJ...697.2057L,2011A&A...528A..88G,2011ApJ...743...86M,2012A&A...539A..91C}. These surveys find that a large fraction of galaxies are rotating disks \citep[$\gtrsim$ 30\%; for a review see][]{2013PASA...30...56G}. In the nearby Universe, several integral field unit (IFU) studies are devoted to analyzing in detail individual systems in a given interaction stage \citep[e.g.,][]{2010A&A...524A..56E,2012MNRAS.425L..46A,2014A&A...567A.132W}. 
Some merging objects have been included in large observational samples devoted to studying specific science cases for local galaxies. For instance  the stellar kinematics in  early type galaxies, \citep[e.g., NGC~5953 included in the SAURON project][]{2006MNRAS.369..529F} or ionized gas kinematics for spiral or irregular galaxies \citep[GHASP,][]{2008MNRAS.388..500E}. More recently, a significant number of studies focus on the characterization of the ionized gas velocity fields of  (Ultra)-Luminous IR Galaxies ([U]LIRGs) with  morphological tidal features or merging remnants \citep[e.g., ][]{2005ApJ...621..725C,2006ApJ...651..835D, 2008A&A...479..687A, 2012A&A...546A..64P}. For a large sample of these galaxies \cite{2013A&A...557A..59B} showed that their ionized gas velocity fields are dominated by rotation. Despite these efforts,  these studies usually lack  the comparison between the stellar and ionized gas components at different stages of the merging event. Since these components could react differently along the merger, it is crucial to understand how the kinematic of these components evolve during a galactic encounter. To distinguish the possible distortions or imprints due to secular processes from those related to the merger, a homogeneous control sample is also required.
%
\begin{figure*}
\begin{centering}
\includegraphics[width=15cm,angle=0,clip=true]{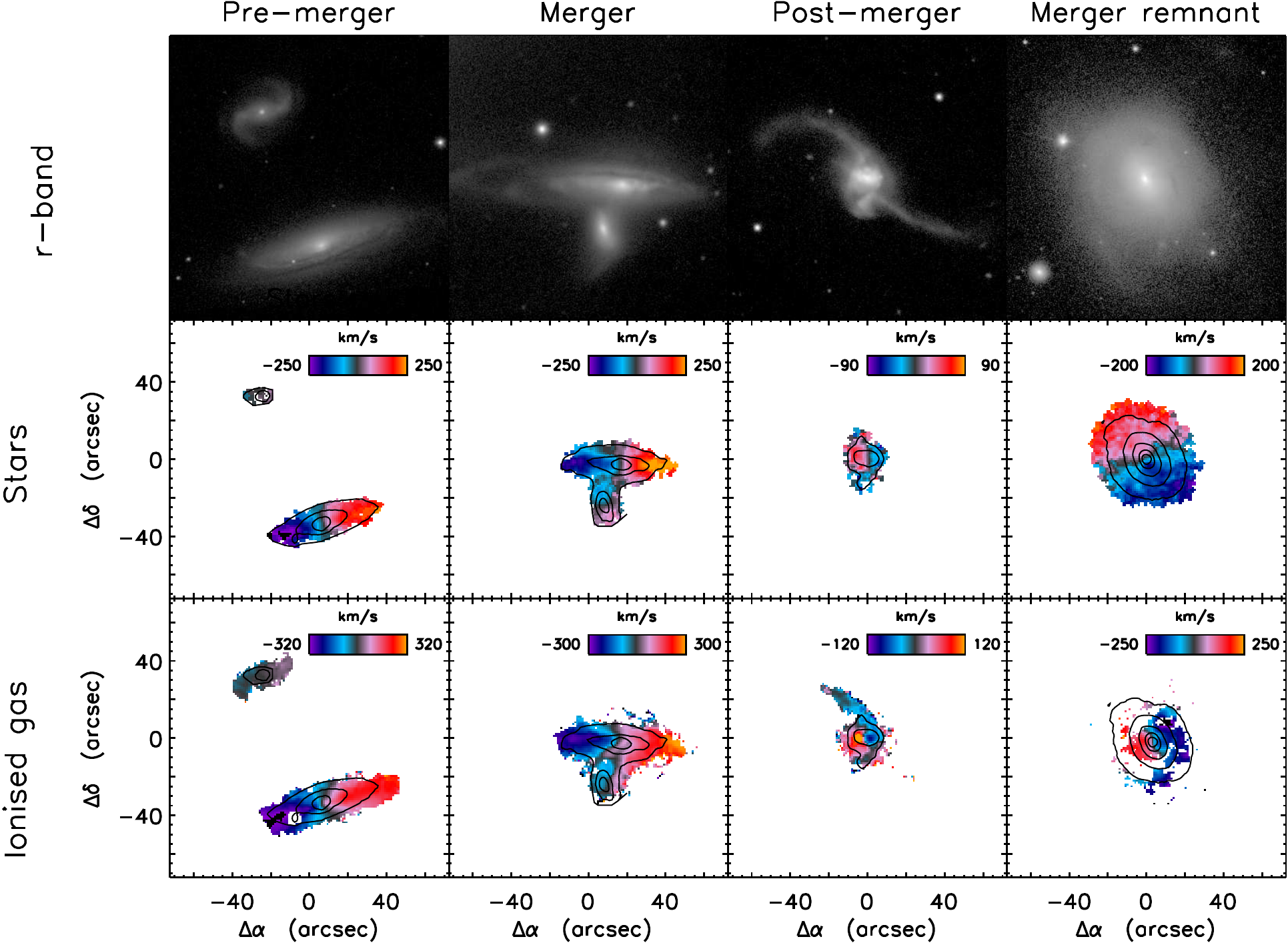}
\caption{\label{inter_stage} Example of the evolutionary scheme for interacting and merging galaxies described in Sec.\, \ref{sec:stage}. Top: SDSS r-band images of galaxies included in the CALIFA survey, from left to right: pre-merger stage, well-defined separation between the galaxies, no evident interaction signatures (IC~0944 and KUG~1349+143); merger stage:  two defined nuclei with evident interaction features (NGC~169 and NGC~169A); post-merger stage, single extended nucleus in optical with prominent tidal features (NGC~2623); merger remnant stage, possible tidal debris (NGC~5739). Middle: stellar velocity fields. Bottom: H$\alpha$+[NII] velocity field for each galaxy example. Contours represent the continuum obtained from the data cubes. We note that for these objects the velocity field distributions are extracted from a single IFU data cube, except for the pairs of galaxies in the pre-merger stage, where each galaxy has its own IFU data cube. The size of each box is 2 arcmin. Top is north and left is east. }
\end{centering}
\end{figure*}   

The aim of this work is to analyze  the stellar and ionized gas velocity distributions as the merger event evolves by studying several galaxies at different stages of this event;  we also want to  compare kinematic properties of these galaxies with those derived from a set of non-interacting objects. An observational study requires both a  homogeneous method and analysis. Moreover, it has to be able to depict kinematics from stellar and ionized gas components not for a single merging stage, but across the entire merging sequence. The CALIFA survey \citep[][]{2012A&A...538A...8S} allows  this kind of study to be carried out. All the interacting/merging galaxies, as well as the non-interacting objects, are observed with the same observational setup. Owing to its generous wavelength range, we can extract 2D velocity maps from the stellar and ionized gas components. It also  includes a large variety of non-interacting objects allowing us to use them as a control sample. In a companion paper \citep[][hereafter Paper I]{2014A&A...568A..70B}, we characterize and study the kinematic properties of a sample of 80 isolated galaxies drawn from the CALIFA survey. We applied the same methods to characterize the velocity fields in both samples.

The layout of this article is as follows. In Section\,\ref{sec:Data_Methods} we describe the classification of our interacting sample. We also present a comparison of this sample with the CALIFA mother and control samples as well as the extraction of the velocity fields from IFU data and the methodology used to determine kinematic properties directly from the velocities distributions. In Section\,\ref{sec:Kin_MIs} we compare the different kinematic parameters derived from the interacting galaxies with those from the control sample, and we study these properties at different interaction stages. In Section\,\ref{sec:Disc} we discuss our results, while in Section \,\ref{sec:conclusions} we summarize our results and present the conclusions from this study. In Appendix A we present the tables with the kinematic properties, and in Appendix B we present the kinematic velocity distributions for the ionized gas and the stars. Throughout this article we use the \mbox{$H_0$ =  73 km sec$^{-1}$ Mpc $^{-1}$}. The values for stellar mass were derived using a Chabrier stellar initial mass function \citep[][for more details see \citealt{2014A&A...569A...1W}]{2003PASP..115..763C}.

\section{Sample and observations}
\label{sec:Data_Methods}
\subsection{Evolutionary scheme of galactic merging}
\label{sec:stage}
 Our aim is to study how the kinematics of the stellar and the ionized gas components evolve along the merging event using their principal kinematic orientations. Therefore, we need to set an evolutionary scheme for the merger. For this study we used a morphological classification introduced by \cite{2002ApJS..143..315V}. This morphological classification is based on n-body simulations of the merger of two spiral disk galaxies \citep[][and references therein]{1998PhDT.........1S}. It was introduced to analyze a sample of (U)LIRGs and it is still used for studies related to these objects \citep[e.g.,][]{2008A&A...479..687A, 2010ApJ...709..884Y, 2011ApJS..197...27H}. This classification cannot be used as a strict time sequence but rather as an indicator of the interaction/merging stage.
 
In Fig. \ref{inter_stage} we show an example of this evolutionary scheme using the r-band Sloan Digital Sky Survey (SDSS) images as well as the stellar and ionized gas velocity fields. The main features of each stage are highlighted as follows.  
In the \emph{pre-merger} stage the morphology in both galaxies of the pair remains unperturbed and separated. There is no clear evidence of tidal tails or any other visual feature of interaction. We note that this stage also includes systems where the galactic disks overlap in the line of sight (e.g., VV488 in Appendix \ref{sec:maps}). 
In the \emph{merger} stage two nuclei are identifiable and well-defined tidal tails and other merging structures like bridges or plumes are readily seen in the optical images. While the pure morphological data do not guarantee that nuclei will necessarily merge, the presence of these strong interacting features indicates an eventual merger. 
The \emph{post-merger} stage occurs after the nuclei have coalesced. These systems have prominent tidal features, but only one nucleus can be clearly seen in the optical images. Moreover, the nuclear emission can be extended and cut by dust lines.
\emph{Merger remnant} galaxies display faint tidal features or show evidence of previous interaction. Any possible tidal feature in these galaxies is close to the detection limit and the main brightness profile is given by a relaxed system such as the one for an elliptical galaxy.

Several remarks have to be taken into account for such evolutionary scheme. First, from numerical simulations, \cite{2008MNRAS.391.1137L} found similar trends in the evolution of the morphology of equal-mass gas rich mergers. However, they note that morphologies are especially disturbed at the first passage of the galaxies and right after the nuclei coalesced. Then, by visual inspection one can classify interacting galaxies after the first passage as a pair of galaxies prior to this event. Second, even though this scheme is drawn from numerical simulations of two-disk spiral galaxies, we also use it for galaxies of different morphology (e.g., early and late-type binary systems); we stress that the above a scheme has to be taken  as a broad indicator of the merger event. 
Third, the structure of the merger remnants can vary in a wide range of morphologies, depending primarily on  the mass-ratios and the gas fraction of the progenitors \citep[][]{2009ApJ...691.1168H,2009MNRAS.397..802H}. For modest gas fractions the merger remnants are likely to resamble spheroidal-like galaxies \citep[e.g.,][]{2006MNRAS.372..839N, 2013LNP...861..327D}, while for gas-rich equal-mass merger remnants their morphology can be similar to disk-like galaxies \citep{2008MNRAS.391.1137L}. We attempt to include both of these morphological types in the merger remnant stage (see Sect.\,\ref{sec:Sample}). 

By construction we are able to distinguish tidal features until a certain surface-brightness limit. In our case this limit is $\sim$ 25 mag arcsec$^{-2}$ (see Sec.\,\ref{sec:Sample}). With this limit we will focus mainly on a major merger event that likely occurs  at $\sim$ 2.5 Gyr ago \citep{2014A&A...566A..97J}. We note that the lack of deeper imaging for the morphological selection of our interacting sample can bias our selection,  in particular for the merger remnant stage for which we select objects with long-lasting tidal features. To detect remnants with weaker features deeper optical imaging is required \citep[e.g.,][]{2013ApJ...765...28A, 2015MNRAS.446..120D}. Although tidal-features can be caused by major mergers, they can also be the result of a minor merger event \cite[e.g.,][]{2013arXiv1312.1643E}. Numerical simulations as well as observational data suggest that morphological signatures like shells in early-type galaxies can be caused by minor mergers \cite[for instance in NGC~3923,][]{2013A&A...559A.110B}.
\subsection{Interacting sample}
\label{sec:Sample}
%
The sample presented in this work was selected from 256 objects observed until March 2013\footnote{it also includes two galaxies observed in June and October of the same year: \object{NGC 5623} and \object{NGC 7711}}. These galaxies are part of the CALIFA mother sample (hereafter CALIFA MS, see Sec.\,\ref{sec:Control}).   
We consider as objects in pairs those objects with companions within a projected distance of 160 kpc, systemic velocity difference smaller than 300\,km\,s$^{-1}$, and a difference in $r$-band magnitude smaller than 2 mag\footnote{except for the systems \object{Arp~178} (\object{NGC~5614}/\object{NGC~5615}) and \object{UGC~10695} where the difference in magnitude is on the order of 3 mag; however, signatures of interaction are evident.} (relative distances, systemic velocities, and magnitudes were taken from NED \footnote{NASA/IPAC Extragalactic database. http://ned.ipac.caltech.edu/}). The above parameters were taken as thresholds from previous galaxy-pair studies. The projected distance is motivated by pair studies suggesting tidal effects of a companion's properties such as star formation rates up to 150 kpc \citep[][]{2013MNRAS.433L..59P}. For the systemic velocity difference criteria, we use the threshold defined on large surveys of galaxy pairs \citep[e.g.,][]{2008AJ....135.1877E,2013MNRAS.435.3627E}. These criteria yield  66 
objects located in the pre-merger and merger stages. In 16 objects, more than one companion of the CALIFA object satisfies the above criteria (i.e., triplets). In these cases we selected as the companion either the closest or the brightest object to the CALIFA source. Since our primary aim is to study properties of galaxies during a major merger we avoid  selecting targets located in compact groups\footnote{\object{NGC\,4816}, \object{UGC\,12127}, \object{NGC\,7194}, \object{NGC\,3160}, \object{NGC\,7436B}, \object{NGC\,6338}, \object{NGC\,2513}, \object{NGC\,6173}, \object{NGC\,3158}, \object{NGC\,5485}, \object{NGC\,6166NED01}, 
\object{NGC\,4874}}. 

\begin{figure}[!htb]
\begin{centering}
\includegraphics[width=\linewidth, angle=0,clip=true]{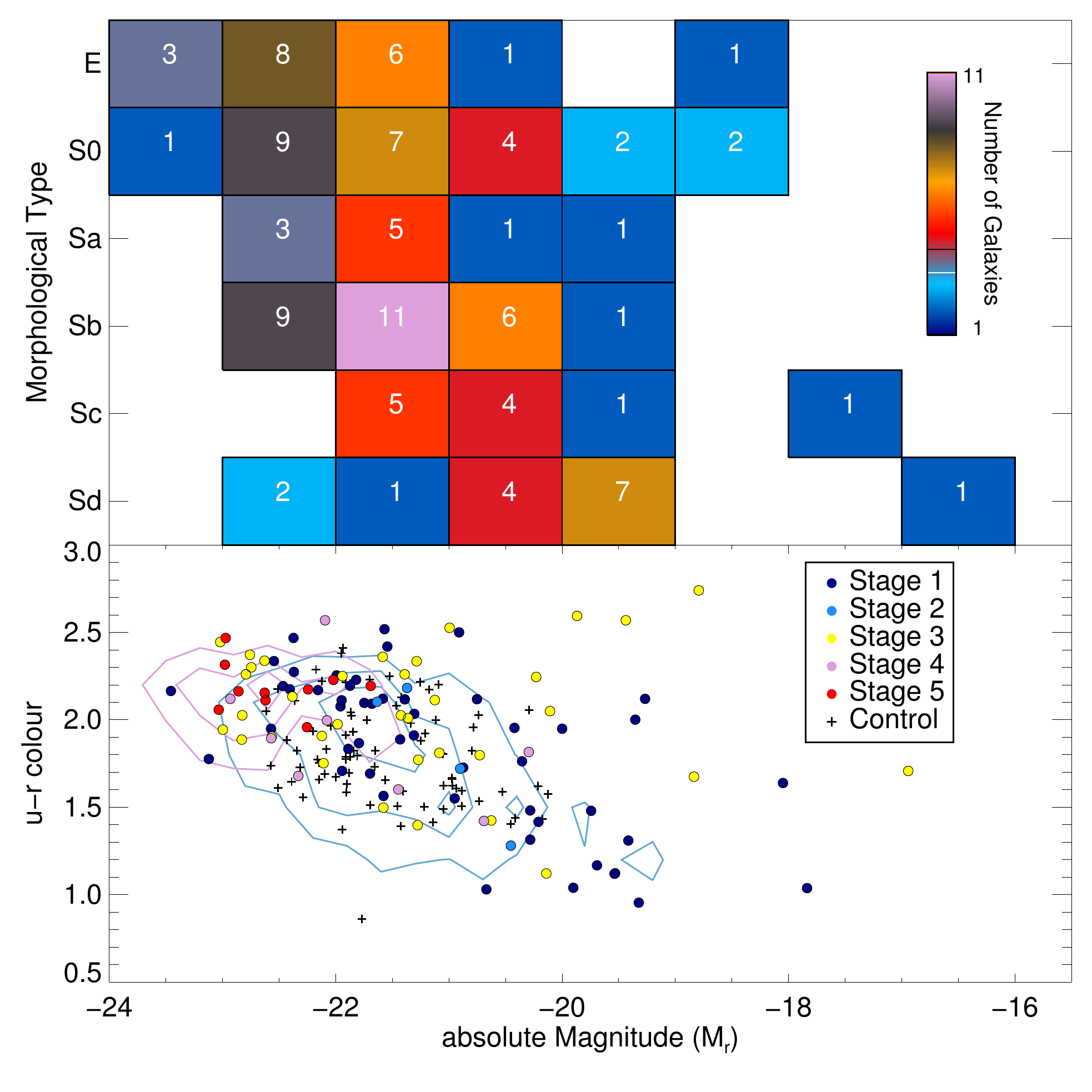}
\caption{\label{Histo2d} Top: Morphological type distribution against SDSS $r$-band absolute magnitude distribution of the interacting sample. The number in each bin represents the  sources included in it. Bottom: Color magnitude diagram for the CALIFA MS, control and interacting samples. Morphologies and absolute magnitudes are taken from \cite{2014A&A...569A...1W}.  Blue, yellow, pink, and red represent pre-merger, merger, post-merger, and merger remnants stages, respectively. Plus symbols represent the control sample. For comparison we plot the density contours of the CALIFA MS. Red and blue contours represent the early- and late-type galaxies. Each of the contours represents  25\%, 50\%, and 75\% of the mother sample. Dotted vertical lines represent the range in absolute magnitudes where the CALIFA MS is representative of the overall galaxy population \citep{2014A&A...569A...1W}.
}
\end{centering}
\end{figure}

From the objects in early interaction stages (pre- and merger) only 12 binary systems have both companions observed, in 6 of the pairs  because both companions are included in the CALIFA sample, and in the other 6 pairs the field of view (FoV) of the IFU is large enough to observe both galaxies. To alleviate this lack of complete pairs, we observed in a CALIFA complementary project those companions not included in the mother sample (P.Is J.K.\,Barrera-Ballesteros and \,G.\,van de Ven). For this study we include 13 additional companions observed and reduced with the same procedure as the CALIFA galaxies (see Sec.\,\ref{sec:Observations}). In total there are 19 additional objects. The final number of objects between the pre-merger and merger stages is 85 (see Table\,\ref{table_sum}). We note that despite our efforts to observe both companions of the binary system, some of them only have IFU data available for one galaxy in the system. 

From the remaining sample of galaxies (i.e., those objects with no close companions of similar brightness) we select the post-merger subsample. We used their SDSS r-band images to classify by visual inspection objects with tidal features like those described above in Sect.\,\ref{sec:stage}. In particular, for the post-merger stage we did not include ``normal'' irregular galaxies \footnote{\object{UGC\,4722} and \object{NGC\,7800}} \citep[i.e., those objects that, even though they are irregular, do not seem to be the result of a merger event, but rather clumpy disks, for  example see Fig.\,2 in][]{2013MNRAS.435.3627E}. We include in our sample the object NGC5623. Even though it does not present any signature of interaction or close companions, its kinematics in both components does not resemble the kinematic expected for a non-interacting object (see Paper I). It shows signatures of a stellar kinematic decouple core (KDC). The complete sample of interacting galaxies included in this study consists of 103 objects. In Table \ref{table_sum} we present the number of individual objects in each interaction stage, while in Table\,\ref{table_morph} we summarize the main properties of our interacting sample.

In the top panel of  Fig.\,\ref{Histo2d} we plot the distribution of the interacting sample according to their morphological type and SDSS $r$-band absolute magnitude. The sample covers a wide range of morphologies and luminosities. Morphologies were obtained by visual inspection from five members of the collaboration \cite[][]{2014A&A...569A...1W}. We note that they assigned regular morphological types to the sample of interacting galaxies. Pre-merger galaxies are expected to have a clear position in the hubble sequence. For merger objects with evident signatures of interaction, they assigned a Hubble type to the central part of the companions. Post-merger objects with distorted morphologies (e.g., NGC\,2623) were tagged as very late-type spirals (Sc-Sd) spanning a wide range of classifications from irregular to elliptical. We use the classification by \cite{2014A&A...569A...1W} and note that for these particular objects there is a large uncertainty in the morphological classification. Absolute magnitudes and stellar masses for the CALIFA objects are obtained using a growth curve analysis based on SDSS-DR7 $ugriz$ images \cite[see details in][]{2014A&A...569A...1W}. The estimation of these quantities for the CALIFA companions in binary systems were derived using a mask to obtain only the flux of the individual target. For companions not included in the CALIFA MS, absolute magnitudes and stellar masses were derived using Petrosian magnitudes from the SDSS-DR7 corrected to obtain similar magnitudes to those derived in the CALIFA MS \citep[see Fig.\,13 in][]{2014A&A...569A...1W}. 
%
\begin{table}[!htb]
\footnotesize
\caption{\label{table_sum}
Summary of the available kinematic data for the sample of interacting galaxies presented in this study. 
}
\begin{center}
\renewcommand{\thefootnote}{\alph{footnote}}
\begin{tabular} {c c c c c c }
\toprule
     &           & \multicolumn{4}{c}{Interaction Stage} \\
     &        & pre    &        & post  &         \\
     &  Total & merger & merger & merger & remnant \\
 \midrule
Individual objects     & 103 & 49 & 36 & 8  & 10  \\ 
Stellar Kinematics     &   85 & 36 & 32 &  7  & 10  \\ 
Ionized gas Kinematics &   82 & 37 & 28 &  8  &  9  \\ 
\bottomrule
\end{tabular}
\tablefoot{First row shows the number of individual objects in each interaction stage. 
The second and third rows present the available stellar and ionized gas velocity fields.
We include in this list the objects where it was not possible to measure the kinematic position angle, see details in Sec.\,\ref{sec:Analysis}
}
\end{center}
\end{table} 
%
\subsection{Parent and control samples}
\label{sec:Control}
\begin{figure}[!htb]
  \includegraphics[width=0.9\linewidth]{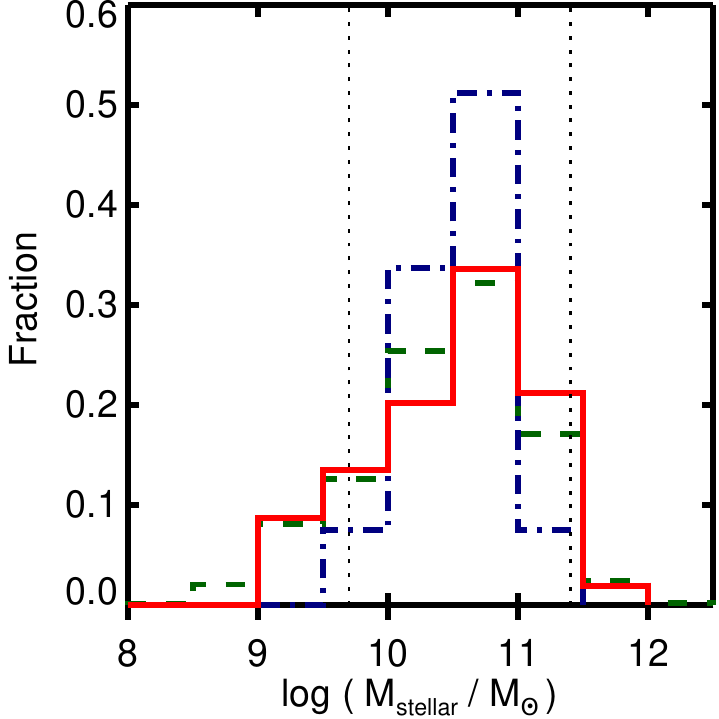}
  \includegraphics[width=0.9\linewidth]{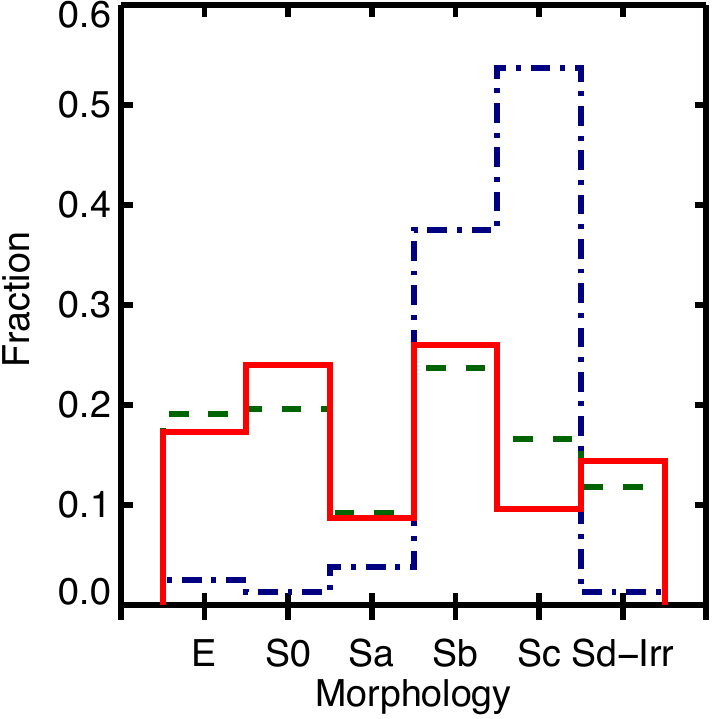}
  \caption{\label{Histo_control} Comparison between CALIFA MS (green dashed lines), control sample (blue dash-dotted lines), and interacting sample  (red solid lines). Top: Stellar mass distributions for these samples. Stellar masses for each galaxy is taken from \cite{2014A&A...569A...1W}. Vertical lines represent the stellar mass range where CALIFA MS is considered to be representative of the nearby galaxy population. Bottom:  Morphological distributions of the three samples.}
  \end{figure}
In order to distinguish possible disturbances in the velocity fields due to internal evolution rather than the merging process, we analyzed in Paper I the stellar and ionized gas velocity fields for 80 non-interacting galaxies (hereafter control sample, see Paper I for details on the selection criteria). 
Both the control and interacting samples are drawn from the CALIFA MS. It includes 939 galaxies, selected from the SDSS DR7 \citep{2009ApJS..182..543A}. The selection criteria is such that galaxies included in the CALIFA MS are in the nearby Universe with redshifts  0.005\,$<z<$\,0.03 and isophotal diameter in the SDSS $r$-band of 45$^{\prime\prime}$\,$\lesssim D_{25}\lesssim$\,80$^{\prime\prime}$. \cite{2014A&A...569A...1W} found that the CALIFA MS is representative of the general galaxy population within the following limits: -19.0 to -23.1 in r-band absolute magnitude, 1.7 to 11.5 kpc in half light radii, and 9.7 to 11.4 in $\log$(M$\mathrm{_{stellar}}$/$ M_{\odot}$).

In the bottom panel of Fig.\,\ref{Histo2d} we compare the CALIFA MS, control, and interacting samples in the color-magnitude diagram. We also mark the range of absolute magnitudes where the CALIFA MS is representative of the galaxy population. A large fraction of the objects from the control and the interacting samples are within this range. In general, the galaxies at different interaction stages spread homogeneously across the color-magnitude diagram except for the late merger remnant. We note  that galaxies at this stage tend to lie in the red sequence rather than the blue cloud. The control sample also shares similar colors with the interacting sample.
 
In the top panel of Fig.\,\ref{Histo_control} we show the stellar mass distribution of the CALIFA MS, control, and interacting samples. Control and interacting distributions peak at similar stellar masses (M$\mathrm{_{stellar}} \sim $ 10$^{10.8}$ M$_{\odot}$). Most of the objects in both samples have their stellar masses within the range where the CALIFA MS is representative for the general galaxy population. However, a Kolmogorov-Schmirnoff test (hereafter KS-test) reveals that these distributions are not likely drawn from the same distribution ($p_{KS}$ = 0.005). Similarly, it is unlikely that control and CALIFA MS stellar mass distributions are drawn from the same distribution ($p_{KS}$ = 0.02). This could be explained by several factors. On the one hand, the observations of interacting galaxies not included in the CALIFA MS allow low-mass galaxies to be part of this sample. Excess in the high-mass tail of this sample is caused by early-type galaxies not present in the isolated sample. On the other hand, isolated galaxies usually display narrow distributions in their physical properties with respect to other samples in different environments \citep[e.g.,][]{2011A&A...532A.117E, 2012A&A...540A..47F}. The latter is also evident when we compare the morphological type of these three different samples (see bottom panel of Fig.\,\ref{Histo_control}). Morphological types of the control sample restricts basically towards late-type spiral galaxies with a clear lack of elliptical, lenticular, and irregular galaxies, while the CALIFA MS and interacting samples spread over a wide range of morphologies. We consider that larger samples, in particular for the isolated galaxies, would be required to probe the kinematic at low/high stellar mass regimes of the interacting sample. 
 \subsection{CALIFA observations and velocity fields extraction}
\label{sec:Observations}
Observations were carried out using the PPAK instrument at Calar Alto Observatory \citep{2005PASP..117..620R}. Its main component consists of 331 fibers of 2\farcs7 diameter each, concentrated in a single hexagon bundle covering a FoV of $74\arcsec\times64\arcsec$, with a filling factor of $\sim$60 \% . Three dithering pointings are taken for each object to reach a filling factor of 100\% across the entire FoV \cite[see details in][]{2013A&A...549A..87H}. The final data cube consists of more than 4000 spectra at a sampling of $1\arcsec\times1\arcsec$ per spaxel \cite[details in][]{2013A&A...549A..87H}. The median spatial resolution of the survey is 2\farcs5 \citep{2015A&A...576A.135G}. The CALIFA survey has two spectral setups, V500 and V1200, low and intermediate resolution, respectively. Objects included in this work have been observed in the V500 setup. It has a nominal resolution of $\lambda$/$\Delta\lambda$\,$\sim$\,850 at $\sim$5000\AA\ and its nominal wavelength range is 3745$-$7300\AA. However, the final data cube has a homogenized spectral resolution (FWHM) over the entire wavelength range of 6.0\AA\ and the wavelength sampling per spaxel is 2.0\AA. The total exposure time per pointing is fixed for all the observed objects to 45 min. The data reduction is performed by a pipeline designed specifically for the CALIFA survey. The reduction process is explained in detail by \cite{2012A&A...538A...8S} and improvements on this pipeline are presented by \cite{2013A&A...549A..87H}. 

For consistency we follow the same kinematic extraction used in the control sample (see details in Paper I). For the stellar kinematic extraction we selected spaxels with continuum S/N\,$>$\,3. To achieve a minimum S/N of 20, we used a Voronoi-binning scheme for optical IFU data implemented by \cite{2003MNRAS.342..345C}. To derive the stellar line-of-sight velocity maps, for each cube, a non-linear combination of a subset of stellar templates from the Indo-U.S. library \citep{2004ApJS..152..251V} is fit to each binned spectra using the penalized pixel-fitting method \citep[pPXF,][]{2004PASP..116..138C}.  Errors for each binned spectrum, determined via Monte Carlo simulations, range from 5 to 20\,km\,s$^{-1}$ for inner to outer spectra, respectively. Figure\,\ref{inter_stage} (middle panels) shows examples of stellar kinematic maps for galaxies at different interaction stages.

The ionized gas emission cube for each object was obtained by subtracting in each spaxel the stellar continuum spectra derived from the best stellar pPXF fitting in its corresponding binned spectrum. No binning was done for the ionized gas. We use a cross-correlation (CC) method to measure the  line-of-sight velocity of the ionized-gas \citep[see][for details on the method]{2013MNRAS.429.2903G}. Briefly, this method compares the spectrum in each spaxel with a template in the wavelength range that includes the H$\alpha$+[\ion{N}{ii}]~$\lambda\lambda$6548,6584 emission lines (6508$-$6623\AA). The template corresponds to a Gaussian model for each emission line in the given wavelength range, shifted to the systemic velocity reported in NED \footnote{this method  also assumes that the templates have a velocity dispersion equal to the instrumental resolution ($\sigma$ $\sim$ 90\,km\,s$^{-1}$ at the H$\alpha$ emission line)}. Figure\,\ref{inter_stage} (bottom panels) shows an example of an ionized-gas velocity maps. Estimated uncertainties in the location of the maximum of the CC function are $\sim$10\,km\,s$^{-1}$.

We avoid  deriving kinematic parameters in objects where the velocity range was smaller than 50 km s$^{-1}$, and the extension in any component was smaller than 5 arcsec. In Table\,\ref{table_sum}, we summarize the number of sources where we are able to determine stellar and ionized gas kinematic parameters.
Both the pPXF and the CC methods are able to determine velocity dispersion maps;  however, owing to the low spectral resolution of the V500 setup we did not attempt any further analysis of these maps. 

\subsection{Robust kinematic properties}
\label{sec:Analysis}

Stellar and ionized gas velocity fields are presented in Appendix \ref{sec:maps}. Several IFU studies find evidence that velocity fields of some interacting galaxies or merger remnants depart from a regular disk-like velocity field \citep[e.g., ][]{2013A&A...557A..59B,2012MNRAS.424..416W, 2005ApJ...621..725C}. In this study we  aim to characterize  the kinematics of the interacting and merging galaxies through parameters derived directly from their stellar and ionized gas velocity distributions. No assumption on the behavior of the galactic components is made. The method we follow to derive these kinematic properties is described in Paper I. In this section we highlight its main features. 

In order to determine the kinematic PA, first we determined the kinematic center. For a rotational velocity field, the kinematic center coincides with its gradient peak (GP), as well as the location of the optical nucleus (ON). To determine the GP we follow the method describe by \cite{2015A&A...573A..59G}. The location of the GP does not coincide with the ON in all of the galaxies. We use half the size of the fiber as a conservative threshold distance to determine an offset between these two locations (1\farcs35). This is slightly larger than the median spatial PSF size (see Sec.\,\ref{sec:Observations}). For the stellar component a small fraction of interacting galaxies, similar to the control sample (see Paper I), display an offset in their kinematic center (17/85). In the case of the ionized gas, a large portion of the interacting sample displays an offset between the GP and ON (30/82). This fraction is almost five times larger than in  the control sample. We note that the orientation of the kinematic PA in not strongly affected by the deviation either of the ON or GP as kinematic center. In particular, for the ionized gas velocity fields only 9 out of 82 galaxies present large differences in their kinematic orientation if one of the two positions is chosen. As kinematic center we choose either ON or GP, depending on the best symmetry of the rotational curve (see details in Paper I).  

We determine the major kinematic axis (hereafter PA$_{\mathrm{kin}}$) in the stellar and ionized gas velocity fields following the same method as explained in Paper I: PA$_{\mathrm{kin}}$ is determined as the average of the polar coordinates of the spaxels with the maximum and minimum projected line-of-sight velocities with respect to the kinematic center. This method  independently measures PA$_{\mathrm{kin}}$ for the receding and the approaching sides. The  standard deviation ($\delta$PA$_{\mathrm{kin}}$) of these coordinates measures their scatter around the straight line defined by PA$_{\mathrm{kin}}$. There are some velocity fields where it is not possible to determine any kinematic feature \footnote{NGC5394, IC1079, and UGC11958 for the stellar component; UGC10650, NGC3303NED01 for the ionized gas component; and NGC5615 for both components.}. For these objects the kinematic perturbations across the velocity fields are large enough to prevent any measurement of the kinematic PA. It can be the case that observations with better spatial and spectral resolution provide a determination of the kinematic PA using the above method. To account for such galaxies in our statistics, we assign them a difference of 10\,$^\circ$ more than the largest kinematic parameter under consideration. We include these objects in our analysis to represent those galaxies where measurements of the kinematic PA cannot be performed using the method and data presented in this study.

To compare the stellar and the ionized gas PA$_{\mathrm{kin}}$ we pick the common maximum radius where both components can be measured (see the  value of `r' in Tables \ref{table_Skin} and \ref{table_Gkin}). We note that for some strongly interacting systems the radius r covers each companion, rather than the entire binary system (e.g. \object{NGC~0169}/\object{NGC~0169A}, \object{NGC~4841A}/\object{NGC~4841B}). We estimate the uncertainties on the determination of PA$_{\mathrm{kin}}$ and $\delta$PA$_{\mathrm{kin}}$ via Monte Carlo simulations \citep{2015A&A...573A..59G}. Typical errors for both the stellar and ionized gas components for PA$_{\mathrm{kin}}$ and $\delta$PA$_{\mathrm{kin}}$ are $\sim$ 5\,$^\circ$ (see individual values in Tables \ref{table_Skin} and \ref{table_Gkin} for the stellar and the ionized gas, respectively). We determine the photometric orientation $\mathrm{PA_{morph}}$ of the galaxies by fitting an ellipse model using the standard task \textit{ellipse} of IRAF\footnote{IRAF is distributed by the Optical Astronomy Observatory, which is operated by the Association of Universities for research in Astronomy (AURA) under cooperative agreement with the National Science Foundation.} on the isophotes of the SDSS $r$-band image at the same distance were PA$_{\mathrm{kin}}$  are derived for both of the components.

In summary, we present three kinematic misalignments to trace the deviations of the velocity fields from  ordered motions: the absolute misalignment between the morphological PA and the kinematic orientation (for both stellar and ionized gas components), the comparison between the approaching and receding orientation and the comparison between the stellar and ionized gas PA$_{\mathrm{kin}}$:
\begin{eqnarray}
\Psi_{\mathrm{morph-kin}} &=& | \mathrm{PA_{morph}} -  \mathrm{PA_{kin}} | \\
\Psi_{\mathrm{kin-kin}}  &=&     |\, | \mathrm{PA_{kin,app}} -  \mathrm{PA_{kin,rec}} | - 180^\circ  | \\
\Psi_{\mathrm{gas-star}}  &=&   | \mathrm{PA_{gas}} -  \mathrm{PA_{star}} | 
.\end{eqnarray}
We note that for each galaxy, in each component, two values of  $\Psi_{\mathrm{morph-kin}}$ and  $\Psi_{\mathrm{gas-star}}$ are obtained (corresponding to the approaching and receding sides). For a given component in each galaxy, we report the largest misalignment between both kinematic sides for $\Psi_{\mathrm{morph-kin}}$ and $\Psi_{\mathrm{gas-star}}$. 
%
\begin{figure}[!htb]
 \minipage{0.24\textwidth}
 \includegraphics[width=\linewidth]{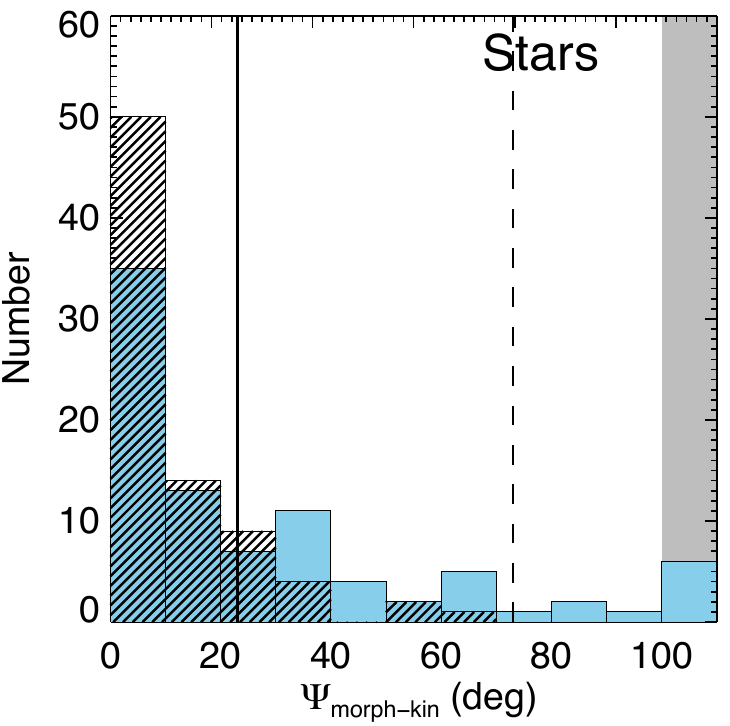}
  \endminipage
  \minipage{0.24\textwidth}
  \includegraphics[width=\linewidth]{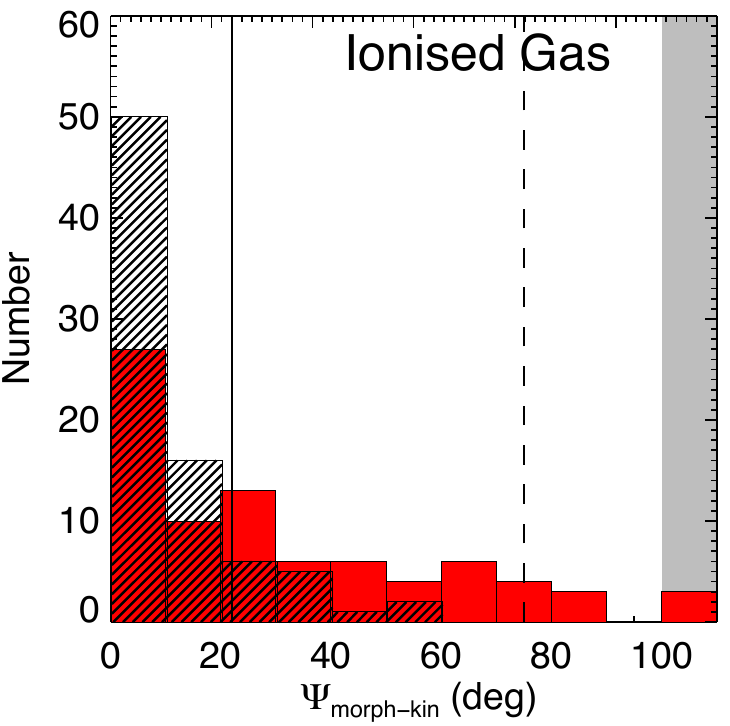}
   \endminipage
  \caption{\label{MK_hist}  Distribution of the misalignment between the photometric and kinematic position angle $\Psi_{\mathrm{morph-kin}}$ for the stellar (left, blue histogram) and the ionized gas (right, red histogram). In each plot line-filled histograms represent the distribution of $\Psi_{\mathrm{morph-kin}}$ for the control sample. Solid lines represent the misalignment which includes 90\% of the control sample. Dashed lines represent the same value for the interacting galaxies. The bins located in the gray areas in both components represent the misalignment for those galaxies where is not possible to determine PA$_{\mathrm{kin}}$}
  \end{figure}
\section{Kinematic (mis)alignments as tracers of  interaction and merger}
\label{sec:Kin_MIs}

Before  presenting the kinematic properties of the interacting sample and contrasting them with those from isolated galaxies, we want to point out the difference between these samples. Even though their absolute magnitudes, stellar masses, and colors are similar (see Figs.\,\ref{Histo2d} and \ref{Histo_control}), there is a clear lack of ellipticals and lenticular galaxies in the control sample compared with the interacting and CALIFA MS (see bottom panel Fig.\,\ref{Histo_control}). Nevertheless, kinematic parameters derived for ellipticals and lenticulars included in the control sample are within the typical values obtained for this sample (see Paper I). For the sake of comparison, we assume that trends found for the galaxies in the control sample are representative of the entire sample regardless their morphology. 
\subsection{Morpho-kinematic misalignments}
\label{sec:MK}  
In Fig.\,\ref{MK_hist} we present the distributions of $\Psi_{\mathrm{morph-kin}}$ for the stars (left panel) and the ionized gas (right panel). To include 90\% of the stellar morpho-kinematic PA misalignments in the control sample, we need to set a limiting value of  21\,$^\circ$. We find that 43\,\% (37/85) of the interacting sample has morpho-kinematic misalignments larger than 21\,$^\circ$ for this component. To include 90\% of the stellar morpho-kinematic misalignments in the interacting sample, we need to set a limit of 72\,$^\circ$. The interacting and control stellar morpho-kinematic misalignments are not likely to be drawn from the same parent distribution ($p_{KS}$ $<$ 0.001). The interacting sample shows a larger median $\Psi_{\mathrm{morph-kin}}$ misalignment and a wider distribution with respect to the control sample (see Table \ref{table:kin_summary}). 

Similar to the stellar component, to include 90\% of the ionized gas morpho-kinematic PA misalignments in the control sample, we need to set a limiting value of  22\,$^\circ$. We find that 52\,\% (43/82) of the interacting sample has morpho-kinematic misalignments larger than 22\,$^\circ$. To include 90\% of the ionized gas morpho-kinematic misalignments in the interacting sample, we need to set a limit of 75\,$^\circ$. As does the stellar component, ionized gas $\Psi_{\mathrm{morph-kin}}$ for the interacting sample shows a larger median misalignment and a wider distribution with respect to the control galaxies (see Table \ref{table:kin_summary}).

The variety of radial velocity distributions is vast. Some galaxies present kinematic PAs that clearly differ from the morphological ones. Although the scope of this study is to present statistical results of the kinematic (mis)alignments of a merging sample in comparison to a isolated sample, for the sake of illustration we highlight one object with  large morpho-kinematic misalignments, the early-type galaxy NGC~5623. This galaxy does not present a close companion, or any other readily observable signature of interaction from the images (we classify this object as a remnant); however, its stellar velocity distribution reveals a clear kinematically decoupled core (KDC). This KDC is aligned with the morphological PA of the galaxy. However, the ionized gas field (which extends $\sim$ 10 arcsec), has an orientation nearly perpendicular to the KDC orientation. We note that this is not the only case of a KDC in our sample \citep[e.g., NGC~5953][]{2006MNRAS.369..529F, 2007spts.conf..111F}. These cases are only an example of how different the kinematic structure of a galaxy can be from what is expected only from its morphology.

\begin{figure}[!htb]
\begin{center}
  \includegraphics[width=\linewidth]{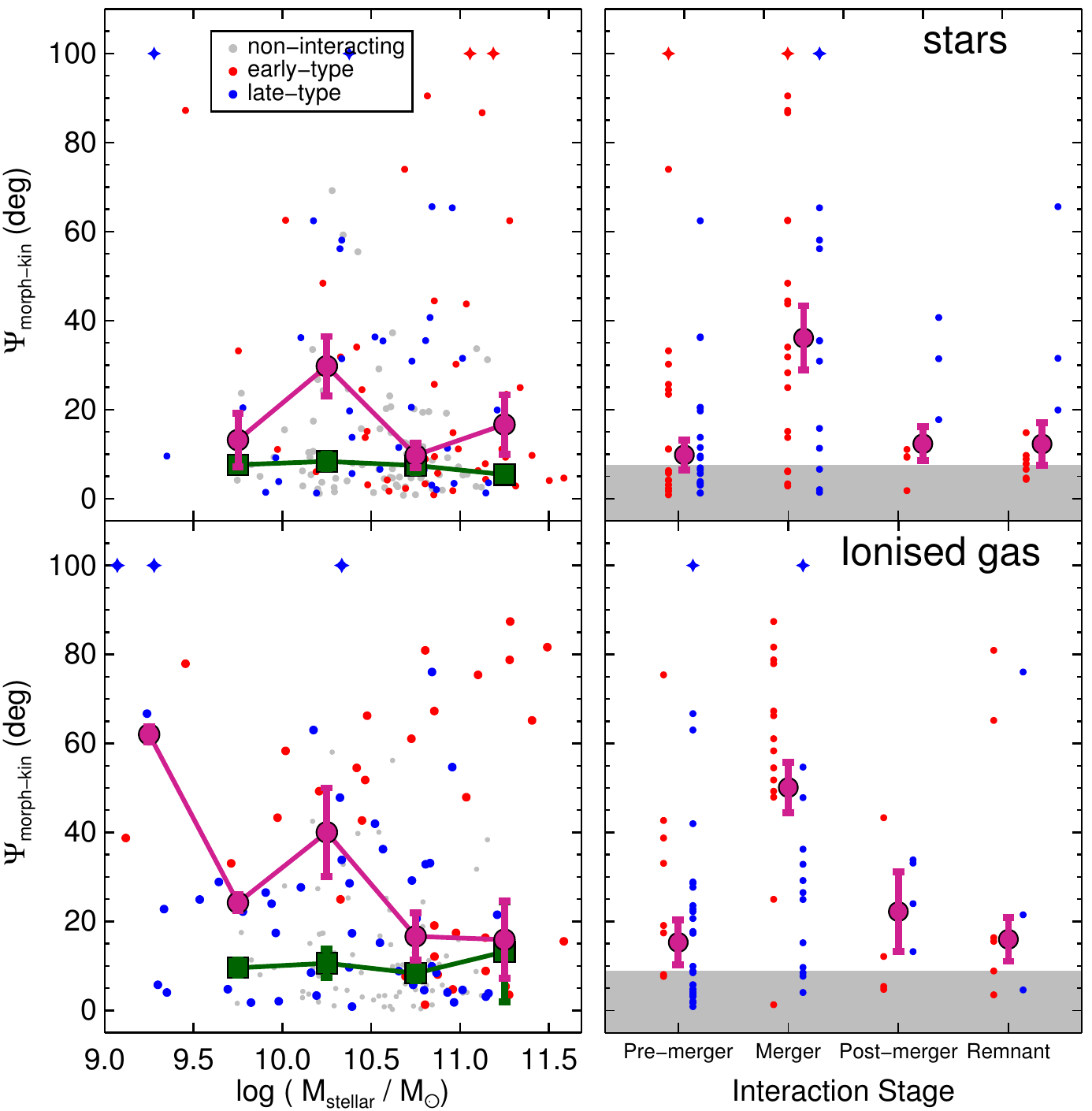}
  \caption{\label{MK_M_stage} Morpho-kinematic misalignment ($\Psi_{\mathrm{morph-kin}}$) for the stellar (top) and ionized gas (bottom) components against the stellar mass (left) and the interacting stage (right). In each of the panels, the red symbols represent early-type galaxies (i.e., E, S0, and Sa) and late-type are shown by blue symbols (i.e., Sb, Sc, Sd). Star symbols represent those objects where it was not possible to determine $\Psi_{\mathrm{morph-kin}}$. For the left panels, gray dots represent the control sample, and  green squares and pink circles represent the median of $\Psi_{\mathrm{morph-kin}}$ in mass bins of 0.5 in units of $\log \mathrm{(M_{stellar}/M_{\odot})}$ for the control and interacting samples, respectively. For the right panels, the pink circles represent the median of $\Psi_{\mathrm{morph-kin}}$ in each interaction stage bin. Gray regions show the median morpho-kinematic misalignment obtained from the control sample. The bar in each bin represents the error determined by bootstrapping method.} 
\end{center}
\end{figure}

In the left panels of Fig.\,\ref{MK_M_stage} we plot the control and interacting morpho-kinematic misalignments with respect to their stellar masses for the stellar (top panel) and ionized gas (bottom panel) components. As we already pointed out in Paper I, the morpho-kinematic misalignments in the control sample (for any component) do not seem to vary strongly at different stellar mass ranges. This is not the case for the interacting sample. For the stellar component,  we find an increment in the morpho-kinematic misalignment at the mass range of 10\,$< \log\mathrm{(M_{stellar}/M_{\odot})} <$\,10.5. The interacting and control sample in this mass range do not seem to be drawn from the same parent distribution (p$_{KS}$ $<$ 0.02). The explanation of this increment in the misalignment at this mass range is non-trivial. It can be due to the distribution of stellar masses of interacting galaxies. Most of the galaxies in this mass bin are pairs of galaxies. According to cosmological simulations, the stellar mass distribution function of galaxy pairs peaks in a similar mass range \citep{2013MNRAS.436.1765M}. We also note that a significant fraction of interacting galaxies included in our sample have their stellar mass within this mass bin (see Fig.\,\ref{Histo_control}). A larger sample of interacting galaxies at different mass bins would be required to confirm this possible difference.

At the same mass bin, the ionized gas $\Psi_{\mathrm{morph-kin}}$ misalignment is also shows an increment comparing with the misalignment observed in the control galaxies. Even more,  each stellar mass bin, the median ionized gas $\Psi_{\mathrm{morph-kin}}$ of the interacting sample is larger compared to the median derived from the control sample (except for the more massive bin). This suggests that ionized gas has a strong reaction to interactions and mergers. The largest ionized gas median misalignment is observed in the less massive bin. This could indicate that lighter companions are likely to show differences in their physical properties induced by their heavy companions \citep[e.g.,][]{2012MNRAS.425L..46A}. We stress that in this particular mass bin there are no galaxies in the control sample to compare with. A detailed study of low-mass interacting galaxies would be required to  test this scenario further.

We study  whether $\Psi_{\mathrm{morph-kin}}$ misalignments are related to a particular stage of the merger event. In the  right panels of Fig.\,\ref{MK_M_stage} we plot the morpho-kinematic misalignments of the interacting sample according to their interaction stages for both, the stellar (upper panel) and ionized gas (bottom panel) components. For the stellar component we find that the merger stage shows the largest median value (and wider distribution) of $\Psi_{\mathrm{morph-kin}}$ with respect to the other interaction stages. In fact, the distribution of the morpho-kinematic misalignments from this stage and the distribution of the control sample misalignments do not seem to be drawn from the same parent distribution ($p_{KS} \sim$ 0). 

Although the spread of $\Psi_{\mathrm{morph-kin}}$ in each stage is rather large, the median value per stage follows a trend across the merger event. Pre-mergers (i.e., pairs of galaxies) present median misalignments slightly larger than median $\Psi_{\mathrm{morph-kin}}$ from the control sample. In the merger stage, its median misalignment is clearly larger than the median from the control sample as well as the median from other interaction stage bins. Post-merger and remnant stages present slightly larger median values than the control sample. 

Ionized gas morpho-kinematic median misalignments at different interaction stages follow a similar trend to those derived from the stellar component with the difference that in each interaction bin the median misalignment from the ionized gas is larger than for the stellar component.The fact that in both components the median $\Psi_{\mathrm{morph-kin}}$ in all the interaction stages is systematically larger than the median misalignment for non-interacting galaxies indicates that interactions and mergers do affect the internal structure of galaxies. Even more, morpho-kinematic misalignments across different merging stages suggest that as the companions merge the velocity distribution of the components departs significantly from those distributions observed in the control sample. After coalescing, remnants tend to show slightly larger morpho-kinematic alignments than those found in non-interacting galaxies. We  also note that comparing $\Psi_{\mathrm{morph-kin}}$ misalignments between both components suggests that ionized gas reacts more easily to interactions and mergers than the stellar component.

Finally, we note that a significant fraction of objects with kinematic misalignment larger than the control sample are early-type galaxies (elliptical and lenticular galaxies). This became evident for the ionized gas in the merger stage (see red points in bottom right panel of Fig.\,\ref{MK_M_stage}). To identify whether this is a signature of the current interaction or a previous one in these galaxies, individual detailed studies are required in each interacting system which is beyond the scope of the present study.

\subsection{Internal Kinematic misalignments}
\label{sec:KK}

\begin{figure}[!htb]
  \minipage{0.24\textwidth}
  \includegraphics[width=\linewidth]{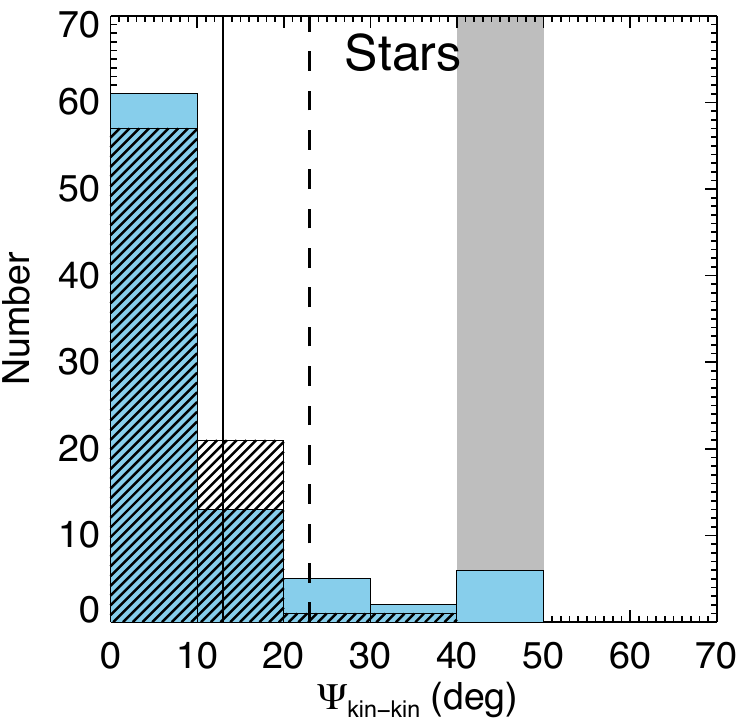}
    \endminipage
  \minipage{0.24\textwidth}
  \includegraphics[width=\linewidth]{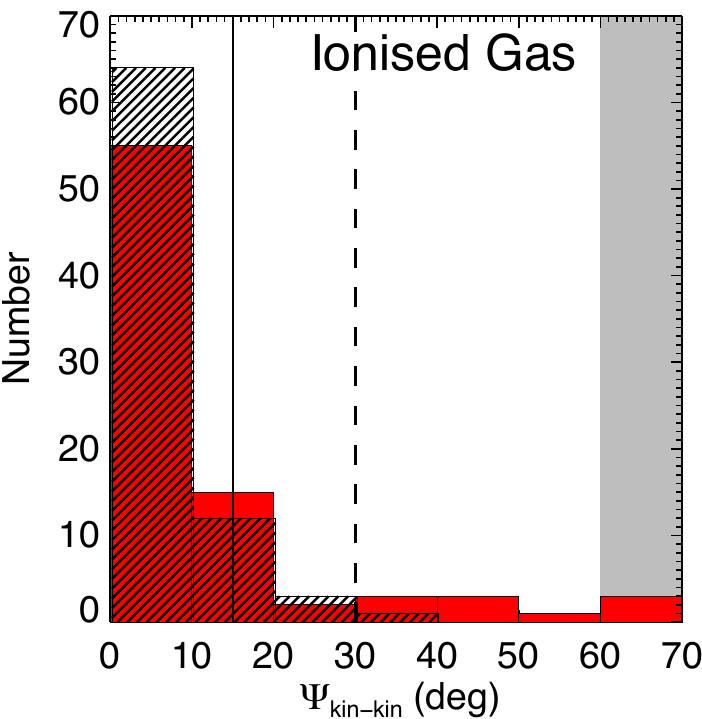}
    \endminipage
  \caption{\label{KK_Histo}Distribution of the kinematic misalignment between the receding and approaching position angles $\Psi_{\mathrm{kin-kin}}$ for the stellar (left, blue histogram) and the ionized gas (right, red histogram) components of the interacting sample. In each plot, line-filled histograms represent the distribution of $\Psi_{\mathrm{kin-kin}}$ for the control sample. Solid lines represent the misalignment which includes 90\% of the control sample. Dashed lines represent the same value for the interacting sample. The bins located in the gray areas represent those galaxies where is not possible to determine $\Psi_{\mathrm{kin-kin}}$.} 
\end{figure}

As we described in Sect.\,\ref{sec:Analysis}, we are able to determine independently the approaching and receding morpho-kinematic PA from the velocity field distributions. Any misalignment between these two sides may indicate possible departures from a pattern of ordered motions. In Fig.\,\ref{KK_Histo} we present for both samples the distributions of this misalignment ($\Psi_{\mathrm{kin-kin}}$) for the stellar (left panel) and ionized gas (right panel) components. To include 90\% of the stellar internal kinematic misalignments in the control sample, we need to set a limiting value of  13\,$^\circ$. We find that 21\,\% (18/85) of the interacting sample has internal kinematic misalignments larger than 13\,$^\circ$.
To include 90\% of the stellar internal kinematic misalignments in the interacting sample, we need to set a limit of 22\,$^\circ$. Medians and standard deviations are similar between these two samples (see Table \ref{table:kin_summary})

Control sample also have small  $\Psi_{\mathrm{kin-kin}}$ values for the ionized gas. To include 90\% of the ionized gas internal kinematic misalignments in the control sample, we need to set a limiting value of  15\,$^\circ$. The interacting sample covers a wider range of $\Psi_{\mathrm{kin-kin}}$ than the control sample in this component (see right panel in Fig.\,\ref{KK_Histo}). We find that 20\% (16/82) of the interacting sample has kinematic misalignments larger than 15\,$^\circ$. To include 90\% of the ionized gas internal kinematic misalignments in the interacting sample, we need to set a limit of 30\,$^\circ$. This suggests that ionized gas could react more easily to mergers than the stellar component (as we already pointed out in Sec.\,\ref{sec:MK}). However, we note that median and standard deviations for this kinematic parameter are similar for both samples (see Table \ref{table:kin_summary}).

These results indicate that interactions and mergers do have an impact on the internal kinematic alignment of galaxies, in particular for the ionized gas component. However, we note that the fraction and median values of these misalignments are smaller than those we find in the morpho-kinematic indicator (see Sec.\,\ref{sec:MK}). A KS-test suggests that control and interacting internal kinematic misalignments can share the same parent distribution.

\begin{figure}[!htb]
\begin{center}
  \includegraphics[width=\linewidth]{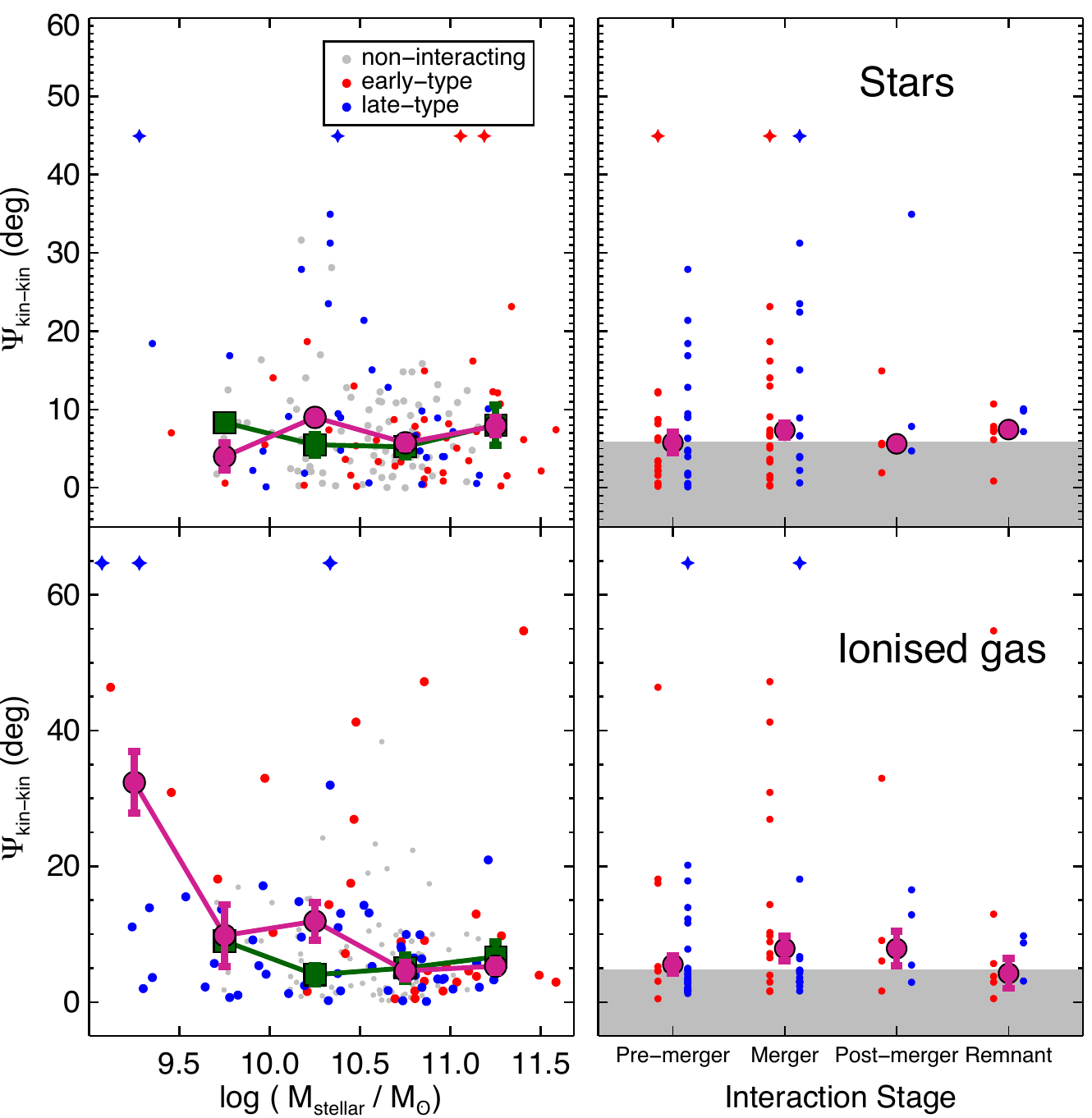}
  \caption{\label{KK_M} Internal kinematic misalignment ($\Psi_{\mathrm{kin-kin}}$) for the stellar (top) and ionized gas (bottom) components against the stellar mass (left) and interaction stage (right). As in Fig.\,\ref{MK_M_stage},  the red symbols represent early-type galaxies (i.e., E, S0, and Sa) and late-type are shown by blue symbols (i.e., Sb, Sc, Sd). Gray dots represent the control sample. For the left panels green squares and pink circles represent the median of $\Psi_{\mathrm{kin-kin}}$ in each mass bin for the control and interacting samples, respectively. For the right panels, the pink circles represent the median of $\Psi_{\mathrm{kin-kin}}$ in each interaction stage bin. The gray region shows the median internal kinematic misalignment obtained from the control sample.  The bar in each bin represents the error determined by bootstrapping method.} 
  \end{center}
 \end{figure}

We also study the dependence of internal kinematic misalignments with respect to the stellar mass. We find that interacting and control samples share similar median internal kinematic alignments in the stellar component at different mass bins (see top left in Fig.\,\ref{KK_M}). This also the case for the ionized gas, except for the stellar mass bin 9\,$< \log\mathrm{(M_{stellar}/M_{\odot})} <$\,9.5 (see bottom left panel in Fig.\,\ref{KK_M}). As does the morpho-kinematic misalignments, ionized gas in interacting galaxies at this mass bin seem to be more kinematically disturbed than more massive galaxies.
When we compare the median values of $\Psi_{\mathrm{kin-kin}}$ for different interaction stages (see right panels in Fig.\,\ref{KK_M}), we find that these misalignments are slightly larger than the median from the control sample. Our results indicate that the signatures of the merger are more subtle in this kinematic tracer than other such as the morpho-kinematic misalignment (see Sec.\,\ref{sec:MK}).

To illustrate the above scenario, we highlight here the well-studied merger remnant NGC~2623 \citep[e.g.,][]{2008ApJ...675L..69E}. Although its stellar velocity gradient is small ($\sim$ 50 km/s), this velocity field has defined approaching and receding sides within the radius we determine PA$_{\mathrm{kin}}$. However, these sides misalign to each other by $\Psi_{\mathrm{kin-kin}}$ $\sim$ 35\,$^\circ$. The orientation of the ionized gas velocity field is similar to the stellar field. It is worth  mentioning that IFU observations in the nuclear region ($\sim$ 600 pc) in the mid-infrared reveals that in the ionized gas (Br$\gamma$) the orientation of the nuclear disk is consistent with that found in our velocity distributions \citep[][]{2014ApJ...784...70M}.

These results indicate that motions of interacting galaxies, probed by the alignment between the receding and approaching kinematic sides, do not show significant differences from those observed in isolated galaxies. This holds for different interacting stages and a significant range of stellar masses. This  suggests that in those interacting systems showing highly distorted morphologies (e.g., post-merger galaxies) the internal kinematics misalignment may not be tracing the actual effect in the motion of the galactic components due to interactions and mergers. A dedicated set of numerical simulations would be required to determine whether the results obtained from this kinematic indicator are representative of the actual motions as the merger evolves. 

It is important to note that a small fraction of interacting galaxies displays small misalignments between their kinematic sides, although their velocity distributions do not resemble a symmetric velocity field. For instance, the interacting galaxy NGC~3991 displays an ionized gas velocity field where its sides seem to be aligned ($\Psi_{\mathrm{kin-kin}}$ = 15 $\pm$ 6\,$^\circ$). However, this object does not show the flattening in the velocity curve expected for a disk-like velocity field (see Appendix \ref{sec:maps}). Other galaxies also display velocity fields with such features: for the stellar component this is the case for Arp~220, NGC~3406NED01, and NGC~3406NED02;  for the ionized gas, NGC~3303, NGC~5394 and NGC~5614. In the following sections we will study other kinematic parameters to characterize as much as possible our sample of interacting galaxies.
%
\begin{table*}[!htb]
\footnotesize
\caption{\label{table:kin_summary}
Summary of the kinematic misalignments of the interacting galaxies compared with those derived from the control sample. 
}
\begin{center}
\renewcommand{\thefootnote}{\alph{footnote}}
\begin{tabular} {c c c c c c c c }
\toprule
 &     & \multicolumn{2}{c}{$\Psi_{\mathrm{morph-kin}}$ ($^\circ$)} & \multicolumn{2}{c}{$\Psi_{\mathrm{kin-kin}}$($^\circ$)}& \multicolumn{2}{c}{$\delta_{\mathrm{kin}}$($^\circ$)}\\
\cmidrule{3-8} 
&   & control & interacting & control & interacting & control & interacting  \\
\midrule 
\multirow{3}{*}{Stellar kinematics} & 90\% of sample & 21 & 72 &13 & 22 & 37 & 41 \\
                                            & median & 7  & 14 & 6 & 6  & 20 & 19  \\ 
                                & standard deviation & 8  & 21 & 5 & 6  & 9  & 9  \\ 
\cmidrule{2-8} 
\multirow{3}{*}{Ionized gas kinematics} & 90\% of sample & 22 & 75 & 15 & 30  & 18 & 42  \\
                                                & median & 8  & 23 &  5 & 6   & 10 & 18 \\ 
                                    & standard deviation & 8  & 28 &  5 & 6   & 4  & 13 \\ 

\bottomrule
\end{tabular}
\end{center}
\end{table*} 
%
\subsection{Kinematic PA deviations}
\label{sec:deltaK}

In addition to the measurement of PA$_{\mathrm{kin}}$, the spatially resolved data also allow us to measure how it changes   across the velocity field by means of its deviation at different radii ($\delta$PA$_{\mathrm{kin}}$, see Sect.\,\ref{sec:Analysis}). This parameter quantifies by how much locations of maximum velocities deviate from a straight line. In a rotational only velocity field this value is zero: all the positions are aligned with the line of nodes for both kinematic sides.

\begin{figure}[!htb]
  \minipage{0.24\textwidth}
  \includegraphics[width=\linewidth]{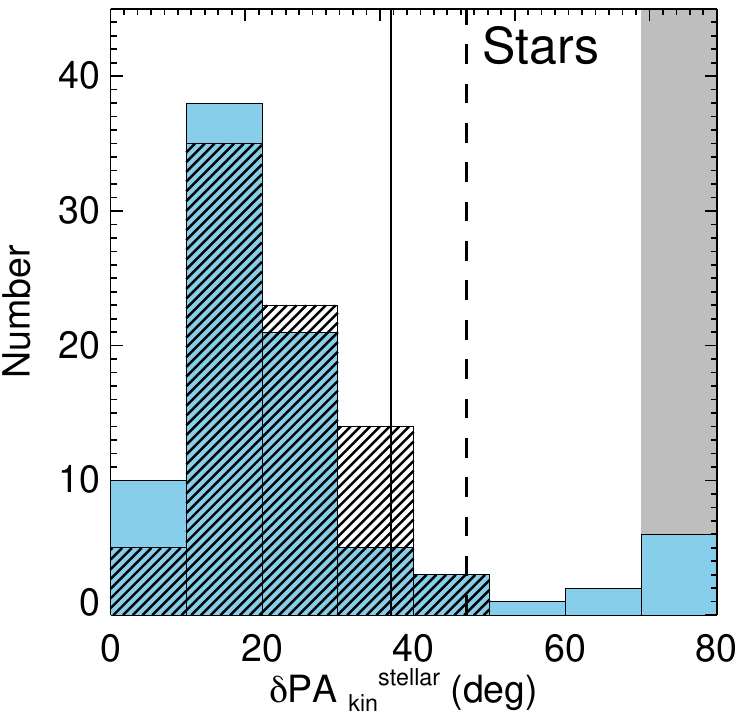}
 \endminipage
  \minipage{0.24\textwidth}
  \includegraphics[width=\linewidth]{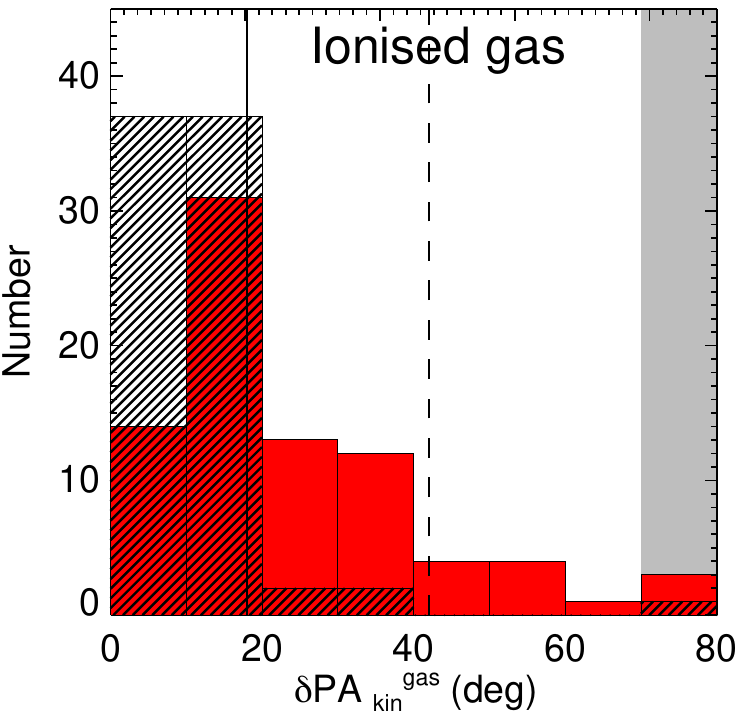}
  \endminipage
  \caption{\label{deltaK_Histo}Distribution of the deviation of the kinematic PA from a straight line ($\delta$PA$_{\mathrm{kin}}$) for the stellar (left, blue histogram) and the ionized gas components (right, red histogram). Similar to Figs.\,\ref{MK_hist} and \ref{KK_Histo}, in each plot line-filled histograms represent the distribution of $\delta$PA$_{\mathrm{kin}}$ for the control sample. Solid lines represent the misalignment which includes 90\% of the control sample. Dashed lines represent the same value for the interacting galaxies. The bins located in the gray areas represent those galaxies where is not possible to determine $\delta$PA$_{\mathrm{kin}}$.} 
\end{figure}
In Fig.\,\ref{deltaK_Histo} we present for both samples the $\delta$PA$_{\mathrm{kin}}$ distributions for the stellar (left panel) and the ionized gas (right panel) components. To include 90\% of the stellar kinematic PA deviations in the control sample, we need to set a limiting value of 37\,$^\circ$. We find that 15\% (11/85) of the interacting sample has stellar $\delta$PA$_{\mathrm{kin}}$ deviations larger than 37\,$^\circ$. To include 90\% of the stellar kinematic PA deviations in the interacting sample, we need to set a limiting value of 41\,$^\circ$. The control and interacting $\delta$PA$_{\mathrm{kin}}$ distributions are likely to be drawn from the same parent distribution ($p_{KS} =$ 0.7). The similarities in both distributions suggests that processes responsible for deviating kinematic axes from a straight line are similar in isolated and interacting objects (see medians and standard deviations for each sample in Table \ref{table:kin_summary}).

For the control sample, ionized gas kinematic PA deviations are smaller than stellar ones. To include 90\% of the ionized gas kinematic PA deviations in the control sample, we need to set a limiting value of 18\,$^\circ$. The interacting sample covers a wider range of ionized gas kinematic PA deviations than the control sample (see right panel of Fig.\,\ref{deltaK_Histo} and Table \ref{table:kin_summary}). We find that 48\% (41/82) of this sample has $\delta$PA$_{\mathrm{kin}}$ deviations larger than 18\,$^\circ$. To include 90\% of the ionized gas kinematic PA deviations in the interacting sample, we need to set a limiting value of 42\,$^\circ$. Contrary to the stellar component, the interacting and control $\delta$PA$_{\mathrm{kin}}$ distributions for the ionized gas are not drawn from the same parent distribution ($p_{KS} \sim$ 0). The kinematic PA deviations for both components support the results from Sec.\,\ref{sec:MK}: interactions and mergers have a larger impact on the ionized gas than on the stellar component.

\begin{figure}[!htb]
\begin{center}
  \includegraphics[width=\linewidth]{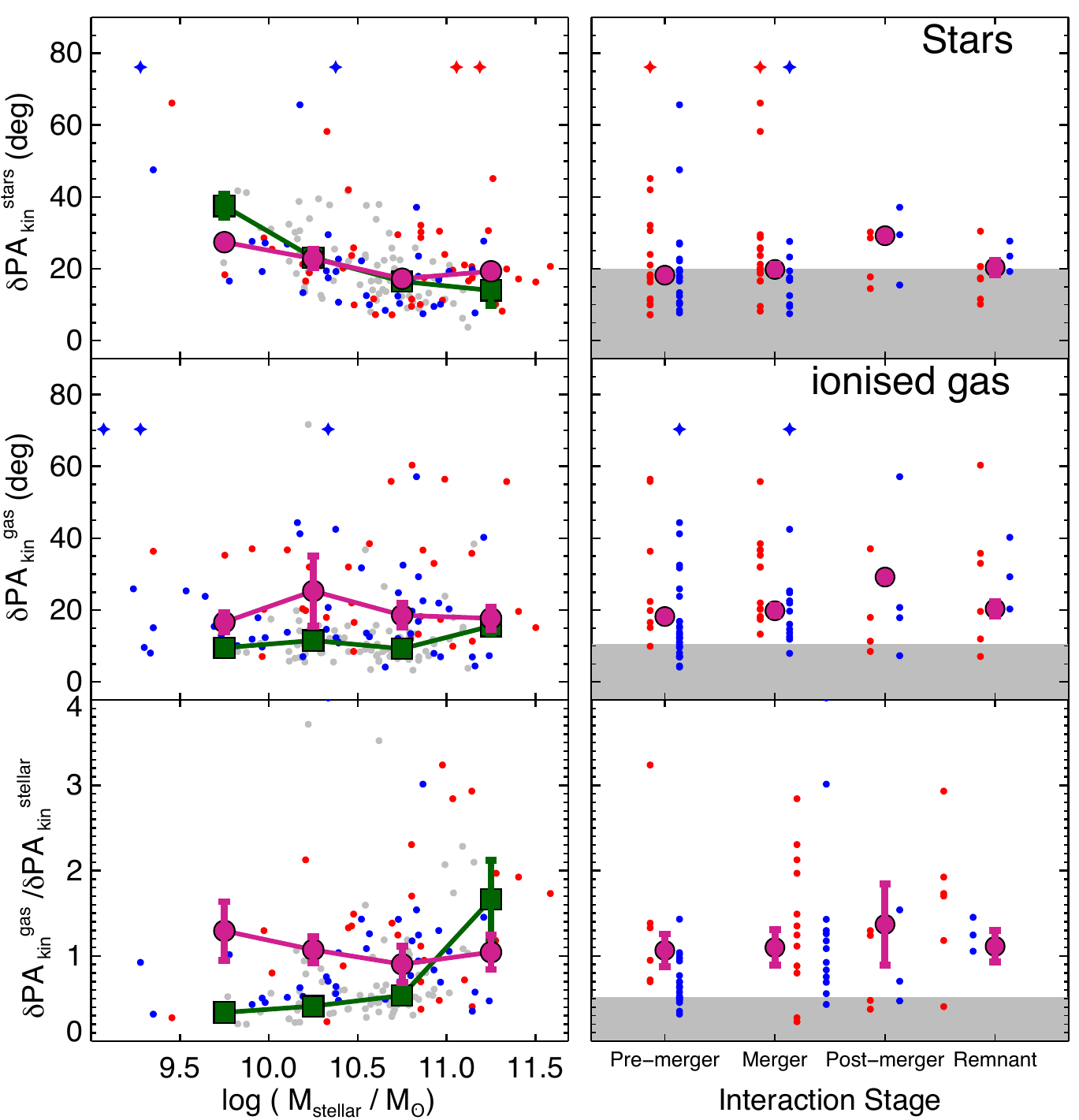}
  \caption{\label{deltaK_MS} Top panels: Kinematic PA deviation from a straight line ($\delta$PA$_{\mathrm{kin}}$) for both the stellar and ionized gas components with respect to the stellar mass (left panels) and interaction stage (right panels). Bottom panels: Ratio between the kinematic PA deviation of the stellar and the ionized gas component. As in Fig.\,\ref{MK_M_stage}, the red symbols represent early-type galaxies (i.e., E, S0, and Sa) and late-type are shown by blue symbols (i.e., Sb, Sc, Sd). Gray dots represent the control sample. For the left panels green squares and pink circles represent the median of $\delta$PA$_{\mathrm{kin}}$ in each mass bin for the control and interacting samples, respectively. For the right panels, the pink circles represent the median of $\delta$PA$_{\mathrm{kin}}$ in each interaction stage bin. The gray regions shows the median kinematic PA deviation obtained from the control sample. The bar in each bin represents the error determined by bootstrapping method.} 
\end{center}
  \end{figure}

As in the previous sections, we study the dependence of the kinematic PA deviations with their stellar masses. We find that these deviations in the stellar component are similar for both the control and interacting samples. Moreover, they tend to decrease for massive galaxies (see top left panel of Fig.\,\ref{deltaK_MS}). The origin for this trend is not evident. Since we observe this trend in both samples, we suggest that it can be produced by an internal process related to the stellar mass of the galaxy rather than an external process such as a merger event. Further studies with larger samples are required to confirm and further explain this possible trend. On the other hand, ionized gas $\delta$PA$_{\mathrm{kin}}$ remains constant at different stellar mass bins for both samples (see middle left panel of Fig.\,\ref{deltaK_MS}). For this component, the medians of $\delta$PA$_{\mathrm{kin}}$ at the different mass bins of the interacting sample are systematically larger than the ones from the control sample.

In an attempt to quantify the impact of the merger in both components, we plot for the interacting sample in the bottom left panel of Fig.\,\ref{deltaK_MS} the ratio between the ionized gas and stellar $\delta$PA$_{\mathrm{kin}}$ deviation with respect to their stellar masses. We find that at low and intermediate stellar mass bins the interacting sample has larger ratios than the control galaxies. While median values for the interacting sample in these mass bins are large or close to unity, the control sample medians are similar to  or lower than 0.5. At the highest mass bin, ratios for both samples are similar within the uncertainties. These results suggest that lighter companions in interacting systems react faster than heavier ones.  In fact, similar results have been found for observations \citep[e.g., NGC~7771+NGC~7770,][]{2012MNRAS.425L..46A} and numerical simulations \citep[e.g.,][]{2003ApJ...597..893N} of minor mergers. A necessary study will be to explore, with a sample covering a wide range in the pairs parameter space, the impact of the mass ratio on the kinematic properties of these systems.

In the right panels of Fig.\,\ref{deltaK_MS} we separate $\delta$PA$_{\mathrm{kin}}$ deviations according to the interaction stage of each galaxy. Except for the post-merger stage, stellar $\delta$PA$_{\mathrm{kin}}$ deviations remain  almost constant for different stages of the merger. Moreover, these values are similar to the median deviation of the control sample. For the ionized gas, the medians for different interaction stage bins are larger than the median kinematic deviation from the control sample (9\,$^\circ$). In particular, the median deviation of the post-merger stage is larger than the rest of stages (see middle right panel of Fig.\,\ref{deltaK_MS}). This suggests that both components display enhanced kinematic PA deviations after the two nuclei coalesced. We have to note that our sample at this interaction stage is modest, a large sample of post-merger galaxies is required to confirm this trend. Ratios between ionized gas and stellar $\delta$PA$_{\mathrm{kin}}$ are similar to the unity at different interaction stages and larger than the median ratio of the control sample (see bottom right panel of Fig.\,\ref{deltaK_MS}).
Assuming that the control galaxies are possible progenitors of the interacting sample, it is worthwhile to explore under what circumstances and at what exact stage of the interaction does the kinematic PA deviation gets increased. Measuring this parameter in velocity fields of simulated galaxies will help us to address these issues. 
  
In some objects showing similar PA$_{\mathrm{kin}}$ at both kinematic sides, we found large $\delta$PA$_{\mathrm{kin}}$ values. These are the cases of the merging galaxies NGC~3303, NGC~5394, or NGC~5614. Therefore, it is necessary to take into account different kinematic parameters to characterize the velocity field of interacting galaxies. In Sec.\,\ref{sec:Disc} we discuss the fraction of interacting galaxies that present deviations in more than one kinematic parameter in comparison to the control sample.
%
%
\subsection{Ionized gas and stellar kinematics}
\label{sec:kin_comp}
\begin{figure}[!htb]
\begin{center}
  \includegraphics[width=0.8\linewidth]{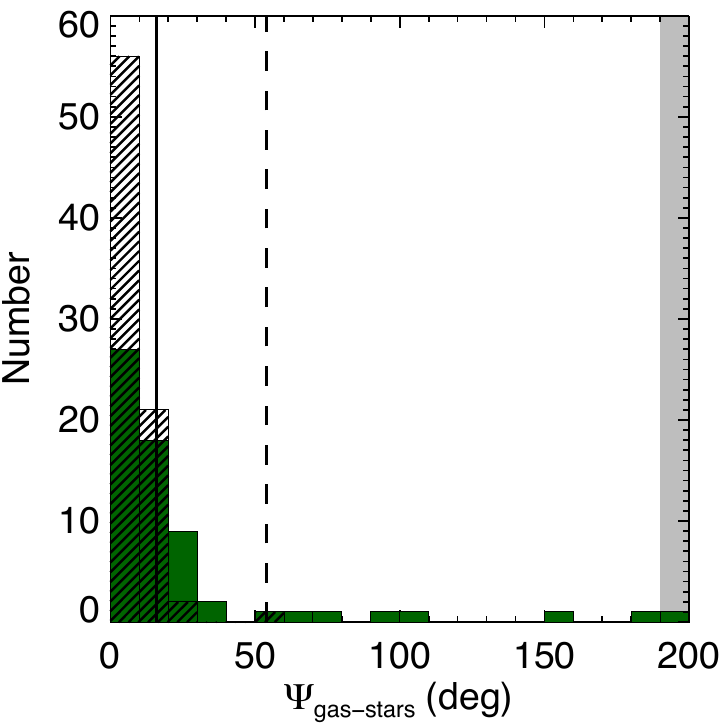}
  \caption{\label{both_Histo} Distribution of the kinematic misalignment between the stellar and ionized gas components ($\Psi$PA$_{\mathrm{gas-stars}}$). Line-filled histograms represent the distribution of $\Psi$PA$_{\mathrm{gas-stars}}$ for the control sample while green histograms show the distribution of these misalignments for the interacting sample.  The solid line represents the largest deviation which includes 90\% measured in the control sample. The dashed line represents the same value for the interacting galaxies.} 
\end{center}
\end{figure}
%
The comparison between spatially-resolved stellar and ionized gas kinematics in the nearby universe is rather scarce. It has been performed for a few individual interacting systems (e.g., NGC~6240 [\citealt{2010A&A...524A..56E}] and NGC~4676 [\citealt{2014A&A...567A.132W}]). Studies including a significant number of targets have been carried out for spiral galaxy bulges or elliptical galaxies (SAURON survey [\citealt{2006MNRAS.369..529F}], ATLAS3D [\citealt{2011MNRAS.417..882D}]). Only recently, the CALIFA survey has allowed  such a comparison to be performed for a sample of isolated galaxies of different morphological types (Paper I). In this section we compare PA$_{\mathrm{kin}}$  between the stars and ionized gas for our sample of interacting galaxies ($\Psi_{\mathrm{gas-stars}}$). Here we contrast the differences with the control sample.

In Fig.\,\ref{both_Histo} we plot the distribution of these differences for the control (80 objects) and interacting sample (where possible, 66 objects). To include 90\% of the $\Psi_{\mathrm{gas-stars}}$ differences in the control sample, we need to set a limiting value of  16\,$^\circ$. We find that 42\% (28/66) of the interacting sample has kinematic differences larger than 16\,$^\circ$. To include 90\% of the $\Psi_{\mathrm{gas-stars}}$ differences in the interacting sample, we need to set a limiting value of  75\,$^\circ$. The distributions of the interacting and control of $\Psi_{\mathrm{gas-stars}}$ are not likely to be drawn from the same parent distribution ($p_{KS} \sim$ 0). We also note that the fraction of interacting galaxies with strong kinematic misalignments (i.e., $\Psi_{\mathrm{gas-stars}}$ > 30\,$^\circ$) reduces to 18\% (12/66).

\begin{figure}[!htb]
  \includegraphics[width=\linewidth]{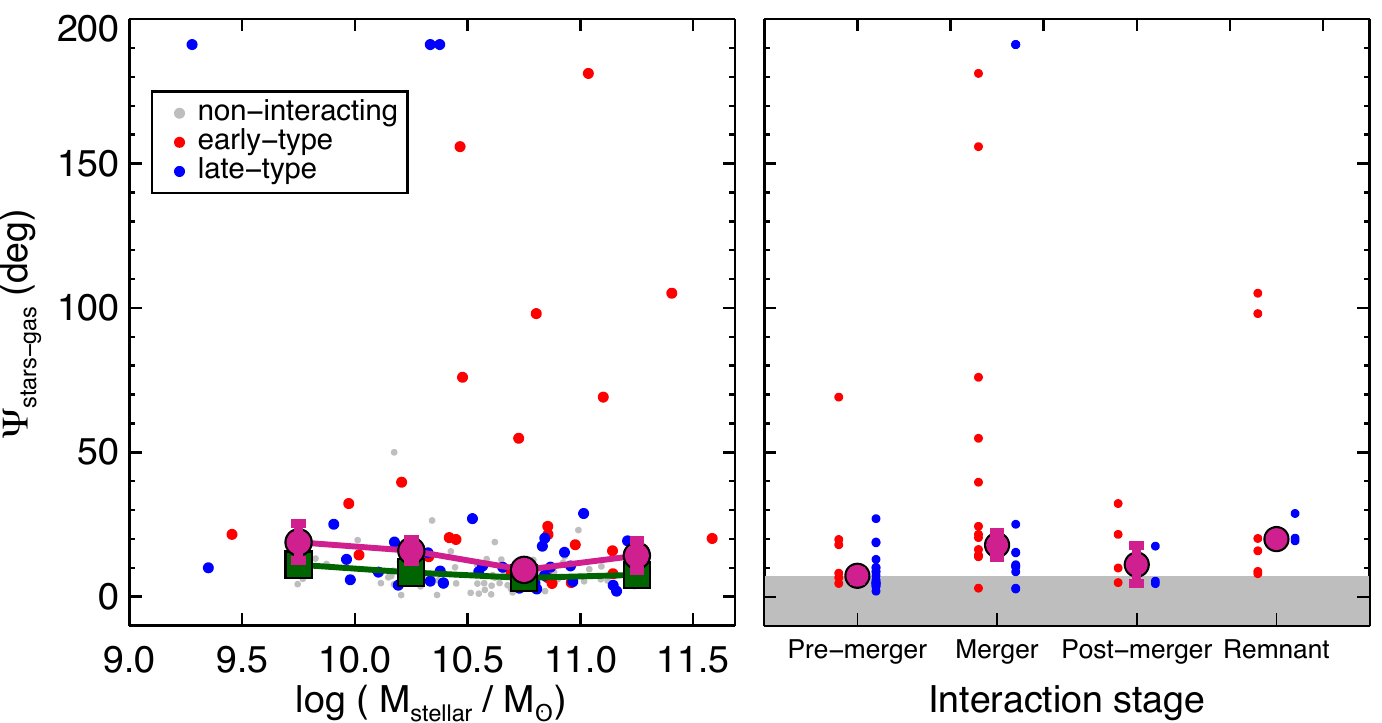}
  \includegraphics[width=\linewidth]{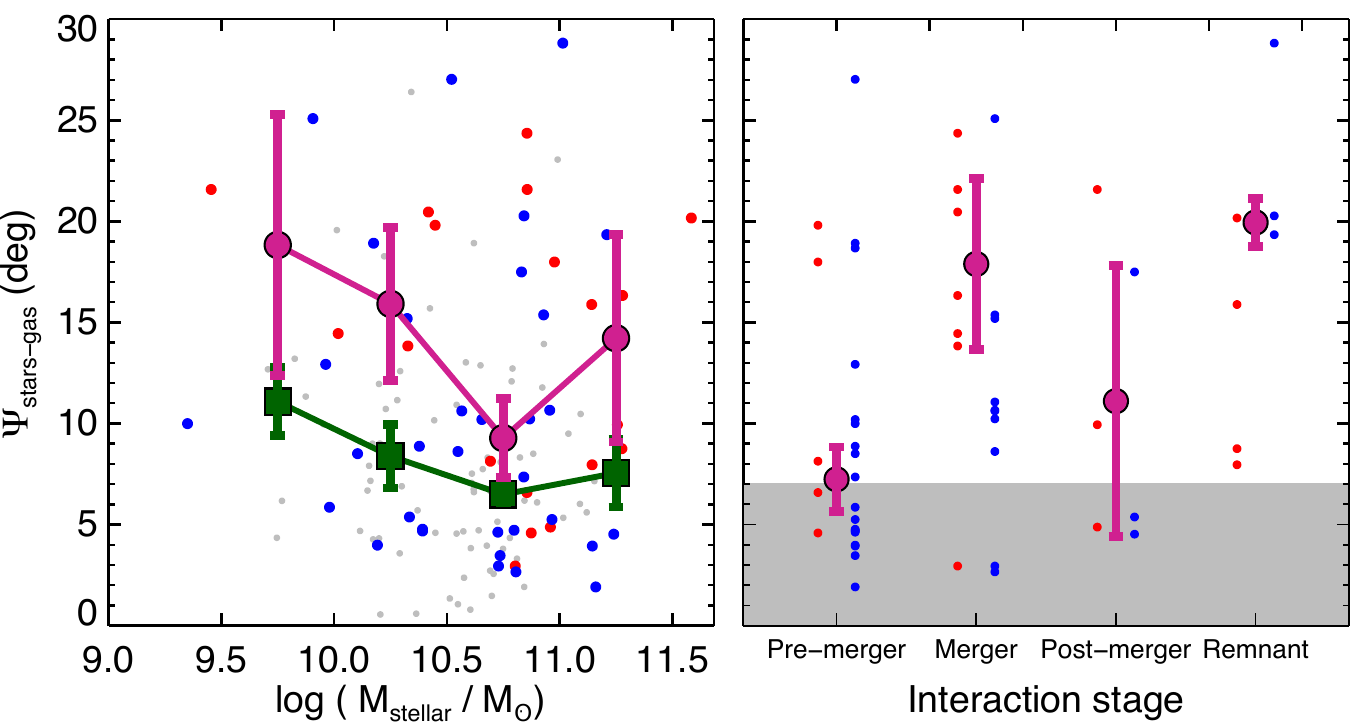}
  \caption{\label{both_comp_all}  Kinematic misalignment between the stellar and the ionized gas component ($\Psi_{\mathrm{stars-gas}}$) against the stellar mass (left) and the interacting stage (right). Top panels show the entire range of $\Psi_{\mathrm{stars-gas}}$ while bottom panels show a zoom to highlight median values. In all the panels the red symbols represent early-type galaxies (i.e., E, S0 and Sa) and late-type are shown by blue symbols (i.e., Sb, Sc, Sd). For the left panels, gray dots represent the control sample while green squares and pink circles represent the median in each stellar mass bin for the control and interacting samples, respectively. For the right panels, the pink circles represent the median of $\Psi_{\mathrm{stars-gas}}$ in each interaction stage bin while the gray regions shows the median $\Psi_{\mathrm{stars-gas}}$ obtained from the control sample. As in previous figures, the bar in each bin represents the error in  determined by a bootstrapping method.} 
  \end{figure}

The left panels of Fig.\,\ref{both_comp_all} show $\Psi_{\mathrm{gas-stars}}$ for the interacting and control samples with respect to their stellar masses. Interacting and control samples share similar trends in their median values for different mass bins (see left panel of Fig.\,\ref{both_comp_all}). However, median misalignments are systematically larger for interacting galaxies. 

In the right panels of Fig.\,\ref{both_comp_all} we separate $\Psi_{\mathrm{gas-stars}}$ of each galaxy according to its interaction stage. On the pre-merger stage, only a few objects show misalignments larger than the median value for isolated galaxies. On the contrary, the median misalignment for merging pairs is significantly larger than the $\Psi_{\mathrm{gas-stars}}$ median for the control sample. Indeed, objects in this stage present a wide range of misalignments and also the largest measured misalignments (e.g., NGC~3303, NGC~5929, NGC~5216, UGC~335NED02). However, in the same stage more than half of the galaxies present alignments similar to those observed in isolated galaxies. Post-merger galaxies present a median misalignment similar to the control sample, despite the large uncertainty and small size of the subsample. Merger remnants show the largest median misalignment. Most of these objects present large misalignments (e.g., NGC~5739, NGC~5623). We also note that all objects with (measured) strong kinematic PA misalignment between the stellar and ionized gas are early-type objects.
The galaxy that shows the strongest distortion is the merging galaxy NGC~3303 (and its companion NGC~3303 NED01). It reaches the counter-rotating case ($\Psi_{\mathrm{gas-stars}} \sim$ 180$^\circ$), and as we already mention in Sect.\,\ref{sec:deltaK}, this system presents a distorted ionized gas velocity field. Although this system is classified morphologically as a pair of early-type galaxies, the close projected separation between the nuclei ($\sim$ 5 kpc) and the fact that they are  embedded in a single disk-like structure with large tidal features suggest that this system is caught right before the two nuclei coalesced \citep{2000MNRAS.318..124G, 2014ApJ...784...16M}.

%
%
%
\section{Discussion}
\label{sec:Disc}

Throughout this study we aim to quantify the impact of the merging on galactic kinematics at different stages of the merging process by comparing different parameters from a sample of interacting galaxies against a sample of isolated objects. Based only on the morpho-kinematic PA misalignments, we find that a significant fraction of interacting galaxies presents large misalignments. Moreover, most of these misalignments are found in pairs with evident signatures of interaction (see Fig.\,\ref{MK_M_stage}), making the morpho-kinematic misalignment a good way to gauge the effect of interaction at this stage of the interaction. On the other hand, when we explore the kinematic alignments between the receding and approaching sides ($\Psi_{\mathrm{kin-kin}}$), we find that their distributions for the interacting sample resemble those from the control sample in both components (see Table \ref{table:kin_summary} and Fig.\,\ref{KK_Histo}). In general, median $\Psi_{\mathrm{kin-kin}}$ values for different interaction stages are similar to the value derived from the control sample (see Sec.\,\ref{sec:KK}) indicating that interactions and subsequent merger have a more subtle impact on this kinematic parameter than other parameters presented in this study. We note that the velocity field covers in most of the cases the central region of the interacting galaxies. It can be the case that at larger scales the misalignment between the kinematic sides of galaxies increases. However, detailed kinematic modeling for interacting galaxies is required in order to understand the cause of the small impact of interactions on this specific parameter. 

Further similarities between the interacting and the control sample come from the stellar kinematic PA deviations ($\delta$PA$_{\mathrm{kin}}$, see left panels of Fig.\ref{deltaK_Histo}). In both samples we find large stellar $\delta$PA$_{\mathrm{kin}}$ medians (see Sec.\,\ref{sec:deltaK}). This suggests that the mechanisms responsible for producing these large values could be similar. Such mechanisms could include radial motions or dynamical heating. We note that ionized gas kinematic deviations in nearly half of the interacting sample are larger than those obtained from the control sample. This indicates that ionized gas is more likely to react to the interaction/merger than  the stellar component. Moreover, ionized gas kinematic deviations are systematically larger at different interaction stages than those found in the control sample, in particular for the post-merger stage where its median deviation is the largest among other interaction stages (see Fig.\ref{deltaK_Histo}).

Comparing the stellar and the ionized gas kinematic PA, we find that nearly half of the interacting galaxies show misalignments larger than those observed in the sample of isolated galaxies (see Fig.\,\ref{both_Histo}). This suggests that a merging event does have an impact on the kinematics of the galaxies. At different interaction stages these median values change significantly. For pre-merger galaxies, median misalignment is similar to the value derived from the control sample, suggesting that interactions have a similar impact on the kinematics of both components to secular processes in isolated galaxies.

As the merger evolves the systems in the merging phase show a larger median misalignment between the two components in comparison to the median from the control sample. This indicates that at this phase, tidal effects do have a differential impact in the kinematic orientation of the components. Detailed numerical studies will be required to explore the initial parameters that lead to observed kinematic configuration of the stars and the ionized gas major kinematic PA.

Owing to its large error bar, the median kinematic misalignment from the post-merger stage ranges from the value obtained from the control sample to the that derived from the merger stage. Studies on numerical simulated mergers suggest that velocity fields should present strong distortions after the two nuclei have coalesced \citep[e.g.,][]{2007A&A...473..761K}. However, spatially resolved observations of central regions of ULIRGs including post-merger galaxies, find that at these scales the kinematic orientation of both the stellar and the ionized gas components are consistent with each other \citep{2014ApJ...784...70M}. In this scenario  young stars formed in situ follow the kinematics of the ionized gas rather than the kinematics of those stars from the progenitor galaxies. A detailed study of the resolved stellar populations in our sample is required to constrain the ages of the stellar population in these merger remnants. Furthermore,  a larger sample of post-merger galaxies is required to understand which type of motion predominates between these two components.

In the remnant subsample we find the largest median misalignment between the stars and the ionized gas. Six out of nine galaxies are early-type objects. Assuming that these galaxies are the remnants of a merger event, according to recent numerical simulations, the kinematic orientation of the stars and ionized gas components is not likely to change significantly \citep{2015MNRAS.448.1271L}. Several factors can explain this apparent dichotomy. First, the size of the remnant sample is rather modest in comparison to the pre-merger and merger samples. Although this is expected \citep{2010MNRAS.401.1043D,2013MNRAS.435.3627E}, our subsamples may not cover homogeneously the space parameters of the merging. Second, as we note in Sec.\,\ref{sec:stage} our  ability to determine whether a galaxy is a merger remnant is based only on the observation of low-surface brightness tidal features. These features could be the result of a recent minor merger or a new supply of gas \citep{2011MNRAS.417..882D,2015MNRAS.448.1271L} leading to the observed strong difference between the kinematic orientation of the stars and the ionized gas in these galaxies.

The wealth of  data allows us to compile the different kinematic parameters presented in this study to quantify the impact of interactions and mergers on the galactic kinematics. We find that 63\% (65/103) of the interacting sample has a kinematic misalignment larger than 90\% of the control sample in at least one of the parameters for any component. We also estimate the fraction of interacting galaxies with differences in at least two kinematic indicators larger than 90\% of the control sample in each component. For the stellar and ionized gas components we find that these fractions correspond to 21\% (18/85) and 39\% (32/82), respectively. When we consider the galaxies with an offset between the gradient peak and the optical nucleus larger than 1\farcs35, we find that these fractions increase to 27\% (23/85) and 53\% (43/82). These fractions reveal that interactions and mergers do  have a significant impact on the motions of the galaxy's components in particular for the ionized gas. Furthermore, from the objects in the interacting sample with $\Psi_{\mathrm{gas-stars}}$ larger than the control sample, we explore the fraction of these objects with large misalignments in any other kinematic parameters presented (i.e., $\Psi_{\mathrm{morph-kin}}$, $\Psi_{\mathrm{kin-kin}}$, and $\delta$PA$_{\mathrm{kin}}$) including both components. We find that 16/66 (23\%) objects show misalignments larger than the values expected for isolated galaxies. Although this fraction is not as large as those derived from some individual kinematic parameters, it reveals that in general the interaction has a noteworthy impact on the motion of galaxies.

Most of the statistical studies of spatially-resolved kinematics in merging galaxies have been carry out at high redshift \citep[$z$ < 0.5, for a review see][]{2013PASA...30...56G}. Mergers at high redshift are usually identified by their morphology \citep[e.g.,][]{2003AJ....126.1183C, 2008MNRAS.386..909C, 2012ApJ...747...34B}. Most of these studies found a significant fraction of high-redshift galaxies with regular kinematic patterns despite their irregular morphology, suggesting the presence of gas-rich disks at early stages of the Universe \citep[e.g.,][]{2008ApJ...682..231S}. Kinematic studies are useful for distinguishing between a rotating disk or distorted motions due to a merging at high-z \citep{2008ApJ...685L..27R,2010ApJ...724.1373G, 2015ApJ...803...62H}. One criteria used to distinguish rotating disks at high redshift is the alignment between (modeled) kinematic and morphological PA \citep[e.g.,][]{2006A&A...455..107F, 2012A&A...539A..92E}. Although it is beyond the scope of the present study, this statistical characterization and model-free methodology can be used as a high-resolution benchmark for kinematic properties of merging galaxies that can be compared with those at high redshift.

Finally, studies of large samples of velocity fields of interacting and mergers galaxies in the local Universe are scarce however, \cite{2013A&A...557A..59B} found that a significant portion of ULIRGs are dominated by rotation (76\%). Their sample covers different stages of interaction from interacting galaxies to mergers. Even though we cannot confirm that interacting and merging galaxies are supported by rotation with the kinematic properties presented here, by comparing these galaxies with a control sample (which is expected to be supported by rotation) we find evidence that this could be the case in some of these objects. 
We should point out however, that the alignment in the kinematic parameters presented in this study do not always represent rotational support. For instance, \citep{2010A&A...524A..56E} found that even though  circular-kinematic signatures are present in the velocity distributions of the binary system NGC~6240, the interacting galaxies are not necessarily dynamically supported by rotation. In a future work we will explore the dynamical support  of our interaction sample through better spectral resolution data from CALIFA \citep[V1200 configuration,][]{2013A&A...549A..87H}.

\section{Summary and conclusions}
\label{sec:conclusions}

In this study we analyze the stellar and ionized gas velocity fields of 103 interacting galaxies covering different stages of the merger event -- from close pairs to merger remnants. To differentiate the kinematic signatures due to internal processes from those triggered by the interaction, we measured homogeneously the same kinematic properties in a sample of isolated galaxies \citep{2014A&A...568A..70B}. We measured the major kinematic position angles from both receding and approaching sides directly from the velocity maps in both components, making no assumptions on the internal kinematics of the interacting systems. This method provides the morpho-kinematic misalignment for both sides, the internal (mis)alignment between both kinematic sides, the deviation of the kinematic PA from a straight line, and -- when stellar and ionized gas kinematics are measurable -- the (mis)alignment between these two components.

We find that 43\% (37/85) and 52\% (43/82) of the interacting sample has stellar and ionized gas morpho-kinematic misalignments larger than those found in the control sample. In particular, we find a large fraction of these morpho-kinematic misalignments in galaxies included in binary systems with evident signatures of interaction. On the other hand, median internal kinematic misalignment for interacting galaxies is slightly larger than the value derived from the control galaxies at different stellar masses and interaction stages. Comparison between the stellar and the ionized gas kinematic PA (66 objects) reveals that 42\% (28/66) of the interacting sample has misalignments larger than those presented by the control sample. In particular, median misalignments of mergers and remnants are significantly larger than the median provided by isolated galaxies. Distributions of the stellar kinematic PA deviations are similar in both samples. However, 48\% (41/82) of the interacting galaxies have ionized gas kinematic PA deviations larger than the control sample, in particular in the~post-merger~stage. This suggests that kinematic PA deviations in the ionized gas can be used as a tracer to determine whether a galaxy is or has been undergoing an interaction or recent merger. 

Our study indicates that interactions have a significant impact on the motion of stars and ionized gas in galaxies. Moreover, our results probe the wide range of kinematic stages observed in galaxies under different phases of merger,  from velocity fields with similar properties to those found in isolated galaxies to very distorted velocity fields with large kinematic misalignments.

The CALIFA survey allows us to characterize the spatially resolved properties of galaxies at different stages of interaction. In particular, this study provides a nearby Universe benchmark for kinematic comparisons with high-redshift galaxies. Further studies with simulated high-redshift observations using the current spatially resolved data will allow us to quantify the fraction of rotating disks and compare it with the fraction observed at high redshift. Such simulations for the ionized gas in CALIFA galaxies have been already carried out \citep{2014A&A...561A.129M}.

Our findings encourage the comparison of observational velocity fields with those obtained from numerical simulations. Exploring different configurations of the merging galaxies with different properties of the progenitors will give significant insight and will  reveal which parameters  yield the variety of observed velocity fields in merging galaxies.

\section*{Acknowledgments}
The authors thank  the referee for her/his useful comments and suggestions. This study makes use of the data provided by the Calar Alto Legacy Field Area (CALIFA) survey (http://www.califa.caha.es). Based on observations collected at the Centro Astron\'{o}mico Hispano Alem\'{a}n (CAHA) at Calar Alto, operated jointly by the Max-Planck-Institut f\"{u}r Astronomie and the Instituto de Astrof\'{\i}sica de Andalucia (CSIC). CALIFA is the first legacy survey performed at the Calar Alto. The CALIFA collaboration would like to thank  the IAA-CSIC and MPIA-MPG as major partners of the observatory, and CAHA itself, for the unique access to the telescope time and support  in manpower and infrastructures. The CALIFA collaboration also thanks the CAHA staff for the dedication to this project. J.B-B. and B.G-L acknowledge support from the Plan Nacional de I+D+i (PNAYA) funding programs (AYA2012-39408-C02-02-1 and AYA2013-41656-P) of Spanish Ministry of Economy and Competitiveness (MINECO). J.F.B.  acknowledges support from the Plan Nacional de I+D+i (PNAYA) funding programs from MINECO (AYA2013-48226-03-1-P, RAVET) A.M.-I. acknowledges support from Agence Nationale de la Recherche through the STILISM project (ANR-12-BS05-0016-02) and from BMBF through the Erasmus-F project (grant number 05 A12BA1). I.M. acknowledges financial support by MINECO grant AYA 2010-15169, Junta de Andaluc\'{\i}a
TIC114 and Proyecto de Excelencia de la Junta de Andaluc\'{\i}a P08-TIC-03531. J.M.A. and V.W. acknowledge support from the European Research Council Starting Grant (SEDmorph; P.I. V. Wild) J.I.P.  acknowledges financial support from the Spanish MINECO under grant AYA2013-47742-C04-1 and from Junta de Andaluc\'{\i}a Excellence Project PEX2011-FQM7058. R.A.M is funded by the Spanish program of International Campus of Excellence Moncloa (CEI). L.V.M. acknowledges support from the grant AYA2011-30491-C02-01 co-financed by MICINN and FEDER funds, and the Junta de Andalucia (Spain) grants P08-FQM-4205 and TIC-114.

\bibliographystyle{aa_arxiv} 
\bibliography{final_interKin_jkbb_aanda_corr}

\def\eprinttmppp@#1arXiv:@{#1}
\providecommand{\arxivlink[1]}{\href{http://arxiv.org/abs/#1}{arXiv:#1}}
\def\eprinttmp@#1arXiv:#2 [#3]#4@{\ifthenelse{\equal{#3}{x}}{\ifthenelse{
\equal{#1}{}}{\arxivlink{\eprinttmppp@#2@}}{\arxivlink{#1}}}{\arxivlink{#2}
  [#3]}}
\providecommand{\eprintlink}[1]{\eprinttmp@#1arXiv: [x]@}
\providecommand{\eprint}[1]{\eprintlink{#1}}
\begin{thebibliography}{86}
\expandafter\ifx\csname natexlab\endcsname\relax\def\natexlab#1{#1}\fi

\bibitem[{{Abazajian} {et~al.}(2009){Abazajian}, {Adelman-McCarthy},
  {Ag{\"u}eros}, {Allam}, {Allende Prieto}, {An}, {Anderson}, {Anderson},
  {Annis}, {Bahcall}, \& {et al.}}]{2009ApJS..182..543A}
{Abazajian}, K.~N., {Adelman-McCarthy}, J.~K., {Ag{\"u}eros}, M.~A., {et~al.}
  2009, \apjs, 182, 543, \eprint{0812.0649}

\bibitem[{{Alonso-Herrero} {et~al.}(2012){Alonso-Herrero}, {Rosales-Ortega},
  {S{\'a}nchez}, {Kennicutt}, {Pereira-Santaella}, \&
  {D{\'\i}az}}]{2012MNRAS.425L..46A}
{Alonso-Herrero}, A., {Rosales-Ortega}, F.~F., {S{\'a}nchez}, S.~F., {et~al.}
  2012, \mnras, 425, L46, \eprint{1206.1686}

\bibitem[{{Arp}(1966)}]{1966ApJS...14....1A}
{Arp}, H. 1966, \apjs, 14, 1

\bibitem[{{Arribas} {et~al.}(2008){Arribas}, {Colina}, {Monreal-Ibero},
  {Alfonso}, {Garc{\'\i}a-Mar{\'\i}n}, \&
  {Alonso-Herrero}}]{2008A&A...479..687A}
{Arribas}, S., {Colina}, L., {Monreal-Ibero}, A., {et~al.} 2008, {\aa}p, 479,
  687, \eprint{0710.2761}

\bibitem[{{Atkinson} {et~al.}(2013){Atkinson}, {Abraham}, \&
  {Ferguson}}]{2013ApJ...765...28A}
{Atkinson}, A.~M., {Abraham}, R.~G., \& {Ferguson}, A.~M.~N. 2013, \apj, 765,
  28, \eprint{1301.4275}

\bibitem[{{Barnes}(2011)}]{2011MNRAS.413.2860B}
{Barnes}, J.~E. 2011, \mnras, 413, 2860, \eprint{1101.5671}

\bibitem[{{Barnes} \& {Hibbard}(2009)}]{2009AJ....137.3071B}
{Barnes}, J.~E. \& {Hibbard}, J.~E. 2009, \aj, 137, 3071, \eprint{0811.3039}

\bibitem[{{Barrera-Ballesteros} {et~al.}(2014){Barrera-Ballesteros},
  {Falc{\'o}n-Barroso}, {Garc{\'{\i}}a-Lorenzo}, {van de Ven}, {Aguerri},
  {Mendez-Abreu}, {Spekkens}, {Lyubenova}, {S{\'a}nchez}, {Husemann}, {Mast},
  {Garc{\'{\i}}a-Benito}, {Iglesias-Paramo}, {Del Olmo}, {M{\'a}rquez},
  {Masegosa}, {Kehrig}, {Marino}, {Verdes-Montenegro}, {Ziegler}, {McIntosh},
  {Bland-Hawthorn}, {Walcher}, \& {Califa Collaboration}}]{2014A&A...568A..70B}
{Barrera-Ballesteros}, J.~K., {Falc{\'o}n-Barroso}, J.,
  {Garc{\'{\i}}a-Lorenzo}, B., {et~al.} 2014, \aap, 568, A70,
  \eprint{1405.5222}

\bibitem[{{Bellocchi} {et~al.}(2013){Bellocchi}, {Arribas}, {Colina}, \&
  {Miralles-Caballero}}]{2013A&A...557A..59B}
{Bellocchi}, E., {Arribas}, S., {Colina}, L., \& {Miralles-Caballero}, D. 2013,
  \aap, 557, A59, \eprint{1307.1659}

\bibitem[{{B{\'{\i}}lek} {et~al.}(2013){B{\'{\i}}lek}, {Jungwiert},
  {J{\'{\i}}lkov{\'a}}, {Ebrov{\'a}}, {Barto{\v s}kov{\'a}}, \& {K{\v
  r}{\'{\i}}{\v z}ek}}]{2013A&A...559A.110B}
{B{\'{\i}}lek}, M., {Jungwiert}, B., {J{\'{\i}}lkov{\'a}}, L., {et~al.} 2013,
  \aap, 559, A110, \eprint{1309.1003}

\bibitem[{{Bluck} {et~al.}(2012){Bluck}, {Conselice}, {Buitrago},
  {Gr{\"u}tzbauch}, {Hoyos}, {Mortlock}, \& {Bauer}}]{2012ApJ...747...34B}
{Bluck}, A.~F.~L., {Conselice}, C.~J., {Buitrago}, F., {et~al.} 2012, \apj,
  747, 34, \eprint{1111.5662}

\bibitem[{{Bois} {et~al.}(2012){Bois}, {Emsellem}, {Bournaud}, {Alatalo},
  {Blitz}, {Bureau}, {Cappellari}, {Davies}, {Davis}, {de Zeeuw}, {Duc},
  {Khochfar}, {Krajnovi{\'c}}, {Kuntschner}, {Lablanche}, {McDermid},
  {Morganti}, {Naab}, {Oosterloo}, {Sarzi}, {Scott}, {Serra}, {Weijmans}, \&
  {Young}}]{2012arXiv1201.0885B}
{Bois}, M., {Emsellem}, E., {Bournaud}, F., {et~al.} 2012, ArXiv e-prints,
  \eprint{1201.0885}

\bibitem[{{Cappellari} \& {Copin}(2003)}]{2003MNRAS.342..345C}
{Cappellari}, M. \& {Copin}, Y. 2003, \mnras, 342, 345,
  \eprint{arXiv:astro-ph/0302262}

\bibitem[{{Cappellari} \& {Emsellem}(2004)}]{2004PASP..116..138C}
{Cappellari}, M. \& {Emsellem}, E. 2004, \pasp, 116, 138,
  \eprint{arXiv:astro-ph/0312201}

\bibitem[{{Chabrier}(2003)}]{2003PASP..115..763C}
{Chabrier}, G. 2003, \pasp, 115, 763, \eprint{astro-ph/0304382}

\bibitem[{{Colina} {et~al.}(2005){Colina}, {Arribas}, \&
  {Monreal-Ibero}}]{2005ApJ...621..725C}
{Colina}, L., {Arribas}, S., \& {Monreal-Ibero}, A. 2005, \apj, 621, 725

\bibitem[{{Conselice} {et~al.}(2003){Conselice}, {Bershady}, {Dickinson}, \&
  {Papovich}}]{2003AJ....126.1183C}
{Conselice}, C.~J., {Bershady}, M.~A., {Dickinson}, M., \& {Papovich}, C. 2003,
  \aj, 126, 1183, \eprint{astro-ph/0306106}

\bibitem[{{Conselice} {et~al.}(2008){Conselice}, {Rajgor}, \&
  {Myers}}]{2008MNRAS.386..909C}
{Conselice}, C.~J., {Rajgor}, S., \& {Myers}, R. 2008, \mnras, 386, 909,
  \eprint{0711.2333}

\bibitem[{{Contini} {et~al.}(2012){Contini}, {Garilli}, {Le F{\`e}vre},
  {Kissler-Patig}, {Amram}, {Epinat}, {Moultaka}, {Paioro}, {Queyrel}, {Tasca},
  {Tresse}, {Vergani}, {L{\'o}pez-Sanjuan}, \&
  {Perez-Montero}}]{2012A&A...539A..91C}
{Contini}, T., {Garilli}, B., {Le F{\`e}vre}, O., {et~al.} 2012, \aap, 539,
  A91, \eprint{1111.3631}

\bibitem[{{Cox} {et~al.}(2006){Cox}, {Dutta}, {Di Matteo}, {Hernquist},
  {Hopkins}, {Robertson}, \& {Springel}}]{2006ApJ...650..791C}
{Cox}, T.~J., {Dutta}, S.~N., {Di Matteo}, T., {et~al.} 2006, \apj, 650, 791,
  \eprint{arXiv:astro-ph/0607446}

\bibitem[{{Darg} {et~al.}(2010){Darg}, {Kaviraj}, {Lintott}, {Schawinski},
  {Sarzi}, {Bamford}, {Silk}, {Proctor}, {Andreescu}, {Murray}, {Nichol},
  {Raddick}, {Slosar}, {Szalay}, {Thomas}, \&
  {Vandenberg}}]{2010MNRAS.401.1043D}
{Darg}, D.~W., {Kaviraj}, S., {Lintott}, C.~J., {et~al.} 2010, \mnras, 401,
  1043, \eprint{0903.4937}

\bibitem[{{Dasyra} {et~al.}(2006){Dasyra}, {Tacconi}, {Davies}, {Naab},
  {Genzel}, {Lutz}, {Sturm}, {Baker}, {Veilleux}, {Sanders}, \&
  {Burkert}}]{2006ApJ...651..835D}
{Dasyra}, K.~M., {Tacconi}, L.~J., {Davies}, R.~I., {et~al.} 2006, \apj, 651,
  835, \eprint{arXiv:astro-ph/0607468}

\bibitem[{{Davis} {et~al.}(2011){Davis}, {Alatalo}, {Sarzi}, {Bureau}, {Young},
  {Blitz}, {Serra}, {Crocker}, {Krajnovi{\'c}}, {McDermid}, {Bois}, {Bournaud},
  {Cappellari}, {Davies}, {Duc}, {de Zeeuw}, {Emsellem}, {Khochfar},
  {Kuntschner}, {Lablanche}, {Morganti}, {Naab}, {Oosterloo}, {Scott}, \&
  {Weijmans}}]{2011MNRAS.417..882D}
{Davis}, T.~A., {Alatalo}, K., {Sarzi}, M., {et~al.} 2011, \mnras, 417, 882,
  \eprint{1107.0002}

\bibitem[{{Di Matteo} {et~al.}(2007){Di Matteo}, {Combes}, {Melchior}, \&
  {Semelin}}]{2007A&A...468...61D}
{Di Matteo}, P., {Combes}, F., {Melchior}, A.-L., \& {Semelin}, B. 2007, \aap,
  468, 61, \eprint{astro-ph/0703212}

\bibitem[{{Duc} {et~al.}(2015){Duc}, {Cuillandre}, {Karabal}, {Cappellari},
  {Alatalo}, {Blitz}, {Bournaud}, {Bureau}, {Crocker}, {Davies}, {Davis}, {de
  Zeeuw}, {Emsellem}, {Khochfar}, {Krajnovi{\'c}}, {Kuntschner}, {McDermid},
  {Michel-Dansac}, {Morganti}, {Naab}, {Oosterloo}, {Paudel}, {Sarzi}, {Scott},
  {Serra}, {Weijmans}, \& {Young}}]{2015MNRAS.446..120D}
{Duc}, P.-A., {Cuillandre}, J.-C., {Karabal}, E., {et~al.} 2015, \mnras, 446,
  120, \eprint{1410.0981}

\bibitem[{{Duc} \& {Renaud}(2013)}]{2013LNP...861..327D}
{Duc}, P.-A. \& {Renaud}, F. 2013, in Lecture Notes in Physics, Berlin Springer
  Verlag, Vol. 861, Lecture Notes in Physics, Berlin Springer Verlag, ed.
  J.~{Souchay}, S.~{Mathis}, \& T.~{Tokieda}, 327

\bibitem[{{Ebrova}(2013)}]{2013arXiv1312.1643E}
{Ebrova}, I. 2013, ArXiv e-prints, \eprint{1312.1643}

\bibitem[{{Ellison} {et~al.}(2013){Ellison}, {Mendel}, {Patton}, \&
  {Scudder}}]{2013MNRAS.435.3627E}
{Ellison}, S.~L., {Mendel}, J.~T., {Patton}, D.~R., \& {Scudder}, J.~M. 2013,
  \mnras, 435, 3627, \eprint{1308.3707}

\bibitem[{{Ellison} {et~al.}(2008){Ellison}, {Patton}, {Simard}, \&
  {McConnachie}}]{2008AJ....135.1877E}
{Ellison}, S.~L., {Patton}, D.~R., {Simard}, L., \& {McConnachie}, A.~W. 2008,
  \aj, 135, 1877, \eprint{0803.0161}

\bibitem[{{Engel} {et~al.}(2010){Engel}, {Davies}, {Genzel}, {Tacconi},
  {Hicks}, {Sturm}, {Naab}, {Johansson}, {Karl}, {Max}, {Medling}, \& {van der
  Werf}}]{2010A&A...524A..56E}
{Engel}, H., {Davies}, R.~I., {Genzel}, R., {et~al.} 2010, \aap, 524, A56,
  \eprint{1009.1539}

\bibitem[{{Epinat} {et~al.}(2008){Epinat}, {Amram}, {Marcelin}, {Balkowski},
  {Daigle}, {Hernandez}, {Chemin}, {Carignan}, {Gach}, \&
  {Balard}}]{2008MNRAS.388..500E}
{Epinat}, B., {Amram}, P., {Marcelin}, M., {et~al.} 2008, \mnras, 388, 500,
  \eprint{0805.0976}

\bibitem[{{Epinat} {et~al.}(2012){Epinat}, {Tasca}, {Amram}, {Contini}, {Le
  F{\`e}vre}, {Queyrel}, {Vergani}, {Garilli}, {Kissler-Patig}, {Moultaka},
  {Paioro}, {Tresse}, {Bournaud}, {L{\'o}pez-Sanjuan}, \&
  {Perret}}]{2012A&A...539A..92E}
{Epinat}, B., {Tasca}, L., {Amram}, P., {et~al.} 2012, \aap, 539, A92,
  \eprint{1201.3329}

\bibitem[{{Espada} {et~al.}(2011){Espada}, {Verdes-Montenegro}, {Huchtmeier},
  {Sulentic}, {Verley}, {Leon}, \& {Sabater}}]{2011A&A...532A.117E}
{Espada}, D., {Verdes-Montenegro}, L., {Huchtmeier}, W.~K., {et~al.} 2011,
  \aap, 532, A117, \eprint{1107.0601}

\bibitem[{{Evans} {et~al.}(2008){Evans}, {Vavilkin}, {Pizagno}, {Modica},
  {Mazzarella}, {Iwasawa}, {Howell}, {Surace}, {Armus}, {Petric}, {Spoon},
  {Barnes}, {Suer}, {Sanders}, {Chan}, \& {Lord}}]{2008ApJ...675L..69E}
{Evans}, A.~S., {Vavilkin}, T., {Pizagno}, J., {et~al.} 2008, \apjl, 675, L69

\bibitem[{{Falc{\'o}n-Barroso} {et~al.}(2006){Falc{\'o}n-Barroso}, {Bacon},
  {Bureau}, {Cappellari}, {Davies}, {de Zeeuw}, {Emsellem}, {Fathi},
  {Krajnovi{\'c}}, {Kuntschner}, {McDermid}, {Peletier}, \&
  {Sarzi}}]{2006MNRAS.369..529F}
{Falc{\'o}n-Barroso}, J., {Bacon}, R., {Bureau}, M., {et~al.} 2006, \mnras,
  369, 529, \eprint{arXiv:astro-ph/0603161}

\bibitem[{{Falc{\'o}n-Barroso} {et~al.}(2007){Falc{\'o}n-Barroso}, {Bacon},
  {Bureau}, {Cappellari}, {Davies}, {de Zeeuw}, {Emsellem}, {Fathi},
  {Krajnovi{\'c}}, {Kuntschner}, {McDermid}, {Peletier}, \&
  {Sarzi}}]{2007spts.conf..111F}
{Falc{\'o}n-Barroso}, J., {Bacon}, R., {Bureau}, M., {et~al.} 2007, in Science
  Perspectives for 3D Spectroscopy, ed. M.~{Kissler-Patig}, J.~R. {Walsh}, \&
  M.~M. {Roth}, 111

\bibitem[{{Fern{\'a}ndez Lorenzo} {et~al.}(2012){Fern{\'a}ndez Lorenzo},
  {Sulentic}, {Verdes-Montenegro}, {Ruiz}, {Sabater}, \&
  {S{\'a}nchez}}]{2012A&A...540A..47F}
{Fern{\'a}ndez Lorenzo}, M., {Sulentic}, J., {Verdes-Montenegro}, L., {et~al.}
  2012, \aap, 540, A47, \eprint{1201.5834}

\bibitem[{{Flores} {et~al.}(2006){Flores}, {Hammer}, {Puech}, {Amram}, \&
  {Balkowski}}]{2006A&A...455..107F}
{Flores}, H., {Hammer}, F., {Puech}, M., {Amram}, P., \& {Balkowski}, C. 2006,
  \aap, 455, 107, \eprint{astro-ph/0603563}

\bibitem[{{F{\"o}rster Schreiber} {et~al.}(2009){F{\"o}rster Schreiber},
  {Genzel}, {Bouch{\'e}}, {Cresci}, {Davies}, {Buschkamp}, {Shapiro},
  {Tacconi}, {Hicks}, {Genel}, {Shapley}, {Erb}, {Steidel}, {Lutz},
  {Eisenhauer}, {Gillessen}, {Sternberg}, {Renzini}, {Cimatti}, {Daddi},
  {Kurk}, {Lilly}, {Kong}, {Lehnert}, {Nesvadba}, {Verma}, {McCracken},
  {Arimoto}, {Mignoli}, \& {Onodera}}]{2009ApJ...706.1364F}
{F{\"o}rster Schreiber}, N.~M., {Genzel}, R., {Bouch{\'e}}, N., {et~al.} 2009,
  \apj, 706, 1364, \eprint{0903.1872}

\bibitem[{{Garc{\'{\i}}a-Benito} {et~al.}(2015){Garc{\'{\i}}a-Benito},
  {Zibetti}, {S{\'a}nchez}, {Husemann}, {de Amorim}, {Castillo-Morales}, {Cid
  Fernandes}, {Ellis}, {Falc{\'o}n-Barroso}, {Galbany}, {Gil de Paz},
  {Gonz{\'a}lez Delgado}, {Lacerda}, {L{\'o}pez-Fernandez}, {de
  Lorenzo-C{\'a}ceres}, {Lyubenova}, {Marino}, {Mast}, {Mendoza}, {P{\'e}rez},
  {Vale Asari}, {Aguerri}, {Ascasibar}, {Bekerait*error*{\.e}},
  {Bland-Hawthorn}, {Barrera-Ballesteros}, {Bomans}, {Cano-D{\'{\i}}az},
  {Catal{\'a}n-Torrecilla}, {Cortijo}, {Delgado-Inglada}, {Demleitner},
  {Dettmar}, {D{\'{\i}}az}, {Florido}, {Gallazzi}, {Garc{\'{\i}}a-Lorenzo},
  {Gomes}, {Holmes}, {Iglesias-P{\'a}ramo}, {Jahnke}, {Kalinova}, {Kehrig},
  {Kennicutt}, {L{\'o}pez-S{\'a}nchez}, {M{\'a}rquez}, {Masegosa}, {Meidt},
  {Mendez-Abreu}, {Moll{\'a}}, {Monreal-Ibero}, {Morisset}, {del Olmo},
  {Papaderos}, {P{\'e}rez}, {Quirrenbach}, {Rosales-Ortega}, {Roth},
  {Ruiz-Lara}, {S{\'a}nchez-Bl{\'a}zquez}, {S{\'a}nchez-Menguiano}, {Singh},
  {Spekkens}, {Stanishev}, {Torres-Papaqui}, {van de Ven}, {Vilchez},
  {Walcher}, {Wild}, {Wisotzki}, {Ziegler}, {Alves}, {Barrado}, {Quintana}, \&
  {Aceituno}}]{2015A&A...576A.135G}
{Garc{\'{\i}}a-Benito}, R., {Zibetti}, S., {S{\'a}nchez}, S.~F., {et~al.} 2015,
  \aap, 576, A135, \eprint{1409.8302}

\bibitem[{{Garc{\'{\i}}a-Lorenzo}(2013)}]{2013MNRAS.429.2903G}
{Garc{\'{\i}}a-Lorenzo}, B. 2013, \mnras, 429, 2903, \eprint{1211.6860}

\bibitem[{{Garc{\'{\i}}a-Lorenzo} {et~al.}(2015){Garc{\'{\i}}a-Lorenzo},
  {M{\'a}rquez}, {Barrera-Ballesteros}, {Masegosa}, {Husemann},
  {Falc{\'o}n-Barroso}, {Lyubenova}, {S{\'a}nchez}, {Walcher}, {Mast},
  {Garc{\'{\i}}a-Benito}, {M{\'e}ndez-Abreu}, {van de Ven}, {Spekkens},
  {Holmes}, {Monreal-Ibero}, {del Olmo}, {Ziegler}, {Bland-Hawthorn},
  {S{\'a}nchez-Bl{\'a}zquez}, {Iglesias-P{\'a}ramo}, {Aguerri}, {Papaderos},
  {Gomes}, {Marino}, {Gonz{\'a}lez Delgado}, {Cortijo-Ferrero},
  {L{\'o}pez-S{\'a}nchez}, {Bekerait{\.e}}, {Wisotzki}, \&
  {Bomans}}]{2015A&A...573A..59G}
{Garc{\'{\i}}a-Lorenzo}, B., {M{\'a}rquez}, I., {Barrera-Ballesteros}, J.~K.,
  {et~al.} 2015, \aap, 573, A59, \eprint{1408.5765}

\bibitem[{{Georgakakis} {et~al.}(2000){Georgakakis}, {Forbes}, \&
  {Norris}}]{2000MNRAS.318..124G}
{Georgakakis}, A., {Forbes}, D.~A., \& {Norris}, R.~P. 2000, \mnras, 318, 124,
  \eprint{astro-ph/0006084}

\bibitem[{{Glazebrook}(2013)}]{2013PASA...30...56G}
{Glazebrook}, K. 2013, \pasa, 30, 56, \eprint{1305.2469}

\bibitem[{{Gnerucci} {et~al.}(2011){Gnerucci}, {Marconi}, {Cresci}, {Maiolino},
  {Mannucci}, {Calura}, {Cimatti}, {Cocchia}, {Grazian}, {Matteucci}, {Nagao},
  {Pozzetti}, \& {Troncoso}}]{2011A&A...528A..88G}
{Gnerucci}, A., {Marconi}, A., {Cresci}, G., {et~al.} 2011, \aap, 528, A88,
  \eprint{1007.4180}

\bibitem[{{Gon{\c c}alves} {et~al.}(2010){Gon{\c c}alves}, {Basu-Zych},
  {Overzier}, {Martin}, {Law}, {Schiminovich}, {Wyder}, {Mallery}, {Rich}, \&
  {Heckman}}]{2010ApJ...724.1373G}
{Gon{\c c}alves}, T.~S., {Basu-Zych}, A., {Overzier}, R., {et~al.} 2010, \apj,
  724, 1373, \eprint{1009.4934}

\bibitem[{{Haan} {et~al.}(2011){Haan}, {Armus}, {Laine}, {Charmandaris},
  {Smith}, {Schweizer}, {Brandl}, {Evans}, {Surace}, {Diaz-Santos},
  {Beir{\~a}o}, {Murphy}, {Stierwalt}, {Hibbard}, {Yun}, \&
  {Jarrett}}]{2011ApJS..197...27H}
{Haan}, S., {Armus}, L., {Laine}, S., {et~al.} 2011, \apjs, 197, 27,
  \eprint{1110.3046}

\bibitem[{{Hopkins} {et~al.}(2008){Hopkins}, {Cox}, {Kere{\v s}}, \&
  {Hernquist}}]{2008ApJS..175..390H}
{Hopkins}, P.~F., {Cox}, T.~J., {Kere{\v s}}, D., \& {Hernquist}, L. 2008,
  \apjs, 175, 390, \eprint{0706.1246}

\bibitem[{{Hopkins} {et~al.}(2009{\natexlab{a}}){Hopkins}, {Cox}, {Younger}, \&
  {Hernquist}}]{2009ApJ...691.1168H}
{Hopkins}, P.~F., {Cox}, T.~J., {Younger}, J.~D., \& {Hernquist}, L.
  2009{\natexlab{a}}, \apj, 691, 1168, \eprint{0806.1739}

\bibitem[{{Hopkins} {et~al.}(2006){Hopkins}, {Hernquist}, {Cox}, {Di Matteo},
  {Robertson}, \& {Springel}}]{2006ApJS..163....1H}
{Hopkins}, P.~F., {Hernquist}, L., {Cox}, T.~J., {et~al.} 2006, \apjs, 163, 1,
  \eprint{astro-ph/0506398}

\bibitem[{{Hopkins} {et~al.}(2009{\natexlab{b}}){Hopkins}, {Somerville}, {Cox},
  {Hernquist}, {Jogee}, {Kere{\v s}}, {Ma}, {Robertson}, \&
  {Stewart}}]{2009MNRAS.397..802H}
{Hopkins}, P.~F., {Somerville}, R.~S., {Cox}, T.~J., {et~al.}
  2009{\natexlab{b}}, \mnras, 397, 802, \eprint{0901.4111}

\bibitem[{{Hung} {et~al.}(2015){Hung}, {Rich}, {Yuan}, {Larson}, {Casey},
  {Smith}, {Sanders}, {Kewley}, \& {Hayward}}]{2015ApJ...803...62H}
{Hung}, C.-L., {Rich}, J.~A., {Yuan}, T., {et~al.} 2015, \apj, 803, 62,
  \eprint{1503.05191}

\bibitem[{{Husemann} {et~al.}(2013){Husemann}, {Jahnke}, {S{\'a}nchez},
  {Barrado}, {Bekerait*error*{\.e}}, {Bomans}, {Castillo-Morales},
  {Catal{\'a}n-Torrecilla}, {Cid Fernandes}, {Falc{\'o}n-Barroso},
  {Garc{\'{\i}}a-Benito}, {Gonz{\'a}lez Delgado}, {Iglesias-P{\'a}ramo},
  {Johnson}, {Kupko}, {L{\'o}pez-Fernandez}, {Lyubenova}, {Marino}, {Mast},
  {Miskolczi}, {Monreal-Ibero}, {Gil de Paz}, {P{\'e}rez}, {P{\'e}rez},
  {Rosales-Ortega}, {Ruiz-Lara}, {Schilling}, {van de Ven}, {Walcher}, {Alves},
  {de Amorim}, {Backsmann}, {Barrera-Ballesteros}, {Bland-Hawthorn}, {Cortijo},
  {Dettmar}, {Demleitner}, {D{\'{\i}}az}, {Enke}, {Florido}, {Flores},
  {Galbany}, {Gallazzi}, {Garc{\'{\i}}a-Lorenzo}, {Gomes}, {Gruel}, {Haines},
  {Holmes}, {Jungwiert}, {Kalinova}, {Kehrig}, {Kennicutt}, {Klar}, {Lehnert},
  {L{\'o}pez-S{\'a}nchez}, {de Lorenzo-C{\'a}ceres}, {M{\'a}rmol-Queralt{\'o}},
  {M{\'a}rquez}, {Mendez-Abreu}, {Moll{\'a}}, {del Olmo}, {Meidt}, {Papaderos},
  {Puschnig}, {Quirrenbach}, {Roth}, {S{\'a}nchez-Bl{\'a}zquez}, {Spekkens},
  {Singh}, {Stanishev}, {Trager}, {Vilchez}, {Wild}, {Wisotzki}, {Zibetti}, \&
  {Ziegler}}]{2013A&A...549A..87H}
{Husemann}, B., {Jahnke}, K., {S{\'a}nchez}, S.~F., {et~al.} 2013, \aap, 549,
  A87, \eprint{1210.8150}

\bibitem[{{Jesseit} {et~al.}(2009){Jesseit}, {Cappellari}, {Naab}, {Emsellem},
  \& {Burkert}}]{2009MNRAS.397.1202J}
{Jesseit}, R., {Cappellari}, M., {Naab}, T., {Emsellem}, E., \& {Burkert}, A.
  2009, \mnras, 397, 1202, \eprint{0810.0137}

\bibitem[{{Jesseit} {et~al.}(2007){Jesseit}, {Naab}, {Peletier}, \&
  {Burkert}}]{2007MNRAS.376..997J}
{Jesseit}, R., {Naab}, T., {Peletier}, R.~F., \& {Burkert}, A. 2007, \mnras,
  376, 997, \eprint{arXiv:astro-ph/0606144}

\bibitem[{{Ji} {et~al.}(2014){Ji}, {Peirani}, \& {Yi}}]{2014A&A...566A..97J}
{Ji}, I., {Peirani}, S., \& {Yi}, S.~K. 2014, \aap, 566, A97,
  \eprint{1405.1807}

\bibitem[{{Karl} {et~al.}(2010){Karl}, {Naab}, {Johansson}, {Kotarba}, {Boily},
  {Renaud}, \& {Theis}}]{2010ApJ...715L..88K}
{Karl}, S.~J., {Naab}, T., {Johansson}, P.~H., {et~al.} 2010, \apjl, 715, L88,
  \eprint{1003.0685}

\bibitem[{{Kronberger} {et~al.}(2007){Kronberger}, {Kapferer}, {Schindler}, \&
  {Ziegler}}]{2007A&A...473..761K}
{Kronberger}, T., {Kapferer}, W., {Schindler}, S., \& {Ziegler}, B.~L. 2007,
  \aap, 473, 761, \eprint{0707.2301}

\bibitem[{{Lagos} {et~al.}(2015){Lagos}, {Padilla}, {Davis}, {Lacey}, {Baugh},
  {Gonzalez-Perez}, {Zwaan}, \& {Contreras}}]{2015MNRAS.448.1271L}
{Lagos}, C.~d.~P., {Padilla}, N.~D., {Davis}, T.~A., {et~al.} 2015, \mnras,
  448, 1271, \eprint{1410.5437}

\bibitem[{{Law} {et~al.}(2009){Law}, {Steidel}, {Erb}, {Larkin}, {Pettini},
  {Shapley}, \& {Wright}}]{2009ApJ...697.2057L}
{Law}, D.~R., {Steidel}, C.~C., {Erb}, D.~K., {et~al.} 2009, \apj, 697, 2057,
  \eprint{0901.2930}

\bibitem[{{Lotz} {et~al.}(2008){Lotz}, {Jonsson}, {Cox}, \&
  {Primack}}]{2008MNRAS.391.1137L}
{Lotz}, J.~M., {Jonsson}, P., {Cox}, T.~J., \& {Primack}, J.~R. 2008, \mnras,
  391, 1137, \eprint{0805.1246}

\bibitem[{{Mancini} {et~al.}(2011){Mancini}, {F{\"o}rster Schreiber},
  {Renzini}, {Cresci}, {Hicks}, {Peng}, {Vergani}, {Lilly}, {Carollo},
  {Pozzetti}, {Zamorani}, {Daddi}, {Genzel}, {Maraston}, {McCracken},
  {Tacconi}, {Bouch{\'e}}, {Davies}, {Oesch}, {Shapiro}, {Mainieri}, {Lutz},
  {Mignoli}, \& {Sternberg}}]{2011ApJ...743...86M}
{Mancini}, C., {F{\"o}rster Schreiber}, N.~M., {Renzini}, A., {et~al.} 2011,
  \apj, 743, 86, \eprint{1109.5952}

\bibitem[{{Mast} {et~al.}(2014){Mast}, {Rosales-Ortega}, {S{\'a}nchez},
  {V{\'{\i}}lchez}, {Iglesias-Paramo}, {Walcher}, {Husemann}, {M{\'a}rquez},
  {Marino}, {Kennicutt}, {Monreal-Ibero}, {Galbany}, {de Lorenzo-C{\'a}ceres},
  {Mendez-Abreu}, {Kehrig}, {del Olmo}, {Rela{\~n}o}, {Wisotzki},
  {M{\'a}rmol-Queralt{\'o}}, {Bekerait{\`e}}, {Papaderos}, {Wild}, {Aguerri},
  {Falc{\'o}n-Barroso}, {Bomans}, {Ziegler}, {Garc{\'{\i}}a-Lorenzo},
  {Bland-Hawthorn}, {L{\'o}pez-S{\'a}nchez}, \& {van de
  Ven}}]{2014A&A...561A.129M}
{Mast}, D., {Rosales-Ortega}, F.~F., {S{\'a}nchez}, S.~F., {et~al.} 2014, \aap,
  561, A129, \eprint{1311.3941}

\bibitem[{{Medling} {et~al.}(2014){Medling}, {U}, {Guedes}, {Max}, {Mayer},
  {Armus}, {Holden}, {Ro{\v s}kar}, \& {Sanders}}]{2014ApJ...784...70M}
{Medling}, A.~M., {U}, V., {Guedes}, J., {et~al.} 2014, \apj, 784, 70,
  \eprint{1401.7338}

\bibitem[{{Mezcua} {et~al.}(2014){Mezcua}, {Lobanov}, {Mediavilla}, \&
  {Karouzos}}]{2014ApJ...784...16M}
{Mezcua}, M., {Lobanov}, A.~P., {Mediavilla}, E., \& {Karouzos}, M. 2014, \apj,
  784, 16, \eprint{1401.5920}

\bibitem[{{Moreno} {et~al.}(2013){Moreno}, {Bluck}, {Ellison}, {Patton},
  {Torrey}, \& {Moster}}]{2013MNRAS.436.1765M}
{Moreno}, J., {Bluck}, A.~F.~L., {Ellison}, S.~L., {et~al.} 2013, \mnras, 436,
  1765, \eprint{1309.1191}

\bibitem[{{Naab} \& {Burkert}(2003)}]{2003ApJ...597..893N}
{Naab}, T. \& {Burkert}, A. 2003, \apj, 597, 893,
  \eprint{arXiv:astro-ph/0110179}

\bibitem[{{Naab} {et~al.}(2006){Naab}, {Jesseit}, \&
  {Burkert}}]{2006MNRAS.372..839N}
{Naab}, T., {Jesseit}, R., \& {Burkert}, A. 2006, \mnras, 372, 839,
  \eprint{astro-ph/0605155}

\bibitem[{{Patton} {et~al.}(2013){Patton}, {Torrey}, {Ellison}, {Mendel}, \&
  {Scudder}}]{2013MNRAS.433L..59P}
{Patton}, D.~R., {Torrey}, P., {Ellison}, S.~L., {Mendel}, J.~T., \& {Scudder},
  J.~M. 2013, \mnras, 433, L59, \eprint{1305.1595}

\bibitem[{{Piqueras L{\'o}pez} {et~al.}(2012){Piqueras L{\'o}pez}, {Colina},
  {Arribas}, {Alonso-Herrero}, \& {Bedregal}}]{2012A&A...546A..64P}
{Piqueras L{\'o}pez}, J., {Colina}, L., {Arribas}, S., {Alonso-Herrero}, A., \&
  {Bedregal}, A.~G. 2012, \aap, 546, A64, \eprint{1209.4181}

\bibitem[{{Porter} {et~al.}(2012){Porter}, {Somerville}, {Croton}, {Covington},
  {Graves}, {Faber}, \& {Primack}}]{2012arXiv1201.5918P}
{Porter}, L.~A., {Somerville}, R.~S., {Croton}, D.~J., {et~al.} 2012, ArXiv
  e-prints, \eprint{1201.5918}

\bibitem[{{Privon} {et~al.}(2013){Privon}, {Barnes}, {Evans}, {Hibbard}, \&
  {Mazzarella}}]{2013AAS...22140501P}
{Privon}, G.~C., {Barnes}, J.~E., {Evans}, A.~S., {Hibbard}, J.~E., \&
  {Mazzarella}, J.~M. 2013, in {American Astronomical Society Meeting
  Abstracts}, Vol. 221, {American Astronomical Society Meeting Abstracts}, 405

\bibitem[{{Robertson} \& {Bullock}(2008)}]{2008ApJ...685L..27R}
{Robertson}, B.~E. \& {Bullock}, J.~S. 2008, \apjl, 685, L27,
  \eprint{0808.1100}

\bibitem[{{Roth} {et~al.}(2005){Roth}, {Kelz}, {Fechner}, {Hahn}, {Bauer},
  {Becker}, {B{\"o}hm}, {Christensen}, {Dionies}, {Paschke}, {Popow}, {Wolter},
  {Schmoll}, {Laux}, \& {Altmann}}]{2005PASP..117..620R}
{Roth}, M.~M., {Kelz}, A., {Fechner}, T., {et~al.} 2005, \pasp, 117, 620,
  \eprint{arXiv:astro-ph/0502581}

\bibitem[{{S{\'a}nchez} {et~al.}(2012){S{\'a}nchez}, {Kennicutt}, {Gil de Paz},
  {van de Ven}, {V{\'\i}lchez}, {Wisotzki}, {Walcher}, {Mast}, {Aguerri},
  {Albiol-P{\'e}rez}, {Alonso-Herrero}, {Alves}, {Bakos}, {Bart{\'a}kov{\'a}},
  {Bland-Hawthorn}, {Boselli}, {Bomans}, {Castillo-Morales}, {Cortijo-Ferrero},
  {de Lorenzo-C{\'a}ceres}, {Del Olmo}, {Dettmar}, {D{\'\i}az}, {Ellis},
  {Falc{\'o}n-Barroso}, {Flores}, {Gallazzi}, {Garc{\'\i}a-Lorenzo},
  {Gonz{\'a}lez Delgado}, {Gruel}, {Haines}, {Hao}, {Husemann},
  {Igl{\'e}sias-P{\'a}ramo}, {Jahnke}, {Johnson}, {Jungwiert}, {Kalinova},
  {Kehrig}, {Kupko}, {L{\'o}pez-S{\'a}nchez}, {Lyubenova}, {Marino},
  {M{\'a}rmol-Queralt{\'o}}, {M{\'a}rquez}, {Masegosa}, {Meidt},
  {Mendez-Abreu}, {Monreal-Ibero}, {Montijo}, {Mour{\~a}o}, {Palacios-Navarro},
  {Papaderos}, {Pasquali}, {Peletier}, {P{\'e}rez}, {P{\'e}rez}, {Quirrenbach},
  {Rela{\~n}o}, {Rosales-Ortega}, {Roth}, {Ruiz-Lara},
  {S{\'a}nchez-Bl{\'a}zquez}, {Sengupta}, {Singh}, {Stanishev}, {Trager},
  {Vazdekis}, {Viironen}, {Wild}, {Zibetti}, \&
  {Ziegler}}]{2012A&A...538A...8S}
{S{\'a}nchez}, S.~F., {Kennicutt}, R.~C., {Gil de Paz}, A., {et~al.} 2012,
  {\aa}p, 538, A8, \eprint{1111.0962}

\bibitem[{{Shapiro} {et~al.}(2008){Shapiro}, {Genzel}, {F{\"o}rster Schreiber},
  {Tacconi}, {Bouch{\'e}}, {Cresci}, {Davies}, {Eisenhauer}, {Johansson},
  {Krajnovi{\'c}}, {Lutz}, {Naab}, {Arimoto}, {Arribas}, {Cimatti}, {Colina},
  {Daddi}, {Daigle}, {Erb}, {Hernandez}, {Kong}, {Mignoli}, {Onodera},
  {Renzini}, {Shapley}, \& {Steidel}}]{2008ApJ...682..231S}
{Shapiro}, K.~L., {Genzel}, R., {F{\"o}rster Schreiber}, N.~M., {et~al.} 2008,
  \apj, 682, 231, \eprint{0802.0879}

\bibitem[{{Springel}(2005)}]{2005MNRAS.364.1105S}
{Springel}, V. 2005, \mnras, 364, 1105, \eprint{arXiv:astro-ph/0505010}

\bibitem[{{Surace}(1998)}]{1998PhDT.........1S}
{Surace}, J.~A. 1998, PhD thesis, Institute for Astronomy University of Hawaii
  2680 Woodlawn Dr.~Honolulu, HI 96822

\bibitem[{{Swinbank} {et~al.}(2006){Swinbank}, {Bower}, {Smith}, {Smail},
  {Kneib}, {Ellis}, {Stark}, \& {Bunker}}]{2006MNRAS.368.1631S}
{Swinbank}, A.~M., {Bower}, R.~G., {Smith}, G.~P., {et~al.} 2006, \mnras, 368,
  1631, \eprint{astro-ph/0603042}

\bibitem[{{Toomre} \& {Toomre}(1972)}]{Tom_72}
{Toomre}, A. \& {Toomre}, J. 1972, \apj, 178, 623

\bibitem[{{Valdes} {et~al.}(2004){Valdes}, {Gupta}, {Rose}, {Singh}, \&
  {Bell}}]{2004ApJS..152..251V}
{Valdes}, F., {Gupta}, R., {Rose}, J.~A., {Singh}, H.~P., \& {Bell}, D.~J.
  2004, \apjs, 152, 251, \eprint{arXiv:astro-ph/0402435}

\bibitem[{{Veilleux} {et~al.}(2002){Veilleux}, {Kim}, \&
  {Sanders}}]{2002ApJS..143..315V}
{Veilleux}, S., {Kim}, D.-C., \& {Sanders}, D.~B. 2002, \apjs, 143, 315,
  \eprint{arXiv:astro-ph/0207401}

\bibitem[{{Walcher} {et~al.}(2014){Walcher}, {Wisotzki}, {Bekerait{\'e}},
  {Husemann}, {Iglesias-P{\'a}ramo}, {Backsmann}, {Barrera Ballesteros},
  {Catal{\'a}n-Torrecilla}, {Cortijo}, {del Olmo}, {Garcia Lorenzo},
  {Falc{\'o}n-Barroso}, {Jilkova}, {Kalinova}, {Mast}, {Marino},
  {M{\'e}ndez-Abreu}, {Pasquali}, {S{\'a}nchez}, {Trager}, {Zibetti},
  {Aguerri}, {Alves}, {Bland-Hawthorn}, {Boselli}, {Castillo Morales}, {Cid
  Fernandes}, {Flores}, {Galbany}, {Gallazzi}, {Garc{\'{\i}}a-Benito}, {Gil de
  Paz}, {Gonz{\'a}lez-Delgado}, {Jahnke}, {Jungwiert}, {Kehrig}, {Lyubenova},
  {M{\'a}rquez Perez}, {Masegosa}, {Monreal Ibero}, {P{\'e}rez}, {Quirrenbach},
  {Rosales-Ortega}, {Roth}, {Sanchez-Blazquez}, {Spekkens}, {Tundo}, {van de
  Ven}, {Verheijen}, {Vilchez}, \& {Ziegler}}]{2014A&A...569A...1W}
{Walcher}, C.~J., {Wisotzki}, L., {Bekerait{\'e}}, S., {et~al.} 2014, \aap,
  569, A1, \eprint{1407.2939}

\bibitem[{{Westmoquette} {et~al.}(2012){Westmoquette}, {Clements}, {Bendo}, \&
  {Khan}}]{2012MNRAS.424..416W}
{Westmoquette}, M.~S., {Clements}, D.~L., {Bendo}, G.~J., \& {Khan}, S.~A.
  2012, \mnras, 424, 416, \eprint{1205.0203}

\bibitem[{{Wild} {et~al.}(2014){Wild}, {Rosales-Ortega}, {Falc{\'o}n-Barroso},
  {Garc{\'{\i}}a-Benito}, {Gallazzi}, {Gonz{\'a}lez Delgado}, {Bekerait{\'e}},
  {Pasquali}, {Johansson}, {Garc{\'{\i}}a Lorenzo}, {van de Ven}, {Pawlik},
  {Per{\'e}z}, {Monreal-Ibero}, {Lyubenova}, {Cid Fernandes},
  {M{\'e}ndez-Abreu}, {Barrera-Ballesteros}, {Kehrig}, {Iglesias-P{\'a}ramo},
  {Bomans}, {M{\'a}rquez}, {Johnson}, {Kennicutt}, {Husemann}, {Mast},
  {S{\'a}nchez}, {Walcher}, {Alves}, {Aguerri}, {Alonso Herrero},
  {Bland-Hawthorn}, {Catal{\'a}n-Torrecilla}, {Florido}, {Gomes}, {Jahnke},
  {L{\'o}pez-S{\'a}nchez}, {de Lorenzo-C{\'a}ceres}, {Marino},
  {M{\'a}rmol-Queralt{\'o}}, {Olden}, {del Olmo}, {Papaderos}, {Quirrenbach},
  {V{\'{\i}}lchez}, \& {Ziegler}}]{2014A&A...567A.132W}
{Wild}, V., {Rosales-Ortega}, F., {Falc{\'o}n-Barroso}, J., {et~al.} 2014,
  \aap, 567, A132, \eprint{1405.7814}

\bibitem[{{Yuan} {et~al.}(2010){Yuan}, {Kewley}, \&
  {Sanders}}]{2010ApJ...709..884Y}
{Yuan}, T.-T., {Kewley}, L.~J., \& {Sanders}, D.~B. 2010, \apj, 709, 884,
  \eprint{0911.3728}

\end{thebibliography}

 \begin{appendix}
 \section[Tables]{Table of kinematic properties}
 \label{sec:tables}
 \begin{table*}[!htb]
\tiny
\caption{\label{table_morph} Morphological parameters of the sample of interacting CALIFA galaxies used in this study.
} 
\begin{center}
\renewcommand{\thefootnote}{\alph{footnote}}
\begin{tabular} {c c c c c}
\toprule
 CALIFA id & Name & Morphological& Interaction & Stellar mass       \\
           &      &     type     &    stage    & log(M/M$_{\odot}$) \\
    (1)    &  (2) &     (3)      &     (4)     &     (5)            \\
\midrule
  14 & UGC00312 & Sd & 1\phantom{000} & \phantom{0}9.64 \\
  17 & UGC00335NED02 & E4 & 2\phantom{000} & 10.73 \\
  22 & NGC0169 & Sab & 2\phantom{000} & 10.87 \\
  26 & NGC0192 & Sab & 1\phantom{000} & 10.73 \\
  39 & NGC0444 & Scd & 1\phantom{000} & \phantom{0}9.82 \\
  42 & NGC0477 & Sbc & 1 (2) & 10.38 \\
  44 & NGC0499 & E5 & 1\phantom{000} & 11.26 \\
 119 & NGC1167 & S0 & 4\phantom{000} & 11.58 \\
 127 & NGC1349 & E6 & 4\phantom{000} & 11.14 \\
 155 & UGC03995 & Sb & 2\phantom{000} & 10.81 \\
 156 & NGC2449 & Sab & 1\phantom{000} & 10.74 \\
 165 & UGC04132 & Sbc & 1\phantom{000} & 10.65 \\
 186 & IC2247 & Sab & 1\phantom{000} & 10.39 \\
 213 & NGC2623 & Scd & 3\phantom{000} & 10.33 \\
 274 & IC0540 & Sab & 1\phantom{000} & \phantom{0}9.78 \\
 311 & NGC3106 & Sab & 4\phantom{000} & 11.21 \\
 314 & UGC05498NED01 & Sa & 1\phantom{000} & 10.69 \\
 340 & NGC3303 & S0a & 2\phantom{000} & 11.03 \\
 360 & NGC3406NED01 & S0 & 2\phantom{000} & 11.34 \\
 475 & NGC3991 & Sm & 2\phantom{000} & \phantom{0}9.53 \\
 479 & NGC4003 & S0a & 1\phantom{000} & 10.98 \\
 486 & UGC07012 & Scd & 1\phantom{000} & \phantom{0}9.35 \\
 520 & NGC4211NED02 & S0a & 2\phantom{000} & 10.42 \\
 577 & NGC4676 & Sdm & 2\phantom{000} & 10.73 \\
 589 & NGC4841A & E3 & 2 (1) & 10.94 \\
 593 & UGC08107 & Sa & 3\phantom{000} & 10.96 \\
 633 & NGC5216 & E0 & 2\phantom{000} & 10.48 \\
 634 & NGC5218 & Sab & 2\phantom{000} & 10.57 \\
 663 & IC0944 & Sab & 1\phantom{000} & 11.16 \\
 676 & NGC5378 & Sb & 1\phantom{000} & 10.52 \\
 680 & NGC5394 & Sbc & 2\phantom{000} & 10.38 \\
 740 & NGC5614 & Sa & 2\phantom{000} & 11.28 \\
 746 & NGC5623 & E7 & 4 (0) & 10.80 \\
 758 & NGC5682 & Scd & 1\phantom{000} & \phantom{0}9.33 \\
 766 & NGC5730 & Scd & 1\phantom{000} & \phantom{0}9.61 \\
 767 & NGC5731 & Sd & 1\phantom{000} & \phantom{0}9.30 \\
 769 & UGC09476 & Sbc & 1\phantom{000} & 10.10 \\
 770 & NGC5739 & S0a & 4\phantom{000} & 11.40 \\
 778 & NGC5784 & S0 & 4\phantom{000} & 11.14 \\
 780 & NGC5797 & E7 & 1\phantom{000} & 10.59 \\
 781 & IC1079 & E4 & 1\phantom{000} & 11.06 \\
 785 & UGC09711 & Sab & 1\phantom{000} & 10.76 \\
 795 & NGC5930 & Sab & 2\phantom{000} & 10.55 \\
 796 & NGC5934 & Sb & 2\phantom{000} & 10.93 \\
 797 & UGC09873 & Sb & 1\phantom{000} & \phantom{0}9.96 \\
 801 & NGC5953 & Sa & 2\phantom{000} & 10.33 \\
 802 & ARP220 & Sd & 3\phantom{000} & 10.83 \\
 803 & NGC5957 & Sb & 1\phantom{000} & 10.18 \\
 806 & NGC5966 & E4 & 1\phantom{000} & 11.10 \\
 807 & IC4566 & Sb & 1\phantom{000} & 10.84 \\
 816 & NGC6021 & E5 & 1\phantom{000} & 10.87 \\
\bottomrule
\end{tabular}
\tablefoot{
(1) CALIFA ID number. For the objects not included in the CALIFA mother sample, we tag them as the ID of the CALIFA companion plus 1000.
(2) Name of the galaxy.
(3) Morphological type from visual clasification \cite[see][for details]{2014A&A...569A...1W}.
(4) Interaction stage: 0, 1, 2, 3, 4 represent non-interacting galaxies, objects include in a pair, galaxy in a binary system with clear signatures of interaction, merger remnant with evident tidal features, and merger remnant, respectively.
For galaxies with uncertain interaction stage the alternative stage is present in parenthesis 
(see details of the definition in Sec.\,\ref{sec:stage}).
(5) Stellar masses \cite[see][for details]{2014A&A...569A...1W}.
}
\end{center}
\begin{minipage}{\textwidth}
\end{minipage}
\end{table*} 
   
\begin{table*}[!htb]
\tiny
\addtocounter{table}{-1}
\caption{continue} \label{table_kin_stellar_int}
\begin{center}
\renewcommand{\thefootnote}{\alph{footnote}}
\begin{tabular} {c c c c c}
\toprule
 CALIFA id & Name & Morphological& Interaction & Stellar mass       \\
           &      &     type     &    stage    & log(M/M$_{\odot}$) \\
    (1)    &  (2) &     (3)      &     (4)     &     (5)            \\
\midrule
 822 & UGC10205 & S0a & 3 (0) & 10.86 \\
 828 & UGC10331 & Sc & 1 (1) & \phantom{0}9.73 \\
 832 & NGC6146 & E5 & 1\phantom{000} & 11.50 \\
 833 & NGC6154 & Sab & 4\phantom{000} & 10.84 \\
 843 & UGC10650 & Scd & 1\phantom{000} & \phantom{0}9.07 \\
 844 & NGC6278 & S0a & 1\phantom{000} & 10.48 \\
 846 & UGC10695 & E5 & 2 (1) & 11.28 \\
 850 & NGC6314 & Sab & 1\phantom{000} & 11.15 \\
 852 & UGC10796 & Scd & 1 (2) & \phantom{0}9.24 \\
 858 & UGC10905 & S0a & 4\phantom{000} & 11.28 \\
 860 & NGC6427 & S0 & 1\phantom{000} & 10.60 \\
 871 & NGC6978 & Sb & 1\phantom{000} & 10.97 \\
 873 & UGC11680NED01 & Sb & 2\phantom{000} & 10.96 \\
 874 & NGC7025 & S0a & 3 (0) & 11.26 \\
 877 & UGC11717 & Sab & 1 (2) & 10.80 \\
 882 & NGC7236 & S0 & 2\phantom{000} & 11.12 \\
 883 & UGC11958 & S0 & 2\phantom{000} & 11.19 \\
 892 & VV488NED02 & Sb & 1\phantom{000} & 10.19 \\
 900 & NGC7550 & E4 & 2\phantom{000} & 11.49 \\
 901 & NGC7549 & Sbc & 2\phantom{000} & 10.32 \\
 903 & NGC7562 & E4 & 1\phantom{000} & 11.23 \\
 905 & UGC12494 & Sd & 1\phantom{000} & \phantom{0}9.34 \\
 907 & NGC7608 & Sbc & 1\phantom{000} & \phantom{0}9.98 \\
 913 & NGC7625 & Sa & 3 (0) & \phantom{0}9.97 \\
 915 & NGC7653 & Sb & 1\phantom{000} & 10.39 \\
 916 & NGC7671 & S0 & 1\phantom{000} & 10.85 \\
 922 & UGC12688 & Scd & 2 (4) & \phantom{0}9.91 \\
 923 & NGC7711 & E7 & 4\phantom{000} & 10.96 \\
 925 & NGC7722 & Sab & 3 (0) & 11.24 \\
 927 & NGC7738 & Sb & 4 (0) & 11.01 \\
 932 & NGC7783NED01 & Sa & 2 (1) & 11.31 \\
 935 & UGC12864 & Sc & 3 (0) & \phantom{0}9.94 \\
 939 & NGC4676B & S0a & 2\phantom{000} & 10.02 \\
1014 & UGC00312NOTES01 & S0a & 1\phantom{000} & \phantom{0}9.71 \\
1017 & UGC00335NED01 & E4 & 2\phantom{000} & 10.19 \\
1022 & NGC0169A & S0 & 2\phantom{000} & 10.21 \\
1026 & NGC0197 & S0a & 1\phantom{000} & \phantom{0}9.12 \\
1042 & CGCG536-030 & Sc & 1 (2) & \phantom{0}9.69 \\
1044 & NGC0495 & S0a & 1\phantom{000} & \phantom{0}9.75 \\
1340 & NGC3303NED01 & Sb & 2\phantom{000} & 10.33 \\
1360 & NGC3406NED02 & S0 & 2\phantom{000} & 10.23 \\
1520 & NGC4211NED01 & S0a & 2\phantom{000} & 10.85 \\
1589 & NGC4841B & E0 & 2 (1) & 10.82 \\
1663 & KUG1349+143 & Sb & 1\phantom{000} & 10.16 \\
1740 & NGC5615 & Ir & 2\phantom{000} & \phantom{0}9.28 \\
1780 & NGC5794 & S0 & 1\phantom{000} & 10.69 \\
1781 & IC1078 & Sa & 1\phantom{000} & 10.45 \\
1795 & NGC5929 & Sa & 2\phantom{000} & 10.47 \\
1796 & NGC5935 & S0a & 2\phantom{000} & 10.80 \\
1801 & NGC5954 & Sc & 2\phantom{000} & \phantom{0}9.96 \\
1871 & NGC6977 & Sa & 1\phantom{000} & 10.85 \\
1873 & UGC11680NED02 & E0 & 2\phantom{000} & \phantom{0}9.45 \\
\bottomrule
\end{tabular}
\end{center}
\begin{minipage}{\textwidth}
\end{minipage}
\end{table*} 
\newpage

 \label{table_morph}
 \clearpage
  \begin{table*}[!htb]
\tiny
\caption{\label{table_Skin}
 Stellar Kinematic properties of the interacting sample selected for this study included in the CALIFA survey.
}
\begin{center}
\renewcommand{\thefootnote}{\alph{footnote}}
\begin{tabular} {c c c c c c c c c c c}
\toprule
    id  & V$\mathrm{_{sys}}$  &         r          &  PA$\mathrm{_{morph}}$ (r) &  \multicolumn{2}{c}{PA approaching}                 & \multicolumn{2}{c}{PA receding}                     & \multicolumn{2}{c}{$\Psi_{\mathrm{morph-kin}}$} & $\Psi_{\mathrm{kin-kin}}$ \\
        &                     &                    &                            &  PA$\mathrm{_{kin}}$  & $\delta$PA$\mathrm{_{kin}}$ & PA$\mathrm{_{kin}}$  & $\delta$PA$\mathrm{_{kin}}$  & approaching  & receding                         &                           \\
        &    (km s$^{-1}$)    &      (arcsec)      &            ($^{\circ}$)    &  ($^{\circ}$)         & ($^{\circ}$)                & ($^{\circ}$)         & ($^{\circ}$)                 & ($^{\circ}$) & ($^{\circ}$)                     & ($^{\circ}$)              \\
    (1) &         (2)         &        (3)         &            (4)             &          (5)          &            (6)              &          (7)        &       (8)                     &   (9)       &       (10)                       &   (11)                     \\
\midrule
  14& -- & -- & -- & -- & -- & -- & -- & -- & -- & -- \\
  17&  5409 $\pm$    5 &  5 & 148.8 $\pm$  0.5 & 142.6 $\pm$ 16.6 & \phantom{0}29.5 $\pm$ 15.5 & 145.9 $\pm$ 11.8 & \phantom{0}15.7 $\pm$ 8.3\phantom{0} & \phantom{00}6.2 $\pm$ 16.6 & \phantom{00}2.9 $\pm$ 11.8 & \phantom{00}3.3 $\pm$ 20.4 \\
  22&  4562 $\pm$    8 & 25 & \phantom{0}87.5 $\pm$  0.5 & \phantom{0}85.4 $\pm$ 1.7\phantom{0} & \phantom{00}4.6 $\pm$ 1.2\phantom{0} & \phantom{0}89.3 $\pm$ 2.0\phantom{0} & \phantom{00}7.5 $\pm$ 2.2\phantom{0} & \phantom{00}2.0 $\pm$ 1.7\phantom{0} & \phantom{00}1.9 $\pm$ 2.1\phantom{0} & \phantom{00}3.9 $\pm$ 2.6\phantom{0} \\
  26&  4180 $\pm$    1 & 25 & 164.0 $\pm$  0.2 & 178.2 $\pm$ 3.0\phantom{0} & \phantom{0}10.3 $\pm$ 1.6\phantom{0} & \phantom{00}4.6 $\pm$ 2.4\phantom{0} & \phantom{0}12.4 $\pm$ 2.2\phantom{0} & \phantom{0}14.2 $\pm$ 3.0\phantom{0} & \phantom{0}20.5 $\pm$ 2.4\phantom{0} & \phantom{00}6.3 $\pm$ 3.8\phantom{0} \\
  39& -- & -- & -- & -- & -- & -- & -- & -- & -- & -- \\
  42&  5828 $\pm$    6 & 15 & 115.5 $\pm$  1.3 & 125.7 $\pm$ 5.2\phantom{0} & \phantom{0}18.2 $\pm$ 4.8\phantom{0} & 135.2 $\pm$ 5.8\phantom{0} & \phantom{0}19.2 $\pm$ 3.9\phantom{0} & \phantom{0}10.2 $\pm$ 5.4\phantom{0} & \phantom{0}19.7 $\pm$ 5.9\phantom{0} & \phantom{00}9.5 $\pm$ 7.8\phantom{0} \\
  44&  4375 $\pm$    1 & 25 & \phantom{0}73.5 $\pm$  0.3 & \phantom{0}97.0 $\pm$ 9.0\phantom{0} & \phantom{0}45.1 $\pm$ 11.0 & \phantom{0}84.8 $\pm$ 6.2\phantom{0} & \phantom{0}35.0 $\pm$ 9.4\phantom{0} & \phantom{0}23.4 $\pm$ 9.0\phantom{0} & \phantom{0}11.3 $\pm$ 6.2\phantom{0} & \phantom{0}12.1 $\pm$ 11.0 \\
 119&  4919 $\pm$    4 & 10 & \phantom{0}70.3 $\pm$  1.2 & \phantom{0}65.6 $\pm$ 10.4 & \phantom{0}20.7 $\pm$ 6.0\phantom{0} & \phantom{0}73.1 $\pm$ 6.8\phantom{0} & \phantom{0}14.9 $\pm$ 4.9\phantom{0} & \phantom{00}4.6 $\pm$ 10.5 & \phantom{00}2.8 $\pm$ 6.9\phantom{0} & \phantom{00}7.4 $\pm$ 12.4 \\
 127&  6534 $\pm$    3 & 17 & \phantom{0}35.6 $\pm$  2.3 & \phantom{0}36.1 $\pm$ 5.2\phantom{0} & \phantom{0}19.5 $\pm$ 4.0\phantom{0} & \phantom{0}43.4 $\pm$ 4.7\phantom{0} & \phantom{0}20.6 $\pm$ 4.4\phantom{0} & \phantom{00}0.5 $\pm$ 5.7\phantom{0} & \phantom{00}7.8 $\pm$ 5.2\phantom{0} & \phantom{00}7.4 $\pm$ 7.0\phantom{0} \\
 155&  4753 $\pm$    6 & 22 & 109.2 $\pm$  0.8 & \phantom{0}80.4 $\pm$ 2.9\phantom{0} & \phantom{0}13.5 $\pm$ 3.2\phantom{0} & \phantom{0}73.7 $\pm$ 3.9\phantom{0} & \phantom{0}16.8 $\pm$ 2.2\phantom{0} & \phantom{0}28.8 $\pm$ 3.0\phantom{0} & \phantom{0}35.5 $\pm$ 3.9\phantom{0} & \phantom{00}6.7 $\pm$ 4.8\phantom{0} \\
 156&  4892 $\pm$    4 & 25 & 127.7 $\pm$  0.2 & 134.7 $\pm$ 2.4\phantom{0} & \phantom{0}10.4 $\pm$ 1.3\phantom{0} & 130.1 $\pm$ 2.1\phantom{0} & \phantom{00}8.9 $\pm$ 1.6\phantom{0} & \phantom{00}6.9 $\pm$ 2.4\phantom{0} & \phantom{00}2.3 $\pm$ 2.1\phantom{0} & \phantom{00}4.6 $\pm$ 3.2\phantom{0} \\
 165&  5154 $\pm$    6 & 22 & \phantom{0}23.4 $\pm$  0.8 & \phantom{0}34.9 $\pm$ 2.4\phantom{0} & \phantom{00}7.6 $\pm$ 1.6\phantom{0} & \phantom{0}22.1 $\pm$ 2.9\phantom{0} & \phantom{00}8.4 $\pm$ 1.8\phantom{0} & \phantom{0}11.5 $\pm$ 2.6\phantom{0} & \phantom{00}1.3 $\pm$ 3.0\phantom{0} & \phantom{0}12.8 $\pm$ 3.8\phantom{0} \\
 186&  4257 $\pm$    2 & 20 & 149.0 $\pm$  0.6 & 152.4 $\pm$ 2.2\phantom{0} & \phantom{00}7.5 $\pm$ 1.7\phantom{0} & 143.4 $\pm$ 2.9\phantom{0} & \phantom{0}10.7 $\pm$ 2.1\phantom{0} & \phantom{00}3.3 $\pm$ 2.3\phantom{0} & \phantom{00}5.6 $\pm$ 2.9\phantom{0} & \phantom{00}9.0 $\pm$ 3.6\phantom{0} \\
 213&  5469 $\pm$    5 &  7 & \phantom{0}81.3 $\pm$  1.6 & \phantom{0}77.8 $\pm$ 10.0 & \phantom{0}24.8 $\pm$ 13.5 & 112.7 $\pm$ 14.3 & \phantom{0}29.5 $\pm$ 14.4 & \phantom{00}3.5 $\pm$ 10.1 & \phantom{0}31.4 $\pm$ 14.4 & \phantom{0}34.9 $\pm$ 17.5 \\
 274&  2068 $\pm$    3 & 12 & 169.8 $\pm$  0.7 & 173.3 $\pm$ 4.6\phantom{0} & \phantom{0}16.6 $\pm$ 5.5\phantom{0} & \phantom{0}10.2 $\pm$ 5.5\phantom{0} & \phantom{0}14.5 $\pm$ 6.2\phantom{0} & \phantom{00}3.5 $\pm$ 5.1\phantom{0} & \phantom{0}20.4 $\pm$ 5.8\phantom{0} & \phantom{0}16.9 $\pm$ 7.3\phantom{0} \\
 311&  6155 $\pm$    2 & 20 & 132.2 $\pm$  2.4 & 142.0 $\pm$ 4.6\phantom{0} & \phantom{0}27.7 $\pm$ 4.1\phantom{0} & 152.1 $\pm$ 5.3\phantom{0} & \phantom{0}22.8 $\pm$ 4.9\phantom{0} & \phantom{00}9.8 $\pm$ 5.2\phantom{0} & \phantom{0}19.9 $\pm$ 5.8\phantom{0} & \phantom{0}10.1 $\pm$ 7.0\phantom{0} \\
 314&  6248 $\pm$    4 & 20 & \phantom{0}59.4 $\pm$  0.8 & \phantom{0}61.7 $\pm$ 2.6\phantom{0} & \phantom{00}7.2 $\pm$ 1.5\phantom{0} & \phantom{0}59.0 $\pm$ 2.4\phantom{0} & \phantom{00}7.1 $\pm$ 1.5\phantom{0} & \phantom{00}2.3 $\pm$ 2.7\phantom{0} & \phantom{00}0.5 $\pm$ 2.5\phantom{0} & \phantom{00}2.8 $\pm$ 3.5\phantom{0} \\
 340&  6138 $\pm$    7 & 12 & 153.2 $\pm$  0.8 & \phantom{0}16.9 $\pm$ 7.6\phantom{0} & \phantom{0}18.1 $\pm$ 3.3\phantom{0} & \phantom{0}11.9 $\pm$ 6.7\phantom{0} & \phantom{0}19.6 $\pm$ 5.8\phantom{0} & \phantom{0}43.7 $\pm$ 7.6\phantom{0} & \phantom{0}38.7 $\pm$ 6.8\phantom{0} & \phantom{00}5.1 $\pm$ 10.1 \\
 360&  7422 $\pm$    4 & 15 & \phantom{0}73.0 $\pm$  1.8 & \phantom{0}48.0 $\pm$ 2.2\phantom{0} & \phantom{0}10.0 $\pm$ 2.7\phantom{0} & \phantom{0}71.1 $\pm$ 5.5\phantom{0} & \phantom{0}19.9 $\pm$ 4.0\phantom{0} & \phantom{0}25.0 $\pm$ 2.8\phantom{0} & \phantom{00}1.8 $\pm$ 5.8\phantom{0} & \phantom{0}23.1 $\pm$ 5.9\phantom{0} \\
 475& -- & -- & -- & -- & -- & -- & -- & -- & -- & -- \\
 479&  6527 $\pm$    4 & 22 & 146.0 $\pm$  0.4 & 169.8 $\pm$ 3.0\phantom{0} & \phantom{0}10.5 $\pm$ 2.1\phantom{0} & 176.2 $\pm$ 2.5\phantom{0} & \phantom{0}11.2 $\pm$ 2.3\phantom{0} & \phantom{0}23.8 $\pm$ 3.0\phantom{0} & \phantom{0}30.2 $\pm$ 2.5\phantom{0} & \phantom{00}6.4 $\pm$ 3.9\phantom{0} \\
 486&  3093 $\pm$    1 & 18 & \phantom{0}14.0 $\pm$  1.0 & \phantom{00}5.1 $\pm$ 12.0 & \phantom{0}47.5 $\pm$ 11.8 & \phantom{0}23.5 $\pm$ 10.6 & \phantom{0}25.1 $\pm$ 6.8\phantom{0} & \phantom{00}8.9 $\pm$ 12.0 & \phantom{00}9.6 $\pm$ 10.7 & \phantom{0}18.4 $\pm$ 16.0 \\
 520&  6607 $\pm$    4 &  7 & 170.0 $\pm$  0.0 & \phantom{0}24.1 $\pm$ 10.2 & \phantom{0}20.2 $\pm$ 5.7\phantom{0} & \phantom{0}20.4 $\pm$ 9.0\phantom{0} & \phantom{0}18.6 $\pm$ 10.1 & \phantom{0}34.1 $\pm$ 10.2 & \phantom{0}30.4 $\pm$ 9.0\phantom{0} & \phantom{00}3.6 $\pm$ 13.6 \\
 577&  6591 $\pm$    4 & 10 & \phantom{00}2.8 $\pm$  0.9 & \phantom{0}33.7 $\pm$ 5.7\phantom{0} & \phantom{0}14.9 $\pm$ 4.1\phantom{0} & \phantom{0}27.1 $\pm$ 6.8\phantom{0} & \phantom{0}17.4 $\pm$ 5.8\phantom{0} & \phantom{0}30.9 $\pm$ 5.8\phantom{0} & \phantom{0}24.3 $\pm$ 6.9\phantom{0} & \phantom{00}6.6 $\pm$ 8.9\phantom{0} \\
 589& -- & -- & -- & -- & -- & -- & -- & -- & -- & -- \\
 593&  8218 $\pm$    2 & 20 & \phantom{0}48.0 $\pm$  0.8 & \phantom{0}47.9 $\pm$ 2.4\phantom{0} & \phantom{0}12.2 $\pm$ 2.6\phantom{0} & \phantom{0}49.8 $\pm$ 3.1\phantom{0} & \phantom{0}17.7 $\pm$ 3.6\phantom{0} & \phantom{00}0.1 $\pm$ 2.5\phantom{0} & \phantom{00}1.8 $\pm$ 3.2\phantom{0} & \phantom{00}1.9 $\pm$ 3.9\phantom{0} \\
 633&  2940 $\pm$    1 & 12 & \phantom{0}88.8 $\pm$  2.3 & \phantom{0}98.5 $\pm$ 7.9\phantom{0} & \phantom{0}25.8 $\pm$ 7.4\phantom{0} & 103.9 $\pm$ 5.2\phantom{0} & \phantom{0}12.1 $\pm$ 4.3\phantom{0} & \phantom{00}9.8 $\pm$ 8.2\phantom{0} & \phantom{0}15.1 $\pm$ 5.7\phantom{0} & \phantom{00}5.4 $\pm$ 9.4\phantom{0} \\
 634&  2916 $\pm$    4 & 20 & \phantom{0}91.3 $\pm$  0.7 & \phantom{0}70.9 $\pm$ 2.6\phantom{0} & \phantom{0}10.0 $\pm$ 2.3\phantom{0} & \phantom{0}55.9 $\pm$ 3.2\phantom{0} & \phantom{00}9.5 $\pm$ 1.8\phantom{0} & \phantom{0}20.4 $\pm$ 2.7\phantom{0} & \phantom{0}35.4 $\pm$ 3.2\phantom{0} & \phantom{0}15.1 $\pm$ 4.1\phantom{0} \\
 663&  6901 $\pm$    6 & 20 & 108.3 $\pm$  0.3 & 106.4 $\pm$ 2.1\phantom{0} & \phantom{00}7.7 $\pm$ 2.3\phantom{0} & 104.8 $\pm$ 2.1\phantom{0} & \phantom{00}6.2 $\pm$ 1.6\phantom{0} & \phantom{00}2.0 $\pm$ 2.1\phantom{0} & \phantom{00}3.6 $\pm$ 2.1\phantom{0} & \phantom{00}1.6 $\pm$ 3.0\phantom{0} \\
 676&  2964 $\pm$    2 &  6 & \phantom{0}65.6 $\pm$  0.7 & \phantom{0}80.5 $\pm$ 12.8 & \phantom{0}22.2 $\pm$ 12.0 & 101.9 $\pm$ 13.8 & \phantom{0}21.1 $\pm$ 13.9 & \phantom{0}14.9 $\pm$ 12.8 & \phantom{0}36.3 $\pm$ 13.8 & \phantom{0}21.4 $\pm$ 18.8 \\
 680& -- & -- & -- & -- & -- & -- & -- & -- * & -- * & -- * \\
 740&  3866 $\pm$    3 & 20 & \phantom{00}2.2 $\pm$  2.7 & \phantom{0}64.4 $\pm$ 4.6\phantom{0} & \phantom{0}17.7 $\pm$ 3.5\phantom{0} & \phantom{0}64.7 $\pm$ 4.9\phantom{0} & \phantom{0}18.7 $\pm$ 3.7\phantom{0} & \phantom{0}62.2 $\pm$ 5.3\phantom{0} & \phantom{0}62.4 $\pm$ 5.6\phantom{0} & \phantom{00}0.2 $\pm$ 6.7\phantom{0} \\
 746&  3403 $\pm$    2 & 10 & \phantom{0}26.6 $\pm$  0.3 & \phantom{0}35.5 $\pm$ 3.7\phantom{0} & \phantom{0}11.5 $\pm$ 3.2\phantom{0} & \phantom{0}27.7 $\pm$ 3.4\phantom{0} & \phantom{00}7.8 $\pm$ 3.0\phantom{0} & \phantom{00}8.9 $\pm$ 3.7\phantom{0} & \phantom{00}1.0 $\pm$ 3.4\phantom{0} & \phantom{00}7.8 $\pm$ 5.0\phantom{0} \\
 758& -- & -- & -- & -- & -- & -- & -- & -- & -- & -- \\
 767& -- & -- & -- & -- & -- & -- & -- & -- & -- & -- \\
 769&  3248 $\pm$    2 & 20 & 101.9 $\pm$  1.0 & 128.9 $\pm$ 4.6\phantom{0} & \phantom{0}24.5 $\pm$ 4.0\phantom{0} & 138.0 $\pm$ 6.3\phantom{0} & \phantom{0}26.9 $\pm$ 5.2\phantom{0} & \phantom{0}27.1 $\pm$ 4.7\phantom{0} & \phantom{0}36.2 $\pm$ 6.4\phantom{0} & \phantom{00}9.1 $\pm$ 7.8\phantom{0} \\
 770&  5461 $\pm$    6 & 12 & \phantom{0}30.6 $\pm$  0.6 & \phantom{0}40.3 $\pm$ 5.1\phantom{0} & \phantom{0}14.3 $\pm$ 3.6\phantom{0} & \phantom{0}34.1 $\pm$ 6.8\phantom{0} & \phantom{0}17.2 $\pm$ 3.6\phantom{0} & \phantom{00}9.7 $\pm$ 5.1\phantom{0} & \phantom{00}3.6 $\pm$ 6.8\phantom{0} & \phantom{00}6.2 $\pm$ 8.4\phantom{0} \\
 778&  5459 $\pm$    1 & 15 & \phantom{0}65.8 $\pm$  0.8 & \phantom{0}63.0 $\pm$ 4.6\phantom{0} & \phantom{0}15.7 $\pm$ 2.8\phantom{0} & \phantom{0}70.1 $\pm$ 3.4\phantom{0} & \phantom{0}17.4 $\pm$ 4.1\phantom{0} & \phantom{00}2.9 $\pm$ 4.6\phantom{0} & \phantom{00}4.3 $\pm$ 3.5\phantom{0} & \phantom{00}7.2 $\pm$ 5.7\phantom{0} \\
 780&  4024 $\pm$    3 & 17 & 112.0 $\pm$  0.3 & 107.9 $\pm$ 3.1\phantom{0} & \phantom{0}10.2 $\pm$ 2.4\phantom{0} & 113.9 $\pm$ 4.4\phantom{0} & \phantom{0}11.6 $\pm$ 2.1\phantom{0} & \phantom{00}4.2 $\pm$ 3.2\phantom{0} & \phantom{00}1.9 $\pm$ 4.4\phantom{0} & \phantom{00}6.1 $\pm$ 5.4\phantom{0} \\
 781& -- & -- & -- & -- & -- & -- & -- & -- * & -- * & -- * \\
 785& -- & -- & -- & -- & -- & -- & -- & -- & -- & -- \\
 795&  2629 $\pm$    2 & 15 & 163.9 $\pm$  0.7 & 158.0 $\pm$ 3.4\phantom{0} & \phantom{00}7.7 $\pm$ 1.9\phantom{0} & 157.3 $\pm$ 4.9\phantom{0} & \phantom{0}12.5 $\pm$ 2.5\phantom{0} & \phantom{00}6.0 $\pm$ 3.5\phantom{0} & \phantom{00}6.6 $\pm$ 5.0\phantom{0} & \phantom{00}0.6 $\pm$ 6.0\phantom{0} \\
 796&  5584 $\pm$    7 & 15 & \phantom{00}2.0 $\pm$  0.6 & 170.7 $\pm$ 3.3\phantom{0} & \phantom{00}9.5 $\pm$ 2.7\phantom{0} & 179.6 $\pm$ 3.0\phantom{0} & \phantom{00}9.0 $\pm$ 3.9\phantom{0} & \phantom{0}11.3 $\pm$ 3.4\phantom{0} & \phantom{00}2.4 $\pm$ 3.0\phantom{0} & \phantom{00}8.9 $\pm$ 4.5\phantom{0} \\
 797&  5577 $\pm$    2 & 15 & 121.4 $\pm$  0.4 & 130.6 $\pm$ 3.7\phantom{0} & \phantom{0}11.5 $\pm$ 2.9\phantom{0} & 125.9 $\pm$ 6.9\phantom{0} & \phantom{0}19.2 $\pm$ 4.5\phantom{0} & \phantom{00}9.2 $\pm$ 3.7\phantom{0} & \phantom{00}4.5 $\pm$ 6.9\phantom{0} & \phantom{00}4.7 $\pm$ 7.8\phantom{0} \\
 801&  2010 $\pm$    1 & 11 & \phantom{0}31.2 $\pm$  1.6 & \phantom{0}63.1 $\pm$ 9.6\phantom{0} & \phantom{0}32.5 $\pm$ 18.0 & \phantom{0}55.7 $\pm$ 17.1 & \phantom{0}58.2 $\pm$ 18.0 & \phantom{0}31.8 $\pm$ 9.7\phantom{0} & \phantom{0}24.4 $\pm$ 17.2 & \phantom{00}7.4 $\pm$ 19.6 \\
 802&  5433 $\pm$    2 & 11 & 178.2 $\pm$  2.5 & \phantom{0}34.2 $\pm$ 11.9 & \phantom{0}37.1 $\pm$ 9.0\phantom{0} & \phantom{0}38.9 $\pm$ 19.1 & \phantom{0}35.7 $\pm$ 11.3 & \phantom{0}36.0 $\pm$ 12.2 & \phantom{0}40.7 $\pm$ 19.2 & \phantom{00}4.7 $\pm$ 22.5 \\
 803&  1823 $\pm$    0 & 27 & \phantom{0}96.6 $\pm$  0.9 & \phantom{0}34.2 $\pm$ 4.7\phantom{0} & \phantom{0}29.4 $\pm$ 5.2\phantom{0} & \phantom{0}62.1 $\pm$ 18.6 & \phantom{0}65.6 $\pm$ 40.9 & \phantom{0}62.4 $\pm$ 4.8\phantom{0} & \phantom{0}34.5 $\pm$ 18.6 & \phantom{0}27.9 $\pm$ 19.2 \\
 806&  4483 $\pm$    1 & 12 & \phantom{0}92.1 $\pm$  0.3 & \phantom{0}89.3 $\pm$ 3.9\phantom{0} & \phantom{0}17.8 $\pm$ 7.5\phantom{0} & \phantom{0}85.8 $\pm$ 6.4\phantom{0} & \phantom{0}21.1 $\pm$ 6.5\phantom{0} & \phantom{00}2.8 $\pm$ 3.9\phantom{0} & \phantom{00}6.3 $\pm$ 6.4\phantom{0} & \phantom{00}3.5 $\pm$ 7.5\phantom{0} \\
 807&  5549 $\pm$    1 & 20 & 154.5 $\pm$  0.6 & 151.9 $\pm$ 3.0\phantom{0} & \phantom{0}17.9 $\pm$ 3.7\phantom{0} & 151.5 $\pm$ 3.4\phantom{0} & \phantom{0}14.9 $\pm$ 2.5\phantom{0} & \phantom{00}2.6 $\pm$ 3.0\phantom{0} & \phantom{00}3.0 $\pm$ 3.5\phantom{0} & \phantom{00}0.4 $\pm$ 4.5\phantom{0} \\
 816&  4762 $\pm$    2 & 12 & 149.9 $\pm$  0.5 & 146.4 $\pm$ 3.1\phantom{0} & \phantom{0}10.0 $\pm$ 2.7\phantom{0} & 144.1 $\pm$ 6.6\phantom{0} & \phantom{0}17.5 $\pm$ 5.4\phantom{0} & \phantom{00}3.5 $\pm$ 3.2\phantom{0} & \phantom{00}5.7 $\pm$ 6.6\phantom{0} & \phantom{00}2.2 $\pm$ 7.3\phantom{0} \\
 822&  6499 $\pm$    2 & 15 & 132.4 $\pm$  1.6 & 141.9 $\pm$ 6.1\phantom{0} & \phantom{0}30.2 $\pm$ 5.7\phantom{0} & 127.0 $\pm$ 6.7\phantom{0} & \phantom{0}23.0 $\pm$ 5.2\phantom{0} & \phantom{00}9.4 $\pm$ 6.3\phantom{0} & \phantom{00}5.5 $\pm$ 6.9\phantom{0} & \phantom{0}14.9 $\pm$ 9.0\phantom{0} \\
\bottomrule
\end{tabular}
\tablefoot{
(1) CALIFA id.
(2) Systemic velocity derived from integrated the velocities in a 2.7$\arcsec$ aperture centred in the kinematic centre.
(3) Radius used to average the polar coordinates of the positions from the lines of nodes (see Sec.\,\ref{sec:Analysis} for details).
(4) Morphological PA inferred by fitting an elipse to an isophote at radius r in the r-band SDSS image.
(5) and (6) Kinematic PA and its standard deviation at radius r for the approaching side (see Sec.\,\ref{sec:Analysis} for details).
(7) and (8) Kinematic PA and its standard deviation at radius r for the  receding  side.
(9) and (10) Morpho-kinematic misalignment for the receding and the approaching sides.
(11) Kinematic misalignment between the receding and the approaching sides.
The * symbol tag the objects where it is not posible to determine any kinematic property using our method (see Sec.\,\ref{sec:Analysis} for details).
}
\end{center}
\begin{minipage}{\textwidth}
\end{minipage}
\end{table*} 
   
\begin{table*}[!htb]
\tiny
\addtocounter{table}{-1}
\caption{continue  \ref{table_Skin}}
\begin{center}
\renewcommand{\thefootnote}{\alph{footnote}}
\begin{tabular} {c c c c c c c c c c c}
\toprule
    id  & V$\mathrm{_{sys}}$  &         r          &  PA$\mathrm{_{morph}}$ (r) &  \multicolumn{2}{c}{PA approaching}                 & \multicolumn{2}{c}{PA receding}                     & \multicolumn{2}{c}{$\Psi_{\mathrm{morph-kin}}$} & $\Psi_{\mathrm{kin-kin}}$ \\
        &                     &                    &                            &  PA$\mathrm{_{kin}}$  & $\delta$PA$\mathrm{_{kin}}$ & PA$\mathrm{_{kin}}$  & $\delta$PA$\mathrm{_{kin}}$  & approaching  & receding                         &                           \\
        &    (km s$^{-1}$)    &      (arcsec)      &            ($^{\circ}$)    &  ($^{\circ}$)         & ($^{\circ}$)                & ($^{\circ}$)         & ($^{\circ}$)                 & ($^{\circ}$) & ($^{\circ}$)                     & ($^{\circ}$)              \\
    (1) &         (4)         &        (5)         &            (7)             &          (8)          &            (9)              &          (10)        &       (11)                   &   (10)       &       (11)                       &   (10)                    \\
\midrule
 828& -- & -- & -- & -- & -- & -- & -- & -- & -- & -- \\
 832&  8699 $\pm$    5 & 25 & \phantom{0}74.5 $\pm$  0.5 & \phantom{0}72.6 $\pm$ 3.9\phantom{0} & \phantom{0}16.3 $\pm$ 2.9\phantom{0} & \phantom{0}70.4 $\pm$ 4.5\phantom{0} & \phantom{0}15.6 $\pm$ 2.1\phantom{0} & \phantom{00}1.9 $\pm$ 4.0\phantom{0} & \phantom{00}4.1 $\pm$ 4.6\phantom{0} & \phantom{00}2.2 $\pm$ 6.0\phantom{0} \\
 833&  5931 $\pm$    1 & 20 & 135.6 $\pm$  0.9 & \phantom{0}21.2 $\pm$ 3.0\phantom{0} & \phantom{0}16.8 $\pm$ 4.2\phantom{0} & \phantom{0}11.4 $\pm$ 7.7\phantom{0} & \phantom{0}23.5 $\pm$ 5.1\phantom{0} & \phantom{0}65.6 $\pm$ 3.1\phantom{0} & \phantom{0}55.8 $\pm$ 7.7\phantom{0} & \phantom{00}9.8 $\pm$ 8.2\phantom{0} \\
 843& -- & -- & -- & -- & -- & -- & -- & -- & -- & -- \\
 844&  2837 $\pm$    3 & 25 & 125.6 $\pm$  0.3 & 128.5 $\pm$ 2.3\phantom{0} & \phantom{0} 9.9 $\pm$ 1.4\phantom{0} & 128.7 $\pm$ 2.0\phantom{0} & \phantom{00}9.3 $\pm$ 1.5\phantom{0} & \phantom{00}2.9 $\pm$ 2.4\phantom{0} & \phantom{00}3.1 $\pm$ 2.0\phantom{0} & \phantom{00}0.2 $\pm$ 3.1\phantom{0} \\
 846& -- & -- & -- & -- & -- & -- & -- & -- & -- & -- \\
 850&  6551 $\pm$    1 & 20 & 173.1 $\pm$  0.7 & 173.8 $\pm$ 3.5\phantom{0} & \phantom{0}19.8 $\pm$ 3.4\phantom{0} & 174.4 $\pm$ 2.5\phantom{0} & \phantom{0}11.8 $\pm$ 3.7\phantom{0} & \phantom{00}0.7 $\pm$ 3.5\phantom{0} & \phantom{00}1.3 $\pm$ 2.6\phantom{0} & \phantom{00}0.6 $\pm$ 4.3\phantom{0} \\
 852& -- & -- & -- & -- & -- & -- & -- & -- & -- & -- \\
 858&  7651 $\pm$    6 & 12 & 177.9 $\pm$  0.6 & 171.2 $\pm$ 3.3\phantom{0} & \phantom{0}10.1 $\pm$ 3.2\phantom{0} & \phantom{00}2.0 $\pm$ 2.2\phantom{0} & \phantom{00}8.0 $\pm$ 4.6\phantom{0} & \phantom{00}6.6 $\pm$ 3.3\phantom{0} & \phantom{00}4.1 $\pm$ 2.3\phantom{0} & \phantom{0}10.7 $\pm$ 4.0\phantom{0} \\
 860&  3253 $\pm$    1 & 25 & \phantom{0}35.9 $\pm$  0.3 & \phantom{0}37.5 $\pm$ 1.3\phantom{0} & \phantom{00}7.2 $\pm$ 1.2\phantom{0} & \phantom{0}34.2 $\pm$ 1.3\phantom{0} & \phantom{00}5.9 $\pm$ 1.3\phantom{0} & \phantom{00}1.7 $\pm$ 1.4\phantom{0} & \phantom{00}1.7 $\pm$ 1.4\phantom{0} & \phantom{00}3.3 $\pm$ 1.9\phantom{0} \\
 871&  5923 $\pm$    6 & 20 & 128.4 $\pm$  0.4 & 129.0 $\pm$ 2.3\phantom{0} & \phantom{0}10.1 $\pm$ 2.4\phantom{0} & 125.0 $\pm$ 2.8\phantom{0} & \phantom{00}9.5 $\pm$ 1.8\phantom{0} & \phantom{00}0.6 $\pm$ 2.3\phantom{0} & \phantom{00}3.4 $\pm$ 2.8\phantom{0} & \phantom{00}4.0 $\pm$ 3.6\phantom{0} \\
 873&  7674 $\pm$    7 & 10 & \phantom{0}15.2 $\pm$  1.4 & \phantom{0}76.6 $\pm$ 8.3\phantom{0} & \phantom{0}16.4 $\pm$ 4.3\phantom{0} & \phantom{0}80.6 $\pm$ 6.7\phantom{0} & \phantom{0}16.9 $\pm$ 6.5\phantom{0} & \phantom{0}61.3 $\pm$ 8.4\phantom{0} & \phantom{0}65.3 $\pm$ 6.8\phantom{0} & \phantom{00}4.0 $\pm$ 10.7 \\
 874&  4920 $\pm$    6 & 25 & \phantom{0}49.0 $\pm$  0.7 & \phantom{0}45.4 $\pm$ 1.7\phantom{0} & \phantom{0}14.5 $\pm$ 3.0\phantom{0} & \phantom{0}39.7 $\pm$ 3.4\phantom{0} & \phantom{0}14.5 $\pm$ 2.2\phantom{0} & \phantom{00}3.6 $\pm$ 1.9\phantom{0} & \phantom{00}9.3 $\pm$ 3.5\phantom{0} & \phantom{00}5.7 $\pm$ 3.8\phantom{0} \\
 877&  6294 $\pm$    9 & 13 & \phantom{0}38.5 $\pm$  0.6 & \phantom{0}32.1 $\pm$ 4.6\phantom{0} & \phantom{0}17.5 $\pm$ 3.2\phantom{0} & \phantom{0}38.4 $\pm$ 3.7\phantom{0} & \phantom{00}8.7 $\pm$ 2.5\phantom{0} & \phantom{00}6.5 $\pm$ 4.7\phantom{0} & \phantom{00}0.2 $\pm$ 3.7\phantom{0} & \phantom{00}6.3 $\pm$ 5.9\phantom{0} \\
 882&  7786 $\pm$    2 & 13 & 149.7 $\pm$  3.8 & \phantom{0}40.2 $\pm$ 4.3\phantom{0} & \phantom{0}16.6 $\pm$ 2.7\phantom{0} & \phantom{0}56.4 $\pm$ 2.9\phantom{0} & \phantom{00}9.9 $\pm$ 3.0\phantom{0} & \phantom{0}70.5 $\pm$ 5.7\phantom{0} & \phantom{0}86.7 $\pm$ 4.8\phantom{0} & \phantom{0}16.2 $\pm$ 5.2\phantom{0} \\
 883& -- & -- & -- & -- & -- & -- & -- & -- * & -- * & -- * \\
 892&  4910 $\pm$    4 & 15 & \phantom{0}70.6 $\pm$  0.5 & \phantom{0}69.4 $\pm$ 5.1\phantom{0} & \phantom{0}13.3 $\pm$ 2.7\phantom{0} & \phantom{0}71.2 $\pm$ 3.1\phantom{0} & \phantom{00}7.5 $\pm$ 2.3\phantom{0} & \phantom{00}1.3 $\pm$ 5.1\phantom{0} & \phantom{00}0.6 $\pm$ 3.1\phantom{0} & \phantom{00}1.9 $\pm$ 5.9\phantom{0} \\
 900& -- & -- & -- & -- & -- & -- & -- & -- & -- & -- \\
 901&  4651 $\pm$    3 & 15 & 139.4 $\pm$  1.3 & \phantom{0}15.5 $\pm$ 3.7\phantom{0} & \phantom{0}14.3 $\pm$ 3.4\phantom{0} & 172.0 $\pm$ 4.7\phantom{0} & \phantom{0}19.4 $\pm$ 5.1\phantom{0} & \phantom{0}56.1 $\pm$ 3.9\phantom{0} & \phantom{0}32.6 $\pm$ 4.9\phantom{0} & \phantom{0}23.5 $\pm$ 6.0\phantom{0} \\
 903&  3579 $\pm$    0 & 25 & \phantom{0}81.0 $\pm$  0.3 & \phantom{0}92.1 $\pm$ 2.7\phantom{0} & \phantom{0}27.7 $\pm$ 5.0\phantom{0} & \phantom{0}79.8 $\pm$ 7.1\phantom{0} & \phantom{0}30.6 $\pm$ 6.2\phantom{0} & \phantom{0}11.1 $\pm$ 2.7\phantom{0} & \phantom{00}1.2 $\pm$ 7.1\phantom{0} & \phantom{0}12.3 $\pm$ 7.6\phantom{0} \\
 905& -- & -- & -- & -- & -- & -- & -- & -- & -- & -- \\
 907&  3476 $\pm$    2 & 15 & \phantom{0}19.4 $\pm$  0.6 & \phantom{0}23.2 $\pm$ 5.7\phantom{0} & \phantom{0}23.6 $\pm$ 7.5\phantom{0} & \phantom{0}23.1 $\pm$ 6.3\phantom{0} & \phantom{0}27.2 $\pm$ 7.3\phantom{0} & \phantom{00}3.8 $\pm$ 5.8\phantom{0} & \phantom{00}3.7 $\pm$ 6.3\phantom{0} & \phantom{00}0.1 $\pm$ 8.5\phantom{0} \\
 913&  1632 $\pm$    3 & 23 & \phantom{0}27.8 $\pm$  1.2 & \phantom{0}33.4 $\pm$ 5.2\phantom{0} & \phantom{0}20.9 $\pm$ 4.0\phantom{0} & \phantom{0}38.9 $\pm$ 6.8\phantom{0} & \phantom{0}28.6 $\pm$ 4.8\phantom{0} & \phantom{00}5.6 $\pm$ 5.3\phantom{0} & \phantom{0}11.1 $\pm$ 6.9\phantom{0} & \phantom{00}5.5 $\pm$ 8.6\phantom{0} \\
 915&  4227 $\pm$    3 & 20 & 160.3 $\pm$  3.2 & 169.3 $\pm$ 5.5\phantom{0} & \phantom{0}22.7 $\pm$ 4.2\phantom{0} & 174.1 $\pm$ 4.3\phantom{0} & \phantom{0}17.5 $\pm$ 5.1\phantom{0} & \phantom{00}9.0 $\pm$ 6.3\phantom{0} & \phantom{0}13.8 $\pm$ 5.3\phantom{0} & \phantom{00}4.8 $\pm$ 6.9\phantom{0} \\
 916&  3856 $\pm$    6 & 20 & 134.7 $\pm$  0.2 & 133.9 $\pm$ 2.6\phantom{0} & \phantom{00}9.9 $\pm$ 1.9\phantom{0} & 134.3 $\pm$ 2.5\phantom{0} & \phantom{0}10.0 $\pm$ 1.8\phantom{0} & \phantom{00}0.9 $\pm$ 2.6\phantom{0} & \phantom{00}0.4 $\pm$ 2.5\phantom{0} & \phantom{00}0.4 $\pm$ 3.6\phantom{0} \\
 922&  5222 $\pm$    2 & 15 & \phantom{0}91.7 $\pm$  0.7 & \phantom{0}93.1 $\pm$ 5.1\phantom{0} & \phantom{0}27.6 $\pm$ 10.0 & \phantom{0}90.9 $\pm$ 6.6\phantom{0} & \phantom{0}24.0 $\pm$ 7.4\phantom{0} & \phantom{00}1.4 $\pm$ 5.1\phantom{0} & \phantom{00}0.8 $\pm$ 6.7\phantom{0} & \phantom{00}2.2 $\pm$ 8.3\phantom{0} \\
 923&  4065 $\pm$    3 &  7 & \phantom{0}93.2 $\pm$  0.7 & 108.0 $\pm$ 14.6 & \phantom{0}30.4 $\pm$ 15.6 & 107.1 $\pm$ 13.4 & \phantom{0}27.3 $\pm$ 12.2 & \phantom{0}14.8 $\pm$ 14.6 & \phantom{0}14.0 $\pm$ 13.5 & \phantom{00}0.9 $\pm$ 19.8 \\
 925&  4018 $\pm$    8 & 15 & 129.4 $\pm$ 11.7 & 139.3 $\pm$ 4.6\phantom{0} & \phantom{0}15.5 $\pm$ 3.3\phantom{0} & 147.2 $\pm$ 5.8\phantom{0} & \phantom{0}15.1 $\pm$ 2.7\phantom{0} & \phantom{0} 9.9 $\pm$ 12.6 & \phantom{0}17.7 $\pm$ 13.1 & \phantom{00}7.8 $\pm$ 7.4\phantom{0} \\
 927&  6704 $\pm$    2 & 15 & \phantom{0}43.2 $\pm$  0.7 & \phantom{0}74.8 $\pm$ 4.1\phantom{0} & \phantom{0}14.3 $\pm$ 2.8\phantom{0} & \phantom{0}67.6 $\pm$ 5.3\phantom{0} & \phantom{0}19.3 $\pm$ 3.8\phantom{0} & \phantom{0}31.5 $\pm$ 4.2\phantom{0} & \phantom{0}24.3 $\pm$ 5.3\phantom{0} & \phantom{00}7.2 $\pm$ 6.7\phantom{0} \\
 932&  7747 $\pm$    5 & 15 & 103.9 $\pm$  0.2 & 105.2 $\pm$ 2.6\phantom{0} & \phantom{00}8.2 $\pm$ 3.0\phantom{0} & 106.8 $\pm$ 2.1\phantom{0} & \phantom{00}6.3 $\pm$ 1.7\phantom{0} & \phantom{00}1.3 $\pm$ 2.6\phantom{0} & \phantom{00}2.8 $\pm$ 2.1\phantom{0} & \phantom{00}1.5 $\pm$ 3.3\phantom{0} \\
 935& -- & -- & -- & -- & -- & -- & -- & -- & -- & -- \\
 939&  6483 $\pm$    3 & 10 & \phantom{0}33.2 $\pm$  0.8 & 164.7 $\pm$ 9.4\phantom{0} & \phantom{0}25.5 $\pm$ 5.5\phantom{0} & 150.6 $\pm$ 6.3\phantom{0} & \phantom{0}16.0 $\pm$ 5.0\phantom{0} & \phantom{0}48.5 $\pm$ 9.5\phantom{0} & \phantom{0}62.5 $\pm$ 6.4\phantom{0} & \phantom{0}14.0 $\pm$ 11.4 \\
1014& -- & -- & -- & -- & -- & -- & -- & -- & -- & -- \\
1017&  5428 $\pm$    2 & 10 & 127.1 $\pm$  8.9 & 121.4 $\pm$ 7.0\phantom{0} & \phantom{0}21.2 $\pm$ 6.7\phantom{0} & 121.1 $\pm$ 7.4\phantom{0} & \phantom{0}20.7 $\pm$ 8.6\phantom{0} & \phantom{00}5.7 $\pm$ 11.3 & \phantom{00}6.0 $\pm$ 11.6 & \phantom{00}0.3 $\pm$ 10.2 \\
1022&  4559 $\pm$    7 &  5 & \phantom{0}10.0 $\pm$  0.0 & \phantom{0}38.3 $\pm$ 8.8\phantom{0} & \phantom{0}16.6 $\pm$ 6.3\phantom{0} & \phantom{0}19.6 $\pm$ 13.4 & \phantom{0}12.8 $\pm$ 7.9\phantom{0} & \phantom{0}28.3 $\pm$ 8.8\phantom{0} & \phantom{00}9.6 $\pm$ 13.4 & \phantom{0}18.7 $\pm$ 16.0 \\
1026& -- & -- & -- & -- & -- & -- & -- & -- & -- & -- \\
1042& -- & -- & -- & -- & -- & -- & -- & -- & -- & -- \\
1044&  4067 $\pm$    3 & 15 & 169.9 $\pm$  1.2 & 136.7 $\pm$ 5.4\phantom{0} & \phantom{0}16.2 $\pm$ 3.5\phantom{0} & 137.3 $\pm$ 5.5\phantom{0} & \phantom{0}18.3 $\pm$ 4.1\phantom{0} & \phantom{0}33.2 $\pm$ 5.5\phantom{0} & \phantom{0}32.6 $\pm$ 5.6\phantom{0} & \phantom{00}0.6 $\pm$ 7.7\phantom{0} \\
1340&  6145 $\pm$    1 &  7 & 140.0 $\pm$  0.0 & \phantom{0}81.9 $\pm$ 7.2\phantom{0} & \phantom{0}17.5 $\pm$ 6.9\phantom{0} & 113.2 $\pm$ 7.7\phantom{0} & \phantom{0}17.1 $\pm$ 8.6\phantom{0} & \phantom{0}58.1 $\pm$ 7.2\phantom{0} & \phantom{0}26.8 $\pm$ 7.7\phantom{0} & \phantom{0}31.2 $\pm$ 10.5 \\
1360&  7044 $\pm$    1 &  7 & \phantom{0}10.0 $\pm$  0.0 & \phantom{0}58.4 $\pm$ 9.5\phantom{0} & \phantom{0}16.3 $\pm$ 5.4\phantom{0} & \phantom{0}49.3 $\pm$ 10.0 & \phantom{0}18.9 $\pm$ 7.8\phantom{0} & \phantom{0}48.4 $\pm$ 9.5\phantom{0} & \phantom{0}39.3 $\pm$ 10.0 & \phantom{00}9.1 $\pm$ 13.8 \\
1520&  6534 $\pm$    1 &  7 & 125.0 $\pm$  0.0 & \phantom{0}80.6 $\pm$ 13.1 & \phantom{0}28.7 $\pm$ 8.5\phantom{0} & \phantom{0}81.7 $\pm$ 14.7 & \phantom{0}26.5 $\pm$ 11.9 & \phantom{0}44.4 $\pm$ 13.1 & \phantom{0}43.3 $\pm$ 14.7 & \phantom{00}1.1 $\pm$ 19.7 \\
1589&  6262 $\pm$    1 &  7 & -- & \phantom{0}96.5 $\pm$ 11.1 & \phantom{0}30.4 $\pm$ 13.1 & \phantom{0}89.7 $\pm$ 12.6 & \phantom{0}27.3 $\pm$ 17.7 & \phantom{0}90.5 $\pm$ 11.1 & \phantom{0}83.7 $\pm$ 12.6 & \phantom{00}6.8 $\pm$ 16.8 \\
1663& -- & -- & -- & -- & -- & -- & -- & -- & -- & -- \\
1740& -- & -- & -- & -- & -- & -- & -- & -- * & -- * & -- * \\
1780&  4195 $\pm$    3 & 14 & 109.1 $\pm$  7.1 & \phantom{00}3.1 $\pm$ 3.9\phantom{0} & \phantom{0}16.8 $\pm$ 5.3\phantom{0} & 174.3 $\pm$ 5.9\phantom{0} & \phantom{0}17.8 $\pm$ 4.7\phantom{0} & \phantom{0}74.0 $\pm$ 8.1\phantom{0} & \phantom{0}65.3 $\pm$ 8.1\phantom{0} & \phantom{00}8.7 $\pm$ 5.5\phantom{0} \\
1781&  8508 $\pm$    4 & 12 & \phantom{0}43.6 $\pm$  1.0 & \phantom{0}20.7 $\pm$ 12.9 & \phantom{0}42.0 $\pm$ 10.2 & \phantom{0}19.1 $\pm$ 6.4\phantom{0} & \phantom{0}12.8 $\pm$ 5.0\phantom{0} & \phantom{0}22.9 $\pm$ 12.9 & \phantom{0}24.5 $\pm$ 6.5\phantom{0} & \phantom{00}1.6 $\pm$ 14.4 \\
1795&  2505 $\pm$    2 & 10 & \phantom{0}45.2 $\pm$  4.2 & \phantom{0}59.0 $\pm$ 9.0\phantom{0} & \phantom{0}21.2 $\pm$ 6.5\phantom{0} & \phantom{0}46.0 $\pm$ 8.1\phantom{0} & \phantom{0}23.7 $\pm$ 5.0\phantom{0} & \phantom{0}13.7 $\pm$  9.9 & \phantom{00}0.7 $\pm$ 9.1\phantom{0} & \phantom{0}13.0 $\pm$ 12.1 \\
1796&  5362 $\pm$    8 & 10 & \phantom{0}26.5 $\pm$  0.6 & \phantom{0}29.8 $\pm$ 3.6\phantom{0} & \phantom{00}9.5 $\pm$ 3.1\phantom{0} & \phantom{0}23.2 $\pm$ 4.0\phantom{0} & \phantom{00}9.5 $\pm$ 3.8\phantom{0} & \phantom{00}3.3 $\pm$ 3.6\phantom{0} & \phantom{00}3.3 $\pm$ 4.0\phantom{0} & \phantom{00}6.6 $\pm$ 5.4\phantom{0} \\
1801&  1944 $\pm$    4 & 20 & \phantom{00}1.7 $\pm$  0.9 & \phantom{00}8.4 $\pm$ 4.2\phantom{0} & \phantom{0}11.5 $\pm$ 2.4\phantom{0} & 165.9 $\pm$ 4.6\phantom{0} & \phantom{0}23.3 $\pm$ 10.5 & \phantom{00}6.7 $\pm$ 4.3\phantom{0} & \phantom{0}15.8 $\pm$ 4.7\phantom{0} & \phantom{0}22.4 $\pm$ 6.2\phantom{0} \\
1871&  6072 $\pm$    3 & 20 & 139.0 $\pm$  0.0 & 164.7 $\pm$ 3.6\phantom{0} & \phantom{0}32.1 $\pm$ 4.1\phantom{0} & 156.0 $\pm$ 3.0\phantom{0} & \phantom{0}11.8 $\pm$ 2.3\phantom{0} & \phantom{0}25.7 $\pm$ 3.6\phantom{0} & \phantom{0}17.0 $\pm$ 3.0\phantom{0} & \phantom{00}8.7 $\pm$ 4.7\phantom{0} \\
1873&  7806 $\pm$    6 &  8 & \phantom{0}55.8 $\pm$ 10.3 & 135.9 $\pm$ 20.6 & \phantom{0}66.1 $\pm$ 17.4 & 143.0 $\pm$ 21.4 & \phantom{0}27.8 $\pm$ 12.5 & \phantom{0}80.2 $\pm$ 23.0 & \phantom{0}87.2 $\pm$ 23.8 & \phantom{00}7.0 $\pm$ 29.7 \\
\bottomrule
\end{tabular}
\end{center}
\begin{minipage}{\textwidth}
\end{minipage}
\end{table*}

 \clearpage
  \begin{table*}[!htb]
\tiny
\caption{\label{table_Gkin}
 Ionised Gas Kinematic properties of the interacting sample selected for this study included in the CALIFA survey.
}
\begin{center}
\renewcommand{\thefootnote}{\alph{footnote}}
\begin{tabular} {c c c c c c c c c c c}
\toprule
    id  & V$\mathrm{_{sys}}$  &         r          &  PA$\mathrm{_{morph}}$ (r) &  \multicolumn{2}{c}{PA approaching}                 & \multicolumn{2}{c}{PA receding}                     & \multicolumn{2}{c}{$\Psi_{\mathrm{morph-kin}}$} & $\Psi_{\mathrm{kin-kin}}$ \\
        &                     &                    &                            &  PA$\mathrm{_{kin}}$  & $\delta$PA$\mathrm{_{kin}}$ & PA$\mathrm{_{kin}}$  & $\delta$PA$\mathrm{_{kin}}$  & approaching  & receding                         &                           \\
        &    (km s$^{-1}$)    &      (arcsec)      &            ($^{\circ}$)    &  ($^{\circ}$)         & ($^{\circ}$)                & ($^{\circ}$)         & ($^{\circ}$)                 & ($^{\circ}$) & ($^{\circ}$)                     & ($^{\circ}$)              \\
    (1) &         (4)         &        (5)         &            (7)             &          (8)          &            (9)              &          (10)        &       (11)                   &   (10)       &       (11)                       &   (10)                    \\
\midrule
  14&  4323 $\pm$    0 & 30 & \phantom{00}9.5 $\pm$  0.9 & \phantom{0}38.4 $\pm$ 0.7\phantom{0} & \phantom{0}23.8 $\pm$ 1.6\phantom{0} & \phantom{0}36.2 $\pm$ 1.5\phantom{0} & \phantom{0}13.2 $\pm$ 2.9\phantom{0} & \phantom{0}28.9 $\pm$ 1.1\phantom{0} & \phantom{0}26.7 $\pm$ 1.7\phantom{0} & \phantom{00}2.2 $\pm$ 1.6\phantom{0} \\
  17&  5402 $\pm$    4 &  5 & 148.8 $\pm$  0.5 & \phantom{0}87.7 $\pm$ 15.3 & \phantom{0}26.1 $\pm$ 13.1 & \phantom{0}96.6 $\pm$ 22.5 & \phantom{0}36.7 $\pm$ 18.4 & \phantom{0}61.1 $\pm$ 15.3 & \phantom{0}52.2 $\pm$ 22.5 & \phantom{00}8.9 $\pm$ 27.2 \\
  22&  4552 $\pm$    8 & 25 & \phantom{0}87.5 $\pm$  0.5 & \phantom{0}79.0 $\pm$ 1.4\phantom{0} & \phantom{00}7.4 $\pm$ 1.4\phantom{0} & \phantom{0}79.1 $\pm$ 4.4\phantom{0} & \phantom{0}22.6 $\pm$ 2.4\phantom{0} & \phantom{00}8.5 $\pm$ 1.5\phantom{0} & \phantom{00}8.4 $\pm$ 4.5\phantom{0} & \phantom{00}0.1 $\pm$ 4.7\phantom{0} \\
  26&  4185 $\pm$    0 & 25 & 164.0 $\pm$  0.2 & 173.6 $\pm$ 1.0\phantom{0} & \phantom{00}5.7 $\pm$ 0.7\phantom{0} & \phantom{00}1.8 $\pm$ 1.8\phantom{0} & \phantom{0}12.1 $\pm$ 1.4\phantom{0} & \phantom{00}9.6 $\pm$ 1.1\phantom{0} & \phantom{0}17.8 $\pm$ 1.8\phantom{0} & \phantom{00}8.2 $\pm$ 2.1\phantom{0} \\
  39&  4811 $\pm$    1 & 27 & 160.2 $\pm$  0.3 & 158.4 $\pm$ 2.0\phantom{0} & \phantom{0}10.1 $\pm$ 2.2\phantom{0} & 159.4 $\pm$ 1.7\phantom{0} & \phantom{00}5.8 $\pm$ 1.1\phantom{0} & \phantom{00}1.8 $\pm$ 2.0\phantom{0} & \phantom{00}0.8 $\pm$ 1.7\phantom{0} & \phantom{00}1.0 $\pm$ 2.6\phantom{0} \\
  42&  5860 $\pm$    8 & 15 & 115.5 $\pm$  1.3 & 133.1 $\pm$ 3.3\phantom{0} & \phantom{0}12.3 $\pm$ 3.5\phantom{0} & 144.1 $\pm$ 3.7\phantom{0} & \phantom{0}11.1 $\pm$ 2.8\phantom{0} & \phantom{0}17.6 $\pm$ 3.5\phantom{0} & \phantom{0}28.6 $\pm$ 3.9\phantom{0} & \phantom{0}11.0 $\pm$ 5.0\phantom{0} \\
  44& -- & -- & -- & -- & -- & -- & -- & -- & -- & -- \\
 119&  4906 $\pm$    8 & 10 & \phantom{0}70.3 $\pm$  1.2 & \phantom{0}85.8 $\pm$ 4.9\phantom{0} & \phantom{0}15.7 $\pm$ 6.1\phantom{0} & \phantom{0}82.9 $\pm$ 11.9 & \phantom{0}35.8 $\pm$ 12.6 & \phantom{0}15.5 $\pm$ 5.0\phantom{0} & \phantom{0}12.6 $\pm$ 12.0 & \phantom{00}2.9 $\pm$ 12.9 \\
 127&  6564 $\pm$    3 & 17 & \phantom{0}35.6 $\pm$  2.3 & \phantom{0}52.0 $\pm$ 12.0 & \phantom{0}35.4 $\pm$ 11.8 & \phantom{0}39.0 $\pm$ 6.4\phantom{0} & \phantom{0}60.3 $\pm$ 10.0 & \phantom{0}16.4 $\pm$ 12.2 & \phantom{00}3.4 $\pm$ 6.8\phantom{0} & \phantom{0}13.0 $\pm$ 13.6 \\
 155&  4689 $\pm$    9 & 22 & 109.2 $\pm$  0.8 & \phantom{0}80.3 $\pm$ 3.3\phantom{0} & \phantom{00}9.8 $\pm$ 2.3\phantom{0} & \phantom{0}76.4 $\pm$ 3.2\phantom{0} & \phantom{0}19.7 $\pm$ 3.4\phantom{0} & \phantom{0}28.9 $\pm$ 3.4\phantom{0} & \phantom{0}32.8 $\pm$ 3.3\phantom{0} & \phantom{00}3.9 $\pm$ 4.6\phantom{0} \\
 156&  4883 $\pm$    8 & 25 & 127.7 $\pm$  0.2 & 133.3 $\pm$ 1.8\phantom{0} & \phantom{00}7.2 $\pm$ 1.4\phantom{0} & 133.5 $\pm$ 2.1\phantom{0} & \phantom{00}7.2 $\pm$ 1.2\phantom{0} & \phantom{00}5.6 $\pm$ 1.8\phantom{0} & \phantom{00}5.8 $\pm$ 2.1\phantom{0} & \phantom{00}0.2 $\pm$ 2.8\phantom{0} \\
 165&  5162 $\pm$    3 & 22 & \phantom{0}23.4 $\pm$  0.8 & \phantom{0}30.6 $\pm$ 1.0\phantom{0} & \phantom{00}4.1 $\pm$ 0.6\phantom{0} & \phantom{0}32.3 $\pm$ 0.8\phantom{0} & \phantom{00}3.8 $\pm$ 0.5\phantom{0} & \phantom{00}7.2 $\pm$ 1.3\phantom{0} & \phantom{00}8.9 $\pm$ 1.2\phantom{0} & \phantom{00}1.7 $\pm$ 1.3\phantom{0} \\
 186&  4266 $\pm$    1 & 20 & 149.0 $\pm$  0.6 & 149.8 $\pm$ 1.3\phantom{0} & \phantom{0}11.1 $\pm$ 3.4\phantom{0} & 148.2 $\pm$ 1.0\phantom{0} & \phantom{00}4.9 $\pm$ 1.0\phantom{0} & \phantom{00}0.8 $\pm$ 1.4\phantom{0} & \phantom{00}0.9 $\pm$ 1.2\phantom{0} & \phantom{00}1.7 $\pm$ 1.7\phantom{0} \\
 213&  5528 $\pm$   10 &  7 & \phantom{0}81.3 $\pm$  1.6 & \phantom{0}83.1 $\pm$ 5.9\phantom{0} & \phantom{0}15.6 $\pm$ 7.3\phantom{0} & 115.1 $\pm$ 11.4 & \phantom{0}20.7 $\pm$ 9.0\phantom{0} & \phantom{00}1.9 $\pm$ 6.1\phantom{0} & \phantom{0}33.8 $\pm$ 11.5 & \phantom{0}32.0 $\pm$ 12.9 \\
 274&  2062 $\pm$    1 & 12 & 169.8 $\pm$  0.7 & \phantom{0}12.0 $\pm$ 7.0\phantom{0} & \phantom{0}16.8 $\pm$ 4.2\phantom{0} & \phantom{0}11.3 $\pm$ 3.4\phantom{0} & \phantom{0}10.3 $\pm$ 4.5\phantom{0} & \phantom{0}22.2 $\pm$ 7.0\phantom{0} & \phantom{0}21.5 $\pm$ 3.5\phantom{0} & \phantom{00}0.7 $\pm$ 4.7\phantom{0} \\
 311&  6173 $\pm$    3 & 20 & 132.2 $\pm$  2.4 & 153.7 $\pm$ 8.1\phantom{0} & \phantom{0}40.2 $\pm$ 5.0\phantom{0} & 132.8 $\pm$ 5.4\phantom{0} & \phantom{0}27.2 $\pm$ 5.3\phantom{0} & \phantom{0}21.5 $\pm$ 8.5\phantom{0} & \phantom{00}0.6 $\pm$ 5.9\phantom{0} & \phantom{0}20.9 $\pm$ 9.8\phantom{0} \\
 314&  6279 $\pm$    3 & 20 & \phantom{0}59.4 $\pm$  0.8 & \phantom{0}66.6 $\pm$ 2.0\phantom{0} & \phantom{00}7.4 $\pm$ 1.9\phantom{0} & \phantom{0}67.1 $\pm$ 2.4\phantom{0} & \phantom{0} 9.9 $\pm$ 2.3\phantom{0} & \phantom{00}7.2 $\pm$ 2.1\phantom{0} & \phantom{00}7.7 $\pm$ 2.5\phantom{0} & \phantom{00}0.5 $\pm$ 3.1\phantom{0} \\
 340&  6143 $\pm$    4 & 12 & 153.2 $\pm$  0.8 & \phantom{0}18.2 $\pm$ 8.8\phantom{0} & \phantom{0}41.7 $\pm$ 14.1 & \phantom{0}21.1 $\pm$ 19.0 & \phantom{0}55.8 $\pm$ 13.3 & \phantom{0}45.0 $\pm$ 8.8\phantom{0} & \phantom{0}47.9 $\pm$ 19.0 & \phantom{00}2.9 $\pm$ 20.9 \\
 360& -- & -- & -- & -- & -- & -- & -- & -- & -- & -- \\
 475&  3165 $\pm$    1 & 22 & \phantom{0}31.1 $\pm$  0.5 & \phantom{0}40.5 $\pm$ 3.0\phantom{0} & \phantom{0}12.6 $\pm$ 2.0\phantom{0} & \phantom{0}56.0 $\pm$ 5.6\phantom{0} & \phantom{0}25.3 $\pm$ 2.5\phantom{0} & \phantom{00}9.4 $\pm$ 3.0\phantom{0} & \phantom{0}24.9 $\pm$ 5.6\phantom{0} & \phantom{0}15.5 $\pm$ 6.4\phantom{0} \\
 479&  6589 $\pm$   10 & 22 & 146.0 $\pm$  0.4 & 163.4 $\pm$ 2.3\phantom{0} & \phantom{0}36.4 $\pm$ 9.8\phantom{0} & 158.2 $\pm$ 5.0\phantom{0} & \phantom{0}31.4 $\pm$ 3.9\phantom{0} & \phantom{0}17.4 $\pm$ 2.3\phantom{0} & \phantom{0}12.2 $\pm$ 5.0\phantom{0} & \phantom{00}5.2 $\pm$ 5.5\phantom{0} \\
 486&  3077 $\pm$    1 & 18 & \phantom{0}14.0 $\pm$  1.0 & \phantom{0} 9.9 $\pm$ 2.6\phantom{0} & \phantom{0}12.6 $\pm$ 3.0\phantom{0} & \phantom{0}13.5 $\pm$ 3.1\phantom{0} & \phantom{0}15.1 $\pm$ 2.2\phantom{0} & \phantom{00}4.0 $\pm$ 2.8\phantom{0} & \phantom{00}0.4 $\pm$ 3.2\phantom{0} & \phantom{00}3.6 $\pm$ 4.0\phantom{0} \\
 520&  6627 $\pm$    0 &  7 & 170.0 $\pm$  0.0 & \phantom{0}44.5 $\pm$ 4.0\phantom{0} & \phantom{0}12.3 $\pm$ 6.2\phantom{0} & \phantom{0}37.4 $\pm$ 10.5 & \phantom{0}17.8 $\pm$ 5.1\phantom{0} & \phantom{0}54.5 $\pm$ 4.0\phantom{0} & \phantom{0}47.4 $\pm$ 10.5 & \phantom{00}7.1 $\pm$ 11.2 \\
 577&  6588 $\pm$    1 & 10 & \phantom{00}2.8 $\pm$  0.9 & \phantom{0}32.0 $\pm$ 2.7\phantom{0} & \phantom{00}8.1 $\pm$ 1.2\phantom{0} & \phantom{0}24.1 $\pm$ 3.3\phantom{0} & \phantom{0}24.8 $\pm$ 4.2\phantom{0} & \phantom{0}29.2 $\pm$ 2.9\phantom{0} & \phantom{0}21.3 $\pm$ 3.4\phantom{0} & \phantom{00}7.9 $\pm$ 4.3\phantom{0} \\
 589& -- & -- & -- & -- & -- & -- & -- & -- & -- & -- \\
 593&  8291 $\pm$    0 & 20 & \phantom{0}48.0 $\pm$  0.8 & \phantom{0}52.8 $\pm$ 0.6\phantom{0} & \phantom{00}6.8 $\pm$ 0.9\phantom{0} & \phantom{0}51.2 $\pm$ 0.9\phantom{0} & \phantom{00}8.5 $\pm$ 0.5\phantom{0} & \phantom{00}4.7 $\pm$ 1.0\phantom{0} & \phantom{00}3.1 $\pm$ 1.2\phantom{0} & \phantom{00}1.6 $\pm$ 1.1\phantom{0} \\
 633&  2920 $\pm$    9 & 12 & \phantom{0}88.8 $\pm$  2.3 & \phantom{0}22.5 $\pm$ 11.5 & \phantom{0}24.1 $\pm$ 4.4\phantom{0} & \phantom{0}63.8 $\pm$ 6.5\phantom{0} & \phantom{0}38.5 $\pm$ 12.4 & \phantom{0}66.2 $\pm$ 11.7 & \phantom{0}25.0 $\pm$ 6.9\phantom{0} & \phantom{0}41.3 $\pm$ 13.2 \\
 634&  2876 $\pm$    4 & 20 & \phantom{0}91.3 $\pm$  0.7 & \phantom{0}60.3 $\pm$ 1.8\phantom{0} & \phantom{00}8.1 $\pm$ 2.2\phantom{0} & \phantom{0}55.1 $\pm$ 2.5\phantom{0} & \phantom{0}12.6 $\pm$ 2.1\phantom{0} & \phantom{0}31.0 $\pm$ 1.9\phantom{0} & \phantom{0}36.3 $\pm$ 2.6\phantom{0} & \phantom{00}5.3 $\pm$ 3.1\phantom{0} \\
 663&  6937 $\pm$    2 & 20 & 108.3 $\pm$  0.3 & 104.5 $\pm$ 1.1\phantom{0} & \phantom{00}4.4 $\pm$ 1.0\phantom{0} & 106.7 $\pm$ 1.2\phantom{0} & \phantom{00}4.1 $\pm$ 0.8\phantom{0} & \phantom{00}3.9 $\pm$ 1.2\phantom{0} & \phantom{00}1.7 $\pm$ 1.2\phantom{0} & \phantom{00}2.2 $\pm$ 1.6\phantom{0} \\
 676&  2964 $\pm$    3 &  6 & \phantom{0}65.6 $\pm$  0.7 & 107.5 $\pm$ 7.4\phantom{0} & \phantom{0}15.4 $\pm$ 7.4\phantom{0} & \phantom{0}93.3 $\pm$ 18.7 & \phantom{0}31.8 $\pm$ 11.7 & \phantom{0}42.0 $\pm$ 7.5\phantom{0} & \phantom{0}27.7 $\pm$ 18.8 & \phantom{0}14.3 $\pm$ 20.2 \\
 680&  3438 $\pm$    1 & 20 & \phantom{0}29.5 $\pm$  1.3 & \phantom{0}24.0 $\pm$ 9.4\phantom{0} & \phantom{0}37.6 $\pm$ 6.2\phantom{0} & \phantom{0}19.8 $\pm$ 15.5 & \phantom{0}42.5 $\pm$ 6.8\phantom{0} & \phantom{00}5.5 $\pm$ 9.5\phantom{0} & \phantom{00}9.7 $\pm$ 15.5 & \phantom{00}4.2 $\pm$ 18.1 \\
 740&  3852 $\pm$    2 & 20 & \phantom{00}2.2 $\pm$  2.7 & \phantom{0}74.5 $\pm$ 5.6\phantom{0} & \phantom{0}36.8 $\pm$ 10.3 & \phantom{0}81.0 $\pm$ 6.6\phantom{0} & \phantom{0}30.1 $\pm$ 3.7\phantom{0} & \phantom{0}72.3 $\pm$ 6.2\phantom{0} & \phantom{0}78.8 $\pm$ 7.1\phantom{0} & \phantom{00}6.5 $\pm$ 8.6\phantom{0} \\
 746&  3388 $\pm$   12 & 10 & \phantom{0}26.6 $\pm$  0.3 & 126.2 $\pm$ 4.6\phantom{0} & \phantom{0}19.6 $\pm$ 8.5\phantom{0} & 125.7 $\pm$ 4.6\phantom{0} & \phantom{0}14.3 $\pm$ 5.2\phantom{0} & \phantom{0}80.4 $\pm$ 4.7\phantom{0} & \phantom{0}80.9 $\pm$ 4.6\phantom{0} & \phantom{00}0.5 $\pm$ 6.5\phantom{0} \\
 758&  2281 $\pm$    0 & 20 & 120.1 $\pm$  0.4 & 142.9 $\pm$ 1.2\phantom{0} & \phantom{00}8.0 $\pm$ 1.3\phantom{0} & 129.0 $\pm$ 1.1\phantom{0} & \phantom{00}6.0 $\pm$ 1.9\phantom{0} & \phantom{0}22.8 $\pm$ 1.2\phantom{0} & \phantom{00}8.9 $\pm$ 1.2\phantom{0} & \phantom{0}13.9 $\pm$ 1.6\phantom{0} \\
 767&  2531 $\pm$    0 & 25 & 115.2 $\pm$  0.3 & 109.4 $\pm$ 1.3\phantom{0} & \phantom{00}9.6 $\pm$ 1.5\phantom{0} & 111.4 $\pm$ 1.2\phantom{0} & \phantom{00}6.9 $\pm$ 1.9\phantom{0} & \phantom{00}5.8 $\pm$ 1.3\phantom{0} & \phantom{00}3.8 $\pm$ 1.2\phantom{0} & \phantom{00}2.0 $\pm$ 1.8\phantom{0} \\
 769&  3246 $\pm$    0 & 20 & 101.9 $\pm$  1.0 & 128.3 $\pm$ 2.5\phantom{0} & \phantom{0}13.8 $\pm$ 3.2\phantom{0} & 129.5 $\pm$ 3.4\phantom{0} & \phantom{0}10.8 $\pm$ 1.4\phantom{0} & \phantom{0}26.4 $\pm$ 2.7\phantom{0} & \phantom{0}27.7 $\pm$ 3.5\phantom{0} & \phantom{00}1.3 $\pm$ 4.2\phantom{0} \\
 770&  5505 $\pm$    2 & 12 & \phantom{0}30.6 $\pm$  0.6 & 145.4 $\pm$ 3.1\phantom{0} & \phantom{0}33.0 $\pm$ 9.5\phantom{0} & \phantom{0}90.7 $\pm$ 6.2\phantom{0} & \phantom{0}20.2 $\pm$ 4.0\phantom{0} & \phantom{0}65.2 $\pm$ 3.2\phantom{0} & \phantom{0}60.1 $\pm$ 6.2\phantom{0} & \phantom{0}54.7 $\pm$ 6.9\phantom{0} \\
 778&  5500 $\pm$    3 & 15 & \phantom{0}65.8 $\pm$  0.8 & \phantom{0}70.9 $\pm$ 1.7\phantom{0} & \phantom{00}7.1 $\pm$ 1.4\phantom{0} & \phantom{0}74.7 $\pm$ 1.7\phantom{0} & \phantom{00}5.5 $\pm$ 1.2\phantom{0} & \phantom{00}5.1 $\pm$ 1.9\phantom{0} & \phantom{00}8.9 $\pm$ 1.8\phantom{0} & \phantom{00}3.8 $\pm$ 2.4\phantom{0} \\
 780& -- & -- & -- & -- & -- & -- & -- & -- & -- & -- \\
 781& -- & -- & -- & -- & -- & -- & -- & -- & -- & -- \\
 785&  8404 $\pm$    5 &  7 & 105.5 $\pm$  0.0 & 116.2 $\pm$ 16.0 & \phantom{0}32.5 $\pm$ 10.8 & 126.2 $\pm$ 7.9\phantom{0} & \phantom{0}14.0 $\pm$ 4.5\phantom{0} & \phantom{0}10.7 $\pm$ 16.0 & \phantom{0}20.7 $\pm$ 7.9\phantom{0} & \phantom{0}10.0 $\pm$ 17.8 \\
 795&  2643 $\pm$    0 & 15 & 163.9 $\pm$  0.7 & 161.9 $\pm$ 4.6\phantom{0} & \phantom{0}13.6 $\pm$ 3.3\phantom{0} & 148.7 $\pm$ 1.0\phantom{0} & \phantom{00}4.3 $\pm$ 0.8\phantom{0} & \phantom{00}2.1 $\pm$ 4.6\phantom{0} & \phantom{0}15.2 $\pm$ 1.3\phantom{0} & \phantom{0}13.1 $\pm$ 4.7\phantom{0} \\
 796&  5627 $\pm$    2 & 15 & \phantom{00}2.0 $\pm$  0.6 & \phantom{00}6.0 $\pm$ 1.2\phantom{0} & \phantom{00}8.0 $\pm$ 1.6\phantom{0} & \phantom{00}2.6 $\pm$ 2.0\phantom{0} & \phantom{00}7.0 $\pm$ 4.0\phantom{0} & \phantom{00}4.0 $\pm$ 1.3\phantom{0} & \phantom{00}0.6 $\pm$ 2.1\phantom{0} & \phantom{00}3.4 $\pm$ 2.3\phantom{0} \\
 797&  5592 $\pm$    1 & 15 & 121.4 $\pm$  0.4 & 121.7 $\pm$ 2.0\phantom{0} & \phantom{00}9.7 $\pm$ 3.4\phantom{0} & 138.8 $\pm$ 2.8\phantom{0} & \phantom{00}9.6 $\pm$ 1.9\phantom{0} & \phantom{00}0.3 $\pm$ 2.0\phantom{0} & \phantom{0}17.4 $\pm$ 2.8\phantom{0} & \phantom{0}17.1 $\pm$ 3.4\phantom{0} \\
 801&  2020 $\pm$    0 & 11 & \phantom{0}31.2 $\pm$  1.6 & \phantom{0}56.2 $\pm$ 2.6\phantom{0} & \phantom{0}13.3 $\pm$ 1.2\phantom{0} & \phantom{0}41.9 $\pm$ 0.5\phantom{0} & \phantom{00}8.1 $\pm$ 0.4\phantom{0} & \phantom{0}25.0 $\pm$ 3.0\phantom{0} & \phantom{0}10.6 $\pm$ 1.6\phantom{0} & \phantom{0}14.4 $\pm$ 2.6\phantom{0} \\
 802&  5435 $\pm$   12 & 11 & 178.2 $\pm$  2.5 & \phantom{0}31.3 $\pm$ 16.2 & \phantom{0}57.1 $\pm$ 10.2 & \phantom{0}21.4 $\pm$ 3.3\phantom{0} & \phantom{0}10.5 $\pm$ 3.2\phantom{0} & \phantom{0}33.1 $\pm$ 16.4 & \phantom{0}23.2 $\pm$ 4.2\phantom{0} & \phantom{0} 9.9 $\pm$ 16.5 \\
 803&  1791 $\pm$    1 & 27 & \phantom{0}96.6 $\pm$  0.9 & \phantom{0}33.6 $\pm$ 6.2\phantom{0} & \phantom{0}41.2 $\pm$ 5.8\phantom{0} & \phantom{0}43.2 $\pm$ 12.5 & \phantom{0}34.9 $\pm$ 6.2\phantom{0} & \phantom{0}63.0 $\pm$ 6.3\phantom{0} & \phantom{0}53.4 $\pm$ 12.5 & \phantom{00}9.6 $\pm$ 13.9 \\
 806&  4501 $\pm$    3 & 12 & \phantom{0}92.1 $\pm$  0.3 & \phantom{0}21.3 $\pm$ 4.8\phantom{0} & \phantom{0}15.1 $\pm$ 3.6\phantom{0} & \phantom{0}16.7 $\pm$ 4.5\phantom{0} & \phantom{0}12.3 $\pm$ 3.6\phantom{0} & \phantom{0}70.8 $\pm$ 4.8\phantom{0} & \phantom{0}75.4 $\pm$ 4.5\phantom{0} & \phantom{00}4.6 $\pm$ 6.6\phantom{0} \\
 807&  5561 $\pm$    2 & 20 & 154.5 $\pm$  0.6 & 144.5 $\pm$ 3.8\phantom{0} & \phantom{0}16.9 $\pm$ 2.7\phantom{0} & 150.9 $\pm$ 2.7\phantom{0} & \phantom{0}14.9 $\pm$ 3.6\phantom{0} & \phantom{0} 9.9 $\pm$ 3.9\phantom{0} & \phantom{00}3.6 $\pm$ 2.8\phantom{0} & \phantom{00}6.4 $\pm$ 4.7\phantom{0} \\
 816&  4780 $\pm$    4 & 12 & 149.9 $\pm$  0.5 & 141.8 $\pm$ 4.1\phantom{0} & \phantom{0}14.3 $\pm$ 3.1\phantom{0} & 146.4 $\pm$ 4.6\phantom{0} & \phantom{0}16.6 $\pm$ 3.6\phantom{0} & \phantom{00}8.1 $\pm$ 4.1\phantom{0} & \phantom{00}3.5 $\pm$ 4.6\phantom{0} & \phantom{00}4.6 $\pm$ 6.1\phantom{0} \\
 822&  6520 $\pm$    1 & 15 & 132.4 $\pm$  1.6 & 120.3 $\pm$ 2.7\phantom{0} & \phantom{0}11.3 $\pm$ 1.1\phantom{0} & 129.4 $\pm$ 1.5\phantom{0} & \phantom{00}6.0 $\pm$ 1.2\phantom{0} & \phantom{0}12.1 $\pm$ 3.1\phantom{0} & \phantom{00}3.1 $\pm$ 2.2\phantom{0} & \phantom{00}9.1 $\pm$ 3.1\phantom{0} \\
\bottomrule
\end{tabular}
\tablefoot{
(1) CALIFA id.
(2) Systemic velocity derived from integrated the velocities in a 2.7$\arcsec$ aperture centred in the kinematic centre.
(3) Radius used to average the polar coordinates of the positions from the lines of nodes (see Sec.\,\ref{sec:Analysis} for details).
(4) Morphological PA inferred by fitting an elipse to an isophote at radius r in the r-band SDSS image.
(5) and (6) Kinematic PA and its standard deviation at radius r for the approaching side (see Sec.\,\ref{sec:Analysis} for details).
(7) and (8) Kinematic PA and its standard deviation at radius r for the  receding  side.
(9) and (10) Morpho-kinematic misalignments for the receding and approaching sides, respectively.
(11) Kinematic misalignment between the receding and the approaching sides.
The * symbol tag the objects where it is not posible to determine any kinematic property using our method (see Sec.\,\ref{sec:Analysis} for details).
}
\end{center}
\begin{minipage}{\textwidth}
\end{minipage}
\end{table*} 
   
\begin{table*}[!htb]
\tiny
\addtocounter{table}{-1}
\caption{continue  \ref{table_Gkin} }
\begin{center}
\renewcommand{\thefootnote}{\alph{footnote}}
\begin{tabular} {c c c c c c c c c c c}
\toprule
    id  & V$\mathrm{_{sys}}$  &         r          &  PA$\mathrm{_{morph}}$ (r) &  \multicolumn{2}{c}{PA approaching}                 & \multicolumn{2}{c}{PA receding}                     & \multicolumn{2}{c}{$\Psi_{\mathrm{morph-kin}}$} & $\Psi_{\mathrm{kin-kin}}$ \\
        &                     &                    &                            &  PA$\mathrm{_{kin}}$  & $\delta$PA$\mathrm{_{kin}}$ & PA$\mathrm{_{kin}}$  & $\delta$PA$\mathrm{_{kin}}$  & approaching  & receding                         &                           \\
        &    (km s$^{-1}$)    &      (arcsec)      &            ($^{\circ}$)    &  ($^{\circ}$)         & ($^{\circ}$)                & ($^{\circ}$)         & ($^{\circ}$)                 & ($^{\circ}$) & ($^{\circ}$)                     & ($^{\circ}$)              \\
    (1) &         (4)         &        (5)         &            (7)             &          (8)          &            (9)              &          (10)        &       (11)                   &   (10)       &       (11)                       &   (10)                    \\
\midrule
 828&  4464 $\pm$    0 & 22 & 137.3 $\pm$  0.4 & 147.2 $\pm$ 1.1\phantom{0} & \phantom{00}9.6 $\pm$ 4.2\phantom{0} & 160.8 $\pm$ 1.4\phantom{0} & \phantom{0}13.0 $\pm$ 1.9\phantom{0} & \phantom{0} 9.9 $\pm$ 1.2\phantom{0} & \phantom{0}23.6 $\pm$ 1.4\phantom{0} & \phantom{0}13.7 $\pm$ 1.8\phantom{0} \\
 832& -- & -- & -- & -- & -- & -- & -- & -- & -- & -- \\
 833&  5952 $\pm$    5 & 20 & 135.6 $\pm$  0.9 & \phantom{0}29.5 $\pm$ 4.4\phantom{0} & \phantom{0}29.3 $\pm$ 7.7\phantom{0} & \phantom{0}31.6 $\pm$ 7.5\phantom{0} & \phantom{0}25.9 $\pm$ 6.4\phantom{0} & \phantom{0}73.9 $\pm$ 4.5\phantom{0} & \phantom{0}76.1 $\pm$ 7.6\phantom{0} & \phantom{00}2.2 $\pm$ 8.7\phantom{0} \\
 843& -- & -- & -- & -- & -- & -- & -- & -- * & -- * & -- * \\
 844& -- & -- & -- & -- & -- & -- & -- & -- & -- & -- \\
 846&  8258 $\pm$   14 &  8 & 105.2 $\pm$  0.6 & \phantom{0}12.6 $\pm$ 8.7\phantom{0} & \phantom{0}14.3 $\pm$ 6.7\phantom{0} & \phantom{00}2.9 $\pm$ 8.5\phantom{0} & \phantom{0}17.4 $\pm$ 7.3\phantom{0} & \phantom{0}87.4 $\pm$ 8.7\phantom{0} & \phantom{0}77.6 $\pm$ 8.5\phantom{0} & \phantom{00}9.8 $\pm$ 12.1 \\
 850&  6564 $\pm$    4 & 20 & 173.1 $\pm$  0.7 & 176.2 $\pm$ 2.6\phantom{0} & \phantom{00}7.0 $\pm$ 1.6\phantom{0} & 170.5 $\pm$ 1.5\phantom{0} & \phantom{00}6.2 $\pm$ 1.4\phantom{0} & \phantom{00}3.1 $\pm$ 2.7\phantom{0} & \phantom{00}2.7 $\pm$ 1.6\phantom{0} & \phantom{00}5.8 $\pm$ 3.0\phantom{0} \\
 852&  3066 $\pm$    3 & 12 & \phantom{0}99.4 $\pm$  0.6 & 155.1 $\pm$ 7.5\phantom{0} & \phantom{0}16.7 $\pm$ 6.9\phantom{0} & 166.1 $\pm$ 8.9\phantom{0} & \phantom{0}25.9 $\pm$ 7.2\phantom{0} & \phantom{0}55.6 $\pm$ 7.5\phantom{0} & \phantom{0}66.7 $\pm$ 8.9\phantom{0} & \phantom{0}11.1 $\pm$ 11.6 \\
 858&  7741 $\pm$   19 & 12 & 177.9 $\pm$  0.6 & 180.0 $\pm$ 0.0\phantom{0} & \phantom{00}6.5 $\pm$ 1.9\phantom{0} & 174.3 $\pm$ 3.5\phantom{0} & \phantom{0}12.0 $\pm$ 5.8\phantom{0} & \phantom{00}2.2 $\pm$ 0.6\phantom{0} & \phantom{00}3.5 $\pm$ 3.5\phantom{0} & \phantom{00}5.7 $\pm$ 3.5\phantom{0} \\
 860& -- & -- & -- & -- & -- & -- & -- & -- & -- & -- \\
 871&  5944 $\pm$    3 & 20 & 128.4 $\pm$  0.4 & 126.8 $\pm$ 1.1\phantom{0} & \phantom{00}5.7 $\pm$ 1.1\phantom{0} & 130.3 $\pm$ 1.6\phantom{0} & \phantom{00}7.0 $\pm$ 1.3\phantom{0} & \phantom{00}1.7 $\pm$ 1.1\phantom{0} & \phantom{00}1.8 $\pm$ 1.6\phantom{0} & \phantom{00}3.5 $\pm$ 1.9\phantom{0} \\
 873&  7720 $\pm$    3 & 10 & \phantom{0}15.2 $\pm$  1.4 & \phantom{0}66.6 $\pm$ 5.4\phantom{0} & \phantom{0}22.0 $\pm$ 5.3\phantom{0} & \phantom{0}69.9 $\pm$ 4.0\phantom{0} & \phantom{00}9.6 $\pm$ 2.8\phantom{0} & \phantom{0}51.4 $\pm$ 5.5\phantom{0} & \phantom{0}54.7 $\pm$ 4.2\phantom{0} & \phantom{00}3.3 $\pm$ 6.7\phantom{0} \\
 874&  4901 $\pm$    9 & 25 & \phantom{0}49.0 $\pm$  0.7 & \phantom{0}43.6 $\pm$ 2.4\phantom{0} & \phantom{0}18.0 $\pm$ 1.9\phantom{0} & \phantom{0}49.6 $\pm$ 2.5\phantom{0} & \phantom{0}13.8 $\pm$ 1.1\phantom{0} & \phantom{00}5.4 $\pm$ 2.5\phantom{0} & \phantom{00}0.6 $\pm$ 2.6\phantom{0} & \phantom{00}6.1 $\pm$ 3.5\phantom{0} \\
 877&  6281 $\pm$    0 & 13 & \phantom{0}38.5 $\pm$  0.6 & \phantom{0}36.6 $\pm$ 2.0\phantom{0} & \phantom{0}13.4 $\pm$ 4.8\phantom{0} & \phantom{0}43.1 $\pm$ 0.5\phantom{0} & \phantom{00}9.7 $\pm$ 1.9\phantom{0} & \phantom{00}1.9 $\pm$ 2.1\phantom{0} & \phantom{00}4.6 $\pm$ 0.8\phantom{0} & \phantom{00}6.5 $\pm$ 2.1\phantom{0} \\
 882& -- & -- & -- & -- & -- & -- & -- & -- & -- & -- \\
 883& -- & -- & -- & -- & -- & -- & -- & -- & -- & -- \\
 892&  4920 $\pm$    2 & 15 & \phantom{0}70.6 $\pm$  0.5 & \phantom{0}69.7 $\pm$ 2.5\phantom{0} & \phantom{00}7.0 $\pm$ 2.2\phantom{0} & \phantom{0}67.3 $\pm$ 2.2\phantom{0} & \phantom{00}6.0 $\pm$ 1.8\phantom{0} & \phantom{00}0.9 $\pm$ 2.6\phantom{0} & \phantom{00}3.4 $\pm$ 2.2\phantom{0} & \phantom{00}2.4 $\pm$ 3.3\phantom{0} \\
 900&  5048 $\pm$    7 &  5 & 166.9 $\pm$  2.8 & \phantom{0}68.5 $\pm$ 12.8 & \phantom{0}20.2 $\pm$ 10.1 & \phantom{0}64.6 $\pm$ 14.5 & \phantom{0}19.5 $\pm$ 11.0 & \phantom{0}81.6 $\pm$ 13.1 & \phantom{0}77.7 $\pm$ 14.7 & \phantom{00}4.0 $\pm$ 19.3 \\
 901&  4651 $\pm$    0 & 15 & 139.4 $\pm$  1.3 & \phantom{00}7.0 $\pm$ 2.7\phantom{0} & \phantom{0}14.7 $\pm$ 2.4\phantom{0} & \phantom{00}7.2 $\pm$ 1.0\phantom{0} & \phantom{00}8.4 $\pm$ 3.1\phantom{0} & \phantom{0}47.6 $\pm$ 3.0\phantom{0} & \phantom{0}47.8 $\pm$ 1.6\phantom{0} & \phantom{00}0.2 $\pm$ 2.9\phantom{0} \\
 903& -- & -- & -- & -- & -- & -- & -- & -- & -- & -- \\
 905& -- & -- & -- & -- & -- & -- & -- & -- & -- & -- \\
 907&  3478 $\pm$    1 & 15 & \phantom{0}19.4 $\pm$  0.6 & \phantom{0}17.4 $\pm$ 3.2\phantom{0} & \phantom{0}12.4 $\pm$ 3.6\phantom{0} & \phantom{0}21.5 $\pm$ 3.5\phantom{0} & \phantom{0}11.7 $\pm$ 4.8\phantom{0} & \phantom{00}2.0 $\pm$ 3.2\phantom{0} & \phantom{00}2.1 $\pm$ 3.6\phantom{0} & \phantom{00}4.1 $\pm$ 4.7\phantom{0} \\
 913&  1612 $\pm$    0 & 23 & \phantom{0}27.8 $\pm$  1.2 & \phantom{0}38.2 $\pm$ 2.5\phantom{0} & \phantom{0}20.7 $\pm$ 2.0\phantom{0} & \phantom{0}71.1 $\pm$ 0.5\phantom{0} & \phantom{0}37.1 $\pm$ 1.0\phantom{0} & \phantom{0}10.3 $\pm$ 2.8\phantom{0} & \phantom{0}43.3 $\pm$ 1.3\phantom{0} & \phantom{0}33.0 $\pm$ 2.6\phantom{0} \\
 915&  4259 $\pm$    2 & 20 & 160.3 $\pm$  3.2 & 164.6 $\pm$ 2.0\phantom{0} & \phantom{00}5.5 $\pm$ 1.2\phantom{0} & 177.6 $\pm$ 3.0\phantom{0} & \phantom{0}10.9 $\pm$ 2.1\phantom{0} & \phantom{00}4.3 $\pm$ 3.7\phantom{0} & \phantom{0}17.4 $\pm$ 4.4\phantom{0} & \phantom{0}13.1 $\pm$ 3.6\phantom{0} \\
 916& -- & -- & -- & -- & -- & -- & -- & -- & -- & -- \\
 922&  5212 $\pm$    1 & 15 & \phantom{0}91.7 $\pm$  0.7 & 118.2 $\pm$ 3.0\phantom{0} & \phantom{0}11.9 $\pm$ 1.4\phantom{0} & 109.0 $\pm$ 1.9\phantom{0} & \phantom{0}10.8 $\pm$ 1.3\phantom{0} & \phantom{0}26.5 $\pm$ 3.1\phantom{0} & \phantom{0}17.4 $\pm$ 2.0\phantom{0} & \phantom{00}9.1 $\pm$ 3.6\phantom{0} \\
 923& -- & -- & -- & -- & -- & -- & -- & -- & -- & -- \\
 925&  4021 $\pm$   11 & 15 & 129.4 $\pm$ 11.7 & 139.4 $\pm$ 2.0\phantom{0} & \phantom{00}6.7 $\pm$ 1.5\phantom{0} & 142.6 $\pm$ 2.2\phantom{0} & \phantom{00}7.3 $\pm$ 2.0\phantom{0} & \phantom{0}10.0 $\pm$ 11.9 & \phantom{0}13.2 $\pm$ 11.9 & \phantom{00}3.3 $\pm$ 2.9\phantom{0} \\
 927&  6722 $\pm$    0 & 15 & \phantom{0}43.2 $\pm$  0.7 & \phantom{0}45.9 $\pm$ 2.1\phantom{0} & \phantom{0}20.3 $\pm$ 1.8\phantom{0} & \phantom{0}47.9 $\pm$ 2.3\phantom{0} & \phantom{0}12.6 $\pm$ 5.1\phantom{0} & \phantom{00}2.7 $\pm$ 2.2\phantom{0} & \phantom{00}4.6 $\pm$ 2.4\phantom{0} & \phantom{00}1.9 $\pm$ 3.1\phantom{0} \\
 932& -- & -- & -- & -- & -- & -- & -- & -- & -- & -- \\
 935&  4674 $\pm$    2 & 25 & \phantom{0}80.4 $\pm$  0.4 & \phantom{0}99.0 $\pm$ 3.6\phantom{0} & \phantom{0}17.9 $\pm$ 5.6\phantom{0} & 104.3 $\pm$ 4.1\phantom{0} & \phantom{0}17.7 $\pm$ 2.7\phantom{0} & \phantom{0}18.6 $\pm$ 3.6\phantom{0} & \phantom{0}24.0 $\pm$ 4.1\phantom{0} & \phantom{00}5.4 $\pm$ 5.4\phantom{0} \\
 939&  6540 $\pm$    2 & 10 & \phantom{0}33.2 $\pm$  0.8 & 154.8 $\pm$ 5.8\phantom{0} & \phantom{0}20.4 $\pm$ 4.5\phantom{0} & 165.1 $\pm$ 7.2\phantom{0} & \phantom{0}17.1 $\pm$ 4.1\phantom{0} & \phantom{0}58.3 $\pm$ 5.9\phantom{0} & \phantom{0}48.1 $\pm$ 7.3\phantom{0} & \phantom{0}10.2 $\pm$ 9.3\phantom{0} \\
1014&  4332 $\pm$    2 & 12 & \phantom{0}12.6 $\pm$  6.1 & 177.7 $\pm$ 5.8\phantom{0} & \phantom{0}14.7 $\pm$ 2.7\phantom{0} & 159.6 $\pm$ 10.3 & \phantom{0}19.9 $\pm$ 6.8\phantom{0} & \phantom{0}15.0 $\pm$ 8.4\phantom{0} & \phantom{0}33.1 $\pm$ 12.0 & \phantom{0}18.1 $\pm$ 11.8 \\
1017& -- & -- & -- & -- & -- & -- & -- & -- & -- & -- \\
1022&  4605 $\pm$    3 &  5 & \phantom{0}10.0 $\pm$  0.0 & \phantom{0}57.7 $\pm$ 20.5 & \phantom{0}33.3 $\pm$ 11.5 & \phantom{0}59.3 $\pm$ 24.2 & \phantom{0}35.3 $\pm$ 23.8 & \phantom{0}47.7 $\pm$ 20.5 & \phantom{0}49.3 $\pm$ 24.2 & \phantom{00}1.6 $\pm$ 31.7 \\
1026&  4092 $\pm$    1 & 15 & \phantom{0}68.0 $\pm$  9.6 & \phantom{0}29.3 $\pm$ 23.1 & \phantom{0}56.4 $\pm$ 9.3\phantom{0} & \phantom{0}75.7 $\pm$ 10.7 & \phantom{0}29.1 $\pm$ 5.9\phantom{0} & \phantom{0}38.8 $\pm$ 25.1 & \phantom{00}7.6 $\pm$ 14.4 & \phantom{0}46.4 $\pm$ 25.5 \\
1042&  5927 $\pm$    3 & 17 & \phantom{0}21.1 $\pm$  1.8 & \phantom{0}20.2 $\pm$ 2.8\phantom{0} & \phantom{00}9.4 $\pm$ 2.0\phantom{0} & \phantom{0}25.9 $\pm$ 4.0\phantom{0} & \phantom{0}15.4 $\pm$ 3.4\phantom{0} & \phantom{00}0.9 $\pm$ 3.4\phantom{0} & \phantom{00}4.8 $\pm$ 4.4\phantom{0} & \phantom{00}5.7 $\pm$ 4.9\phantom{0} \\
1044& -- & -- & -- & -- & -- & -- & -- & -- & -- & -- \\
1340& -- & -- & -- & -- & -- & -- & -- & -- * & -- * & -- * \\
1360& -- & -- & -- & -- & -- & -- & -- & -- & -- & -- \\
1520&  6567 $\pm$    3 &  7 & 125.0 $\pm$  0.0 & 104.9 $\pm$ 11.2 & \phantom{0}27.4 $\pm$ 12.3 & \phantom{0}57.7 $\pm$ 15.5 & \phantom{0}32.0 $\pm$ 16.2 & \phantom{0}20.1 $\pm$ 11.2 & \phantom{0}67.3 $\pm$ 15.5 & \phantom{0}47.2 $\pm$ 19.1 \\
1589& -- & -- & -- & -- & -- & -- & -- & -- & -- & -- \\
1663&  7377 $\pm$    0 & 15 & 107.5 $\pm$  4.2 & \phantom{0}99.0 $\pm$ 9.5\phantom{0} & \phantom{0}44.3 $\pm$ 10.2 & 113.8 $\pm$ 7.7\phantom{0} & \phantom{0}28.9 $\pm$ 12.3 & \phantom{00}8.5 $\pm$ 10.4 & \phantom{00}6.3 $\pm$ 8.8\phantom{0} & \phantom{0}14.8 $\pm$ 12.2 \\
1740& -- & -- & -- & -- & -- & -- & -- & -- * & -- * & -- * \\
1780& -- & -- & -- & -- & -- & -- & -- & -- & -- & -- \\
1781&  8626 $\pm$    2 & 12 & \phantom{0}43.6 $\pm$  1.0 & \phantom{00}0.9 $\pm$ 16.5 & \phantom{0}49.2 $\pm$ 10.6 & \phantom{0}18.4 $\pm$ 20.1 & \phantom{0}55.8 $\pm$ 23.3 & \phantom{0}42.7 $\pm$ 16.5 & \phantom{0}25.2 $\pm$ 20.1 & \phantom{0}17.5 $\pm$ 26.0 \\
1795&  2480 $\pm$    0 & 10 & \phantom{0}45.2 $\pm$  4.2 & \phantom{0}97.0 $\pm$ 0.2\phantom{0} & \phantom{0}32.0 $\pm$ 2.8\phantom{0} & \phantom{0}70.1 $\pm$ 0.2\phantom{0} & \phantom{0}20.2 $\pm$ 2.8\phantom{0} & \phantom{0}51.8 $\pm$ 4.2\phantom{0} & \phantom{0}24.9 $\pm$ 4.2\phantom{0} & \phantom{0}26.9 $\pm$ 0.3\phantom{0} \\
1796&  5377 $\pm$   11 & 10 & \phantom{0}26.5 $\pm$  0.6 & \phantom{0}27.8 $\pm$ 4.6\phantom{0} & \phantom{0}12.6 $\pm$ 4.0\phantom{0} & \phantom{0}26.1 $\pm$ 5.0\phantom{0} & \phantom{0}22.0 $\pm$ 6.7\phantom{0} & \phantom{00}1.3 $\pm$ 4.6\phantom{0} & \phantom{00}0.3 $\pm$ 5.0\phantom{0} & \phantom{00}1.6 $\pm$ 6.8\phantom{0} \\
1801&  1891 $\pm$    1 & 20 & \phantom{00}1.7 $\pm$  0.9 & 177.3 $\pm$ 0.5\phantom{0} & \phantom{0}15.4 $\pm$ 1.8\phantom{0} & 174.0 $\pm$ 1.6\phantom{0} & \phantom{0}16.1 $\pm$ 4.9\phantom{0} & \phantom{00}4.4 $\pm$ 1.0\phantom{0} & \phantom{00}7.7 $\pm$ 1.8\phantom{0} & \phantom{00}3.3 $\pm$ 1.6\phantom{0} \\
1871&  6155 $\pm$    5 & 20 & 139.0 $\pm$  0.0 & 158.1 $\pm$ 3.2\phantom{0} & \phantom{0}22.4 $\pm$ 5.2\phantom{0} & 155.0 $\pm$ 2.9\phantom{0} & \phantom{0}15.5 $\pm$ 3.6\phantom{0} & \phantom{0}19.1 $\pm$ 3.2\phantom{0} & \phantom{0}16.0 $\pm$ 2.9\phantom{0} & \phantom{00}3.1 $\pm$ 4.3\phantom{0} \\
1873&  7849 $\pm$    0 &  8 & \phantom{0}55.8 $\pm$ 10.3 & 133.7 $\pm$ 1.0\phantom{0} & \phantom{0}16.1 $\pm$ 4.7\phantom{0} & 164.5 $\pm$ 2.8\phantom{0} & \phantom{0}18.2 $\pm$ 2.0\phantom{0} & \phantom{0}77.9 $\pm$ 10.4 & \phantom{0}71.2 $\pm$ 10.7 & \phantom{0}30.9 $\pm$ 3.0\phantom{0} \\
\bottomrule
\end{tabular}
\end{center}
\begin{minipage}{\textwidth}
\end{minipage}
\end{table*}

\clearpage
  \begin{table}[!htb]
\caption{\label{table_bothkin} Kinematic (mis)alignment between the stellar and the ionised gas components for the galaxies included in this study.
} 
\begin{center}
\renewcommand{\thefootnote}{\alph{footnote}}
\begin{tabular} {c c c }
\toprule
    id  & \multicolumn{2}{c}{$\Psi_{\mathrm{gas-star}}$} \\
        &     approaching     &   receding               \\
        & ($^{\circ}$)             & ($^{\circ}$)        \\
    (1) &   (2)                   &   (3)                \\
\midrule
  14 & -- & -- \\
  17 & \phantom{0} 54.9 $\pm$ 22.6 & \phantom{0} 49.3 $\pm$ 25.4 \\
  22 & \phantom{00} 6.4 $\pm$ 2.2\phantom{0} & \phantom{0} 10.2 $\pm$ 4.9\phantom{0} \\
  26 & \phantom{00} 4.6 $\pm$ 3.2\phantom{0} & \phantom{00} 2.8 $\pm$ 3.0\phantom{0} \\
  39 & -- & -- \\
  42 & \phantom{00} 7.4 $\pm$ 6.2\phantom{0} & \phantom{00} 8.9 $\pm$ 6.9\phantom{0} \\
  44 & -- & -- \\
 119 & \phantom{0} 20.2 $\pm$ 11.5 & \phantom{00} 9.8 $\pm$ 13.7 \\
 127 & \phantom{0} 15.9 $\pm$ 13.1 & \phantom{00} 4.4 $\pm$ 7.9\phantom{0} \\
 155 & \phantom{00} 0.1 $\pm$ 4.4\phantom{0} & \phantom{00} 2.7 $\pm$ 5.0\phantom{0} \\
 156 & \phantom{00} 1.4 $\pm$ 3.1\phantom{0} & \phantom{00} 3.5 $\pm$ 3.0\phantom{0} \\
 165 & \phantom{00} 4.3 $\pm$ 2.6\phantom{0} & \phantom{0} 10.2 $\pm$ 3.0\phantom{0} \\
 186 & \phantom{00} 2.5 $\pm$ 2.6\phantom{0} & \phantom{00} 4.8 $\pm$ 3.0\phantom{0} \\
 213 & \phantom{00} 5.4 $\pm$ 11.6 & \phantom{00} 2.4 $\pm$ 18.3 \\
 274 & \phantom{0} 18.7 $\pm$ 5.8\phantom{0} & \phantom{00} 1.1 $\pm$ 6.8\phantom{0} \\
 311 & \phantom{0} 11.7 $\pm$ 9.3\phantom{0} & \phantom{0} 19.3 $\pm$ 7.6\phantom{0} \\
 314 & \phantom{00} 4.9 $\pm$ 3.3\phantom{0} & \phantom{00} 8.1 $\pm$ 3.4\phantom{0} \\
 340 &  181.2 $\pm$ 11.6 &  170.8 $\pm$ 20.2 \\
 360 & -- & -- \\
 475 & -- & -- \\
 479 & \phantom{00} 6.4 $\pm$ 3.8\phantom{0} & \phantom{0} 18.0 $\pm$ 5.6\phantom{0} \\
 486 & \phantom{00} 4.8 $\pm$ 12.3 & \phantom{0} 10.0 $\pm$ 11.1 \\
 520 & \phantom{0} 20.5 $\pm$ 11.0 & \phantom{0} 17.0 $\pm$ 13.8 \\
 577 & \phantom{00} 1.7 $\pm$ 6.3\phantom{0} & \phantom{00} 2.9 $\pm$ 7.6\phantom{0} \\
 589 & -- & -- \\
 593 & \phantom{00} 4.9 $\pm$ 2.4\phantom{0} & \phantom{00} 1.3 $\pm$ 3.3\phantom{0} \\
 633 & \phantom{0} 76.0 $\pm$ 13.9 & \phantom{0} 40.1 $\pm$ 8.3\phantom{0} \\
 634 & \phantom{0} 10.6 $\pm$ 3.1\phantom{0} & \phantom{00} 0.8 $\pm$ 4.1\phantom{0} \\
 663 & \phantom{00} 1.9 $\pm$ 2.4\phantom{0} & \phantom{00} 1.9 $\pm$ 2.4\phantom{0} \\
 676 & \phantom{0} 27.0 $\pm$ 14.8 & \phantom{00} 8.6 $\pm$ 23.3 \\
 680 & -- & -- \\
 740 & \phantom{0} 10.0 $\pm$ 7.2\phantom{0} & \phantom{0} 16.3 $\pm$ 8.2\phantom{0} \\
 746 & \phantom{0} 90.7 $\pm$ 5.9\phantom{0} & \phantom{0} 98.0 $\pm$ 5.7\phantom{0} \\
 758 & -- & -- \\
 767 & -- & -- \\
 769 & \phantom{00} 0.7 $\pm$ 5.3\phantom{0} & \phantom{00} 8.5 $\pm$ 7.2\phantom{0} \\
 770 &  105.1 $\pm$ 5.9\phantom{0} & \phantom{0} 56.6 $\pm$ 9.2\phantom{0} \\
 778 & \phantom{00} 8.0 $\pm$ 4.9\phantom{0} & \phantom{00} 4.5 $\pm$ 3.8\phantom{0} \\
 780 & -- & -- \\
 781 & -- & -- \\
 785 & -- & -- \\
 795 & \phantom{00} 3.9 $\pm$ 5.7\phantom{0} & \phantom{00} 8.6 $\pm$ 5.0\phantom{0} \\
 796 & \phantom{0} 15.4 $\pm$ 3.5\phantom{0} & \phantom{00} 3.0 $\pm$ 3.6\phantom{0} \\
 797 & \phantom{00} 8.9 $\pm$ 4.2\phantom{0} & \phantom{0} 12.9 $\pm$ 7.5\phantom{0} \\
 801 & \phantom{00} 6.9 $\pm$ 9.9\phantom{0} & \phantom{0} 13.8 $\pm$ 17.1 \\
 802 & \phantom{00} 2.9 $\pm$ 20.1 & \phantom{0} 17.5 $\pm$ 19.4 \\
 803 & \phantom{00} 0.6 $\pm$ 7.8\phantom{0} & \phantom{0} 18.9 $\pm$ 22.4 \\
 806 & \phantom{0} 68.0 $\pm$ 6.2\phantom{0} & \phantom{0} 69.1 $\pm$ 7.9\phantom{0} \\
 807 & \phantom{00} 7.3 $\pm$ 4.8\phantom{0} & \phantom{00} 0.5 $\pm$ 4.4\phantom{0} \\
 816 & \phantom{00} 4.6 $\pm$ 5.1\phantom{0} & \phantom{00} 2.3 $\pm$ 8.0\phantom{0} \\
 822 & \phantom{0} 21.6 $\pm$ 6.6\phantom{0} & \phantom{00} 2.4 $\pm$ 6.8\phantom{0} \\
\bottomrule
\end{tabular}
\tablefoot{
(1) CALIFA id.
(2) and (3) Kinematic misalignment between the stellar and ionised gas kinematic PA for the approaching and receding sides.
} 
\end{center}
\begin{minipage}{\textwidth}
\end{minipage}
\end{table} 
   
\begin{table}[!htb]
\addtocounter{table}{-1}
\caption{continue \ref{table_bothkin}}
\begin{center}
\renewcommand{\thefootnote}{\alph{footnote}}
\begin{tabular} {c c c }
\toprule
    id  & \multicolumn{2}{c}{$\Psi_{\mathrm{gas-star}}$} \\
        &     approaching     &   receding               \\
        & ($^{\circ}$)             & ($^{\circ}$)        \\
    (1) &   (2)                   &   (3)                \\
\midrule
 828 & -- & -- \\
 832 & -- & -- \\
 833 & \phantom{00} 8.3 $\pm$ 5.3\phantom{0} & \phantom{0} 20.3 $\pm$ 10.7 \\
 843 & -- & -- \\
 844 & -- & -- \\
 846 & -- & -- \\
 850 & \phantom{00} 2.4 $\pm$ 4.3\phantom{0} & \phantom{00} 3.9 $\pm$ 2.9\phantom{0} \\
 852 & -- & -- \\
 858 & \phantom{00} 8.8 $\pm$ 3.3\phantom{0} & \phantom{00} 7.6 $\pm$ 4.1\phantom{0} \\
 860 & -- & -- \\
 871 & \phantom{00} 2.2 $\pm$ 2.5\phantom{0} & \phantom{00} 5.2 $\pm$ 3.2\phantom{0} \\
 873 & \phantom{0} 10.0 $\pm$ 9.9\phantom{0} & \phantom{0} 10.7 $\pm$ 7.8\phantom{0} \\
 874 & \phantom{00} 1.8 $\pm$ 3.0\phantom{0} & \phantom{0}  9.9 $\pm$ 4.3\phantom{0} \\
 877 & \phantom{00} 4.5 $\pm$ 5.1\phantom{0} & \phantom{00} 4.7 $\pm$ 3.7\phantom{0} \\
 882 & -- & -- \\
 883 & -- & -- \\
 892 & \phantom{00} 0.3 $\pm$ 5.6\phantom{0} & \phantom{00} 4.0 $\pm$ 3.8\phantom{0} \\
 900 & -- & -- \\
 901 & \phantom{00} 8.5 $\pm$ 4.6\phantom{0} & \phantom{0} 15.2 $\pm$ 4.9\phantom{0} \\
 903 & -- & -- \\
 905 & -- & -- \\
 907 & \phantom{00} 5.9 $\pm$ 6.6\phantom{0} & \phantom{00} 1.6 $\pm$ 7.2\phantom{0} \\
 913 & \phantom{00} 4.8 $\pm$ 5.8\phantom{0} & \phantom{0} 32.3 $\pm$ 6.8\phantom{0} \\
 915 & \phantom{00} 4.7 $\pm$ 5.8\phantom{0} & \phantom{00} 3.6 $\pm$ 5.2\phantom{0} \\
 916 & -- & -- \\
 922 & \phantom{0} 25.1 $\pm$ 5.9\phantom{0} & \phantom{0} 18.2 $\pm$ 6.9\phantom{0} \\
 923 & -- & -- \\
 925 & \phantom{00} 0.1 $\pm$ 5.0\phantom{0} & \phantom{00} 4.5 $\pm$ 6.2\phantom{0} \\
 927 & \phantom{0} 28.8 $\pm$ 4.6\phantom{0} & \phantom{0} 19.7 $\pm$ 5.7\phantom{0} \\
 932 & -- & -- \\
 935 & -- & -- \\
 939 & \phantom{00} 9.8 $\pm$ 11.1 & \phantom{0} 14.4 $\pm$ 9.6\phantom{0} \\
1014 & -- & -- \\
1017 & -- & -- \\
1022 & \phantom{0} 19.4 $\pm$ 22.3 & \phantom{0} 39.7 $\pm$ 27.7 \\
1026 & -- & -- \\
1042 & -- & -- \\
1044 & -- & -- \\
1340 & -- & -- \\
1360 & -- & -- \\
1520 & \phantom{0} 24.4 $\pm$ 17.3 & \phantom{0} 24.0 $\pm$ 21.3 \\
1589 & -- & -- \\
1663 & -- & -- \\
1740 & -- & -- \\
1780 & -- & -- \\
1781 & \phantom{0} 19.8 $\pm$ 20.9 & \phantom{00} 0.7 $\pm$ 21.1 \\
1795 &  141.9 $\pm$ 9.0\phantom{0} &  155.9 $\pm$ 8.1\phantom{0} \\
1796 & \phantom{00} 2.0 $\pm$ 5.8\phantom{0} & \phantom{00} 2.9 $\pm$ 6.4\phantom{0} \\
1801 & \phantom{0} 11.1 $\pm$ 4.2\phantom{0} & \phantom{00} 8.1 $\pm$ 4.9\phantom{0} \\
1871 & \phantom{00} 6.6 $\pm$ 4.8\phantom{0} & \phantom{00} 0.9 $\pm$ 4.2\phantom{0} \\
1873 & \phantom{00} 2.3 $\pm$ 20.6 & \phantom{0} 21.6 $\pm$ 21.6 \\
\bottomrule
\end{tabular}
\end{center}
\end{table}

  \label{table_bothkin}
 \clearpage
 \section{Velocity fields}
\label{sec:maps}
\begin{figure*}[!htb]
  \minipage{0.33\textwidth}
   \includegraphics[width=\linewidth]{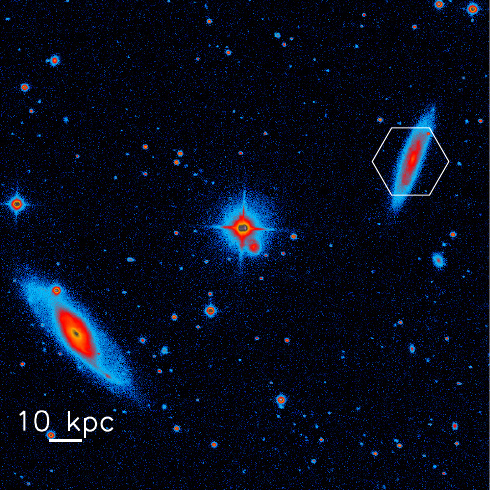}
 \endminipage
  \minipage{0.66\textwidth}
  \includegraphics[width=\linewidth]{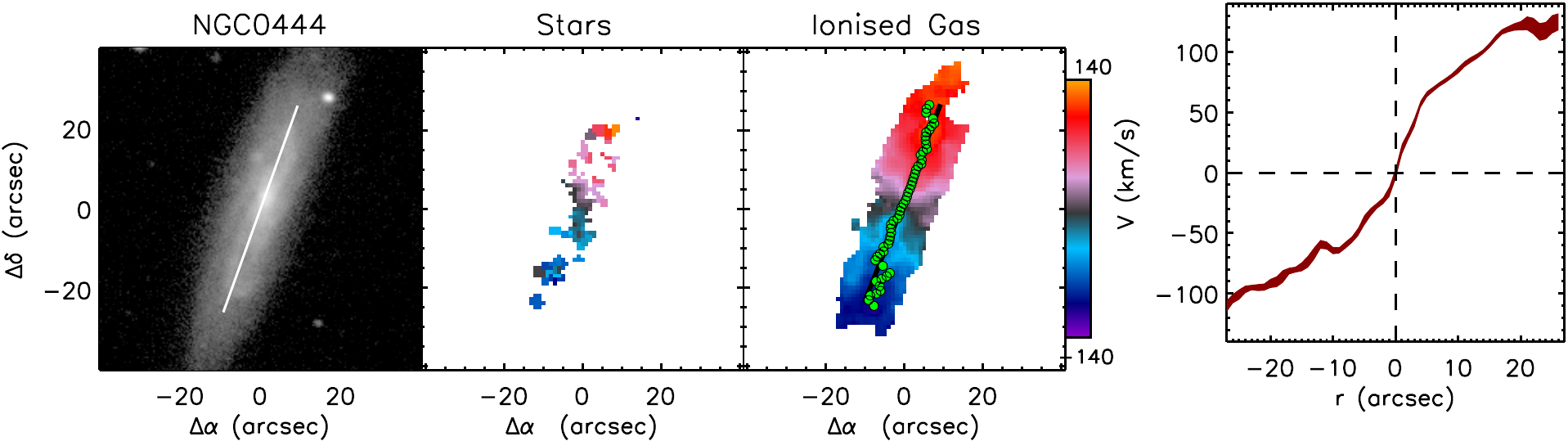}
 \endminipage
  \caption{\label{Early_fig}Objects in the pre-merger stage. Top: UGC~312 and UGC~312NOTES01. Middle:  NGC~192 and NGC~197. Bottom: NGC~444.} 
 \end{figure*}

In this appendix we present the stellar and ionized gas kinematic maps of the sample used in this study. We group from Figs.\,\ref{Early_fig} to B.5 the the galaxies in the different interaction stages described in section \ref{sec:stage}. For each interacting galaxy (system) we show from
left to right: SDSS $r$-band image. For post-merger and merger remnants we enhance the contrast of the images to highlight their morphological features. The white hexagon represents the area covered by the CALIFA FoV. The white line in the bottom left corner represents the 10 kpc scale. In systems where both objects are observed, two hexagons are plotted. The next panel shows the SDSS $r$-band image at the same area for which the kinematic velocity distributions has been observed. The white line represents the morphological PA. The next two panels show the stellar and the ionized gas velocity fields. The green dots in each of these panels represent the position of maximum velocity for that galactocentric distance. The black lines represent the mean kinematic PA derived from these points (see details in section \ref{sec:Analysis}). The last panel shows the velocity curve derived directly from the velocity field (see details in section \ref{sec:Analysis}); the blue and red curves represent the stellar and the ionized gas components, respectively. The thickness of the lines represents the uncertainty in the determination of the maximum velocity for each galactocentric distance by means of Monte Carlo simulations (see details in section \ref{sec:Analysis}). We note that
for galaxies where the kinematic PA was not possible to determine in a given component, neither the velocity curves nor the positions are plotted in the velocity fields.
\clearpage
\begin{figure*}[!htb]
  \minipage{0.33\textwidth}
  \includegraphics[width=\linewidth]{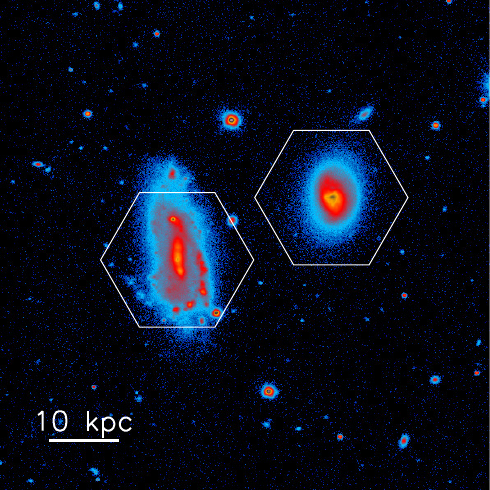}
  \endminipage
  \minipage{0.66\textwidth}
  \includegraphics[width=\linewidth]{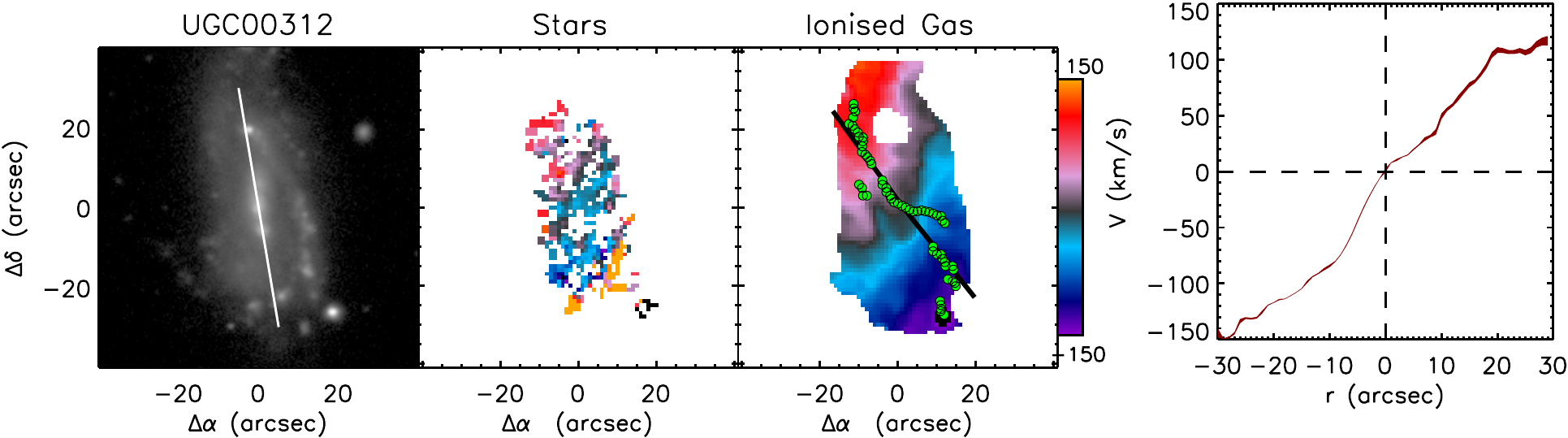}  
   \includegraphics[width=\linewidth]{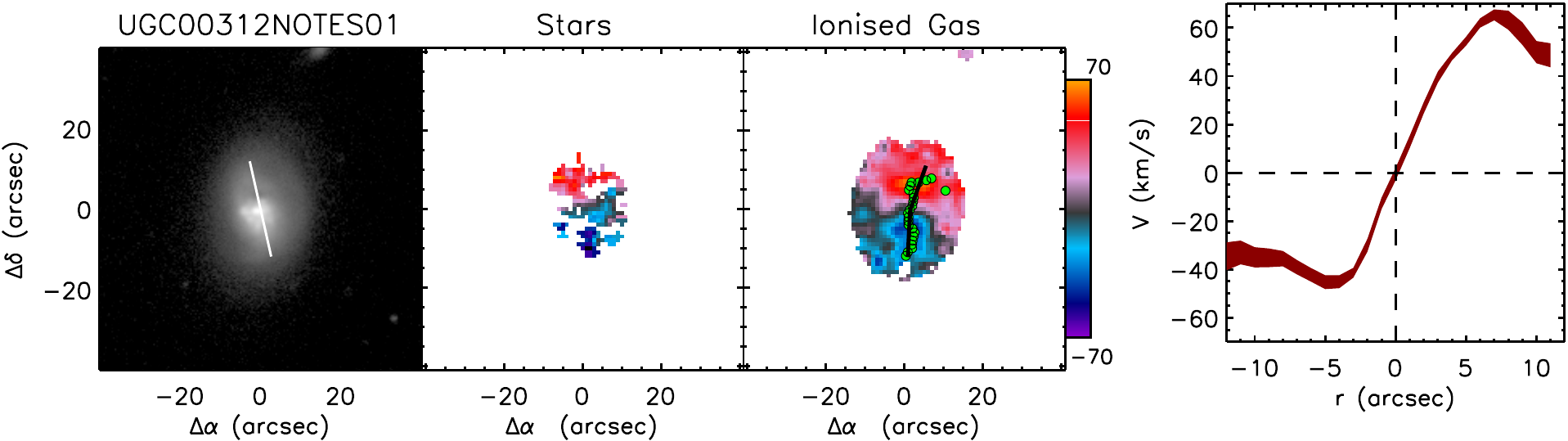}   
  \endminipage
  \end{figure*}
\begin{figure*}[!htb]
  \minipage{0.33\textwidth}
   \includegraphics[width=\linewidth]{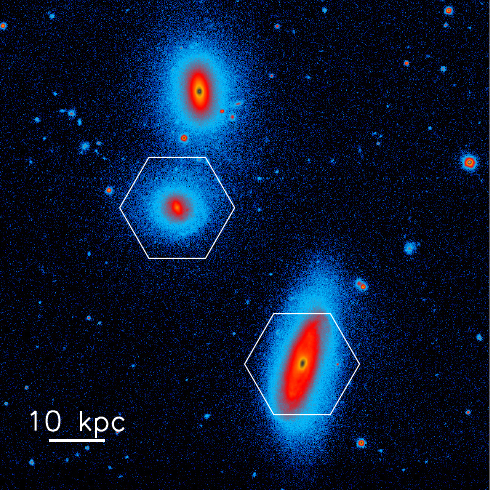}
 \endminipage
  \minipage{0.66\textwidth}
  \includegraphics[width=\linewidth]{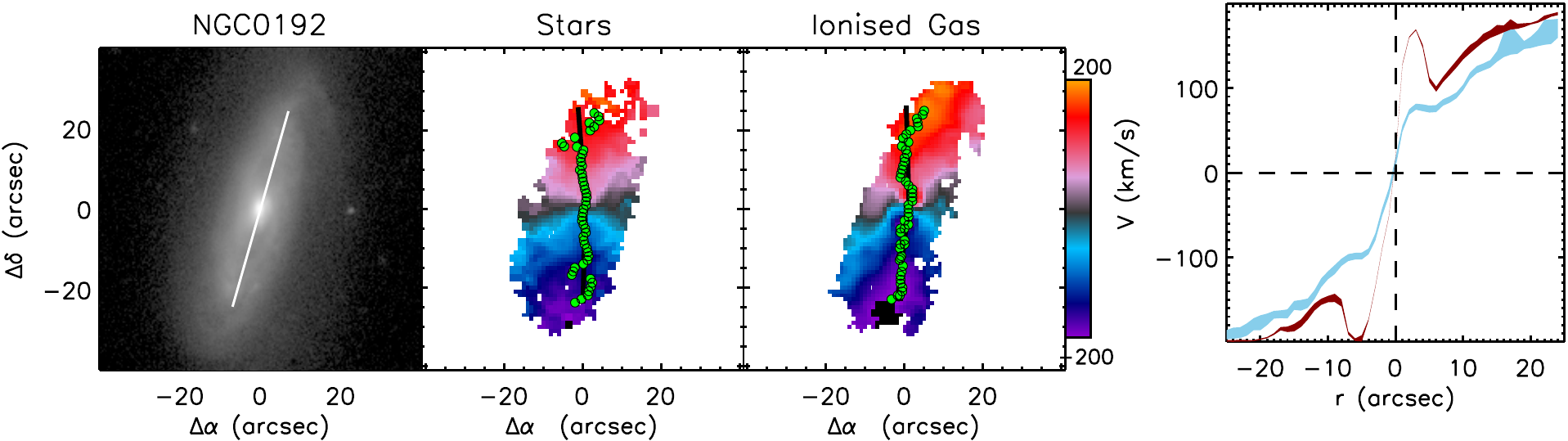}
  \includegraphics[width=\linewidth]{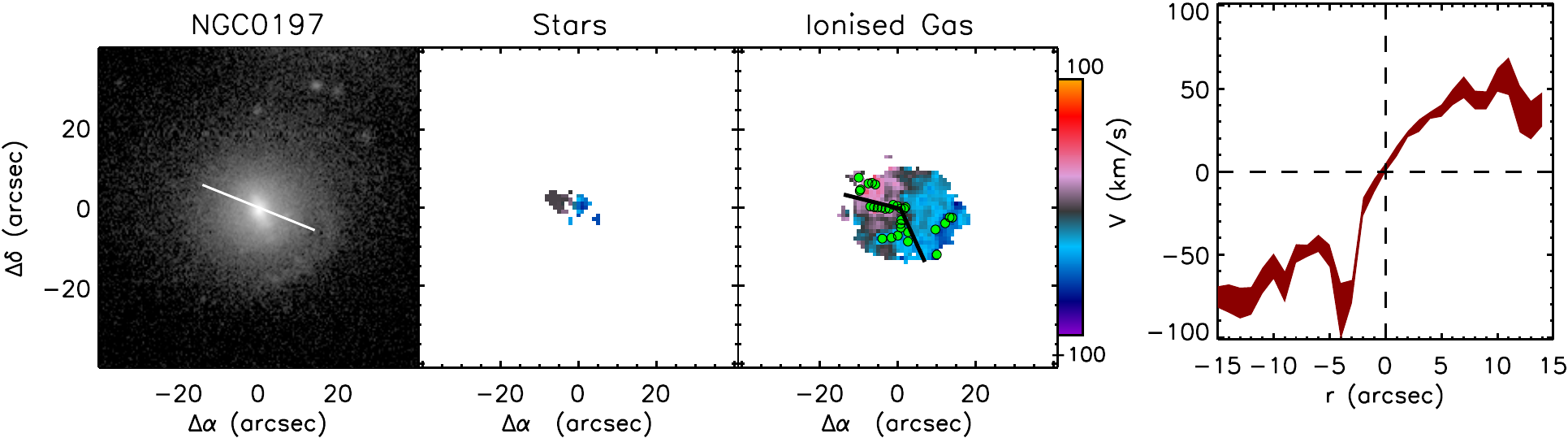}
 \endminipage
 \end{figure*}

\end{appendix}

\end{document}